\documentclass[preprint]{aastex}
\usepackage{epsf}

\def\kms{{\rm km\ s^{-1}}}
\def\cm{{\rm cm}}
\def\K{{\rm K}}
\def\msun{{\rm M}_\odot\ }
\def\pc{{\rm pc}}
\newcommand {\apgt} {\ {\raise-.5ex\hbox{$\buildrel>\over\sim$}}\ }
\newcommand {\aplt} {\ {\raise-.5ex\hbox{$\buildrel<\over\sim$}}\ } 

\def\kB{k_{\rm B}}

\shortauthors{Koyama \& Ostriker}
\begin{document}
\title{Gas Properties and Implications for Galactic Star Formation 
in Numerical Models of 
the Turbulent, Multiphase ISM}
\author{Hiroshi Koyama\altaffilmark{1,2} and Eve C. Ostriker\altaffilmark{1}}
\altaffiltext{1}{Department of Astronomy,
University of Maryland, College Park, MD 20742, USA;
hkoyama@astro.umd.edu, ostriker@astro.umd.edu} 
\altaffiltext{2}{Current address: High-Performance Computing Team,
Integrated Simulation of Living Matter Group, RIKEN,
61-1 Ono-cho, Tsurumi, Yokohama, 230-0046 Japan; hkoyama@riken.jp}

\begin{abstract}
Using 
numerical simulations of galactic disks that resolve scales from
$\sim 1$ to several hundred pc, we investigate dynamical properties of
the multiphase interstellar medium in which turbulence is
driven by feedback from star formation.  We focus on effects of HII
regions by implementing a recipe for intense heating confined within
dense, self-gravitating regions.  Our models are
two-dimensional, representing radial-vertical slices through the disk,
and include sheared
background rotation of the gas, vertical stratification, heating and
cooling to yield temperatures $T\sim 10-10^4\K$, and conduction that
resolves thermal instabilities on our numerical grid.  Each simulation
evolves to reach a quasi-steady state, for which we analyze the
time-averaged properties of the gas. In our suite of models, three
parameters (the gas surface density $\Sigma$, the stellar volume
density $\rho_*$, and the local angular rotation rate $\Omega$) are
separately controlled in order to explore environmental dependencies.
Among other statistical measures, we evaluate turbulent amplitudes,
virial ratios, Toomre $Q$ parameters including turbulence, and the mass
fractions at different densities.  We find that the dense gas ($n>100\
\cm^{-3}$) has turbulence levels similar to those observed in giant
molecular clouds and virial ratios $\sim 1-2$.  Our models show that
the Toomre $Q$ parameter in the dense gas evolves to values near
unity; this demonstrates self-regulation via turbulent feedback.  
We also test how the
star formation rate $\Sigma_{\rm SFR}$ depends on $\Sigma$, $\rho_*$,
and $\Omega$.  Under the assumption that the star formation rate is
proportional to the amount of gas at densities above a threshold
$n_{\rm th}$ divided by the free-fall time at that threshold, we find
that $\Sigma_{\rm SFR} \propto \Sigma ^{1+p}$ with $1+p\sim 1.2-1.4$
when $n_{\rm th}=10^2$ or $10^3 \cm^{-3}$, consistent with observed
Kennicutt-Schmidt relations.  Estimates of star formation rates based
on large-scale properties (the orbital time, the Jeans time, or the
free-fall time at the mean density within a scale height),
however, depart from rates computed using the measured amount of dense
gas, indicating that resolving the ISM structure in galactic disks is necessary
to obtain accurate predictions of the star formation rate.
\end{abstract}

\section{Introduction}

The interstellar medium (ISM) is commonly envisioned as a
self-regulating system in statistical quasi-equilibrium.  Multiple
components of gas with varying densities and temperatures coexist
\citep{1969ApJ...155L.149F,1974ApJ...189L.105C,1977ApJ...218..148M},
animated by turbulence that pervades the whole volume
\citep{2004ARA&A..42..211E}.  
Different components of gas play different roles in the ISM ecosystem,
with the coldest and densest portions responsible for star formation.
Massive stars, when they are born, energize the ISM through the HII
regions and supernova blasts they create \citep{1978ppim.book.....S};
this energy input is important in replenishing continual losses
through turbulent dissipation.  UV radiation from young massive stars
is also crucial in heating the gas.  The rate of star formation is
determined by the available supply of dense gas, which in turn is
regulated by the interplay between dynamics and thermodynamics in the
ISM, and is affected by the galactic environment in which the ISM is
contained \citep{2004RvMP...76..125M,2007ARA&A..45..565M}.  While this
overall framework is generally accepted and is supported by existing theory
and observations, much work remains on both fronts to quantify the
dependence of statistical properties on the global system parameters,
and to establish when and how self-regulated quasi-steady states are
achieved.

Given the importance of time-dependent processes and interdependencies
in the ISM, complex theoretical models are needed in order to address
even rather basic questions. For example, what sets the relative
proportions of the different gas components?  In an idealized
classical picture such as that of \citet{1969ApJ...155L.149F} for the
atomic medium, given a pressure and a mean density $\bar n$, thermal
equilibrium defines a density for each of two stable phases, $n_{\rm
warm}$ and $n_{\rm cold}$, and the ratio of cold to warm gas is given
by a simple algebraic relation: $M_{\rm cold}/M_{\rm warm}= (n_{\rm
warm}^{-1} - \bar n^{-1})/ (\bar n^{-1} - n_{\rm cold}^{-1})$.  In the
real ISM, however, which is a time-dependent system, thermal
equilibrium only holds to the extent that the radiative times are
short compared to dynamical times for compressions and rarefactions.
Furthermore, the value of the mean density $\bar n$ and pressure
(averaged over large scales) are not even known {\it a priori} for a
given ISM surface density, since
the vertical distribution of gas is sensitive to its dynamical state.
This dynamical state itself depends on the (unknown) dense gas
fraction, since more dense gas produces more feedback from star
formation, and hence more turbulence that inflates the disk vertically
to reduce $\bar n$
(and also produces local variations in density and pressure through
compressions and rarefactions).  Multidimensional effects (the ISM is
not simply stratified perpendicular to the galactic plane, but is
composed of filamentary clouds) and self-gravity additionally
complicate the situation.  

In recent years, a number of groups have begun developing models
of the turbulent, multiphase ISM using time-dependent computational
hydrodynamics simulations that include feedback from star formation
\citep[e.g.,][]{
1999ApJ...514L..99K,  
2004A&A...425..899D,  
2005A&A...436..585D,  
2007ApJ...665L..35D,  
2005ApJ...626..864M,  
2006ApJ...653.1266J, 
2006ApJ...638..797D},
self-gravity of the gas
\citep[e.g.,][]{
1999ApJ...516L..13W, 
2002ApJ...577..197W}, 
and both of these effects
\citep[e.g.,][]{
2001ApJ...547..172W,    
2007ApJ...660..276W,    
2005MNRAS.356..737S,   
2006ApJ...641..878T,    
2008ApJ...673..810T,    
2008ApJ...680.1083R}.     
The treatment of feedback in these simulations is to inject thermal energy
in regions identified as sites of star formation; most models focus
on the energy input from supernovae.  
In very large scale simulations that 
have minimum resolution of only 50-100 pc, feedback implemented via
thermal energy deposition is not in 
practice very effective, because the input energy is easily
radiated away.  With finer numerical resolution, feedback regions
expand adiabatically at first to make hot diffuse bubbles, 
driving shocks that sweep up surrounding gas and
ultimately generate turbulence throughout the computational domain.
A number of different issues have been addressed by these recent
simulations, including investigating departures from thermal equilibrium in
density and temperature PDFs, measuring the relative velocity
dispersions of various gas components, and testing whether 
relationships between star formation and gas surface density emerge
that are similar to empirical Kennicutt-Schmidt laws.

Even before the advent of supernovae, massive stars photoionize their
surroundings, creating HII regions within molecular clouds that are
highly overpressured and expand.  HII regions may in fact be the most
important dynamical agents affecting the properties of dense gas in
giant molecular clouds (GMCs), since the original turbulence inherited
from the diffuse ISM is believed to dissipate within a flow crossing
time over the cloud \citep{1998ApJ...508L..99S,1998PhRvL..80.2754M},
while GMCs are thought to live for at least a few crossing times
\citep{2007prpl.conf...81B}.
Analytic and semi-analytic treatments find that 
GMCs with realistic sizes, masses,
and  star formation rates can indeed be maintained by the energy input
from HII regions for a few crossing times, ultimately being destroyed 
through a combination of
photoevaporation and kinetic energy inputs that unbind the remaining mass 
\citep{1979MNRAS.186...59W,
1994ApJ...436..795F,
1997ApJ...476..166W,
2002ApJ...566..302M,2006ApJ...653..361K}.
Recent three-dimensional
numerical studies have begun to address this process in detail 
\citep[e.g.][]{2006ApJ...647..397M,2007ApJ...668..980M,2007ApJ...671..518K},
focusing on 
regions within 
GMCs.

In the present work, we consider how the large-scale dynamical state of the
ISM is affected by star formation feedback in the form of expanding
HII regions.  Our main interests are in exploring how the turbulence
driven by HII regions affects the properties of dense gas 
(we measure statistics of density, temperature, and velocity), in testing
ideas of global self-regulation by feedback (we evaluate Toomre $Q$
parameters and virial ratios), and in exploring how galactic
environment systematically affects the character of the ISM, including
its ability to form stars.  
Complete ISM models should of
course include feedback from supernovae as well as those from HII
regions, and it is our intention to do this in future work.  However,
we consider it useful to adopt a sequential approach, independently
testing the effects of HII region feedback to provide a baseline for
more comprehensive simulations.  
In addition to developing a physical
understanding of the ways in which feedback affects the ISM, another
goal of our work is to investigate the sensitivity of numerical
results to prescriptions that are a necessary -- but not always fully
tested -- aspect of galactic-scale studies of star formation.  In
particular, we examine how the choice of density threshold in
commonly-adopted recipes for star formation affects the resulting
dependence of the star formation rate on ISM surface density.

Our approach to exploring the effects of galactic environment is to
conduct a large suite of local simulations that cover a range of
values for three basic parameters: the total surface density of gas in
the disk ($\Sigma$), 
the local midplane stellar density ($\rho_*$), and the local rate
of galactic rotation ($\Omega$).  The parameter range covers 
a factor of six in gas surface density and galactic angular rotation
rate, and a factor of 30 in stellar density.
Our suite is divided into four 
series, each of which has one independent parameter that is
systematically varied.  
We also include comparisons with hydrostatic
models that are identical in terms of their input parameters to the
fully-dynamic models, but do not include feedback and hence are not
turbulent.  For this first set of pilot studies, we have not
implemented full radiative transfer to evaluate the extent of HII
regions (we intend to do so in the future), but instead introduce a
simple prescription in which the boundaries of HII regions are determined by
the gravitational potential.  Using this approach (rather than, for
example, adopting a single fixed outer radius) has the advantage
that the volume of the heated region expands as the density
surrounding the source drops.  Since our treatment of HII regions does
not attempt to be exact, we do not consider our specific 
results for e.g. velocity dispersions to be more than approximate
(although in fact we find similar values for velocity dispersions in
dense gas to those that are observed in GMCs). 
Instead, we shall emphasize the general properties of a multiphase ISM
system in which turbulence is driven from within the dense phase.

This paper is organized as follows: In \S 2 we describe our numerical
methods, and in particular the recipe for star formation feedback.
The control parameters for our disk models, and the properties of each
model series, are presented in \S 3.  Section 4 gives an overview of
evolution based on our fiducial model.
In \S 5 we present the statistical properties of the gas in each
model, and test environmental influences by
intercomparing the model series.
The implications of our results for star formation, both in real
galaxies and in numerical simulations, is analyzed in \S 6.
We conclude with a summary and discussion in \S \ref{Summary}.

\section{Numerical Methods}

\subsection{Basic Equations}

We study the evolution of rotating, self-gravitating, galactic gas
disks, including local heating and cooling terms. We solve the 
hydrodynamic equations in a local Cartesian reference frame whose
center lies at a galactocentric radius $R_0$ and orbits the galaxy with
a fixed angular velocity $\Omega_0 =\Omega(R_0)$.  In this local frame,
radial, azimuthal, and vertical coordinates are represented by
$x\equiv R-R_0$, $y\equiv R_0(\phi-\Omega_0 t)$, and $z$, respectively,
and terms associated with coordinate curvature are neglected
\citep{1965MNRAS.130..125G,1966ApJ...146..810J}.
The local-frame equilibrium background velocity relative to the center
of the box at $x=y=z=0$ is given by 
${\bf v}_0=-q\Omega_0 x\hat{y}$, where 
\begin{eqnarray}
q=-\left.
\frac{d\ln\Omega}{d\ln R}
\right|_{R_0}
\end{eqnarray}
is the local dimensionless shear rate.
In terms of $q$, the local epicyclic frequency $\kappa$ is given by
\begin{eqnarray}
\kappa^2\equiv \frac{1}{R^3}\frac{d}{dR}(R^4\Omega^2)
=(4-2q)\Omega^2.
\end{eqnarray}
We shall choose $q=1$ for all models, representing a flat background rotation
curve $V_c=\Omega R= const.$ for the unperturbed motion.

In addition to the tidal gravity and Coriolis terms from the
``shearing sheet'' local formulation, we also include 
terms for the vertical gravity of the stellar disk, gas self-gravity, 
radiative heating and cooling, and thermal conduction.  
The resulting equations 
\citep[see e.g.,][]{1995ApJ...440..742H,2004ApJ...601..905P}
are:
\begin{eqnarray}
\frac{\partial \rho}{\partial t}+\nabla\cdot(\rho {\bf v})&=&0,\\
\frac{\partial {\bf v}}{\partial t}+{\bf v}\cdot \nabla {\bf v}
&=&-\frac{1}{\rho}\nabla P+2q\Omega x\hat{\bf x}
-2{\bf \Omega}\times{\bf v}
-\nabla \Phi+{\bf g}_{\ast},\\
\frac{\partial e}{\partial t}+\nabla\cdot(e{\bf v})
&=&-P\nabla\cdot{\bf v}
+n\left[\Gamma(t,{\bf x})-n\Lambda(T)\right] 
+ K\nabla^2T,
\end{eqnarray}
where 
$P=f n \kB T$ and $n =\rho/\mu$.
With $n$ the number of hydrogen nuclei per unit volume, 
$f$ varies from 0.6 to 1.1 depending on whether the gas is
predominantly molecular or atomic; we simply adopt $f=1.1$.
We adopt $\mu=1.4 m_p$ to include the contribution of helium to
the mass density.
Here $e=P/(\gamma-1)$ is the internal energy per unit volume (we adopt
$\gamma=5/3$), $K$ is the thermal conductivity,
$\Phi$ is the self-gravitational potential due to gas,
and the vertical gravitational force due to stellar disk is
\begin{eqnarray}
{\bf g}_{\ast}=-4\pi G\rho_{\ast}z\hat{\bf z},
\end{eqnarray}
where $\rho_{\ast}$ is the stellar density and $z$ is the vertical
coordinate relative to the midplane. 
In the above expressions and elsewhere in the remainder of the text, 
we have dropped the ``0'' subscript on $\Omega$; $\kappa$ also refers
to the value evaluated at the center of the domain.
Computation of the gas self-gravity is discussed below.

In this paper, we present results of two-dimensional simulations of
the above set of equations.  The two independent spatial coordinates
in our models are $x$ and $z$; thus, we follow evolution in
radial-vertical slices through a galactic disk.  Although the
azimuthal ($y$) direction is not an independent spatial
variable for the current set of models, we do include azimuthal
velocities, and their variation with $x$ and $z$.  Inclusion of $v_y$
is important because angular momentum can strongly affect the ability
of self-gravitating perturbations to grow.  Radial motions that are
required for gas to become concentrated are coupled to azimuthal
motions through the Coriolis force; perturbations in the azimuthal
velocity with respect to the mean background shear correspondingly
lead to radial motions via the Coriolis force.  Although our
two-dimensional models do capture some of the effects of galactic
rotation (i.e. epicyclic oscillations), they miss some of the effects
associated with shear.  In three dimensions (or in the
height-integrated $R-\phi$ plane), azimuthal shear can make
it more difficult for self-gravitating concentrations to grow. Of
course, in three dimensions, self-gravity also increases more rapidly
as the density increases, which enhances the ability of dense
concentrations to grow.  Although it will be important to revisit the
present models with fully three-dimensional simulations, we do not
anticipate large changes based on dimensionality.  Previous
three-dimensional simulations of shearing, rotating disks have found
similar (within a factor 2) 
nonlinear instability thresholds for self-gravitating cloud
formation to reduced-dimensional models (see
e.g.
\citealt{2002ApJ...581.1080K,2003ApJ...599.1157K,2007ApJ...660.1232K}, 
and references therein).  Thermal
instability also develops similarly in the two-dimensional and
three-dimensional case to create a cold cloud/warm intercloud
structure (e.g.
\citealt{2004ApJ...601..905P,2005ApJ...629..849P}).

\subsection{Hydrodynamic Code and Boundary Conditions}

The numerical solutions to the two-dimensional 
dynamical equations are obtained using 
a temporally and spatially second order finite volume method
which includes TVD Runge-Kutta integration in time 
\citep{Shu_Osher_Efficient_ENO_JCP1988}, 
with a
directionally unsplit flux update and  
piecewise linear reconstruction with slope limiter 
\citep[see, e.g.,][]{Hirsch_2nd}.
We use Roe's approximate Riemann solver with an entropy fix 
\citep{1981JCoPh..43..357R}.
The heating and cooling terms in the
energy equation are separated out in an operator-split fashion and
updated using implicit time integration 
(see \S \ref{sec:cooling}),
because the
cooling times are frequently much shorter than the other timescales.
The code is parallelized using MPI.

For the 
hydrodynamic update, the time step is set to 
$\Delta t={\rm min}(t_{\rm HD}, t_{\rm cond}, t_{\rm cool},
 t_{\rm heat})$
where 
\begin{eqnarray}
t_{\rm HD}&=&C_{\rm CFL}\,{\rm min}\left(
\frac{1}{\frac{c_s+|v_x|}{\Delta x}+\frac{c_s+|v_z|}{\Delta
z}}\right)_{\rm all\ zones},\\
t_{\rm cond}&=&C_{K}\,{\rm min}\left(
\frac{n \kB(\Delta x)^2}{4K(\gamma-1)}\right)_{\rm all\ zones},\\
t_{\rm cool}&=&C_{T}\,{\rm min}\left(
\frac{\kB T}{(\gamma-1)n\Lambda(T)}\right)_{\rm all\ zones},\\
t_{\rm heat}&=&C_{T}\,{\rm min}\left(
\frac{\kB T}{(\gamma-1)\Gamma({\bf x})}\right)_{\rm all\ zones},
\end{eqnarray}
and we adopt $C_{\rm CFL}=0.8$, $C_{K}=0.5$ and $C_{T}=50$. 
Here, $c_s=(\gamma kT/\mu)^{1/2}$ is the sound speed in any zone.
With a large value of $C_{T}$, the explicit hydrodynamic timestep is
not strongly limited by the cooling time in dense gas.
The adopted $C_{T}$ is chosen such that the solution
agrees with tests of individual expanding ``HII regions'' (for our
feedback model) that have $C_T\sim 1$ (equivalent to explicit
cooling); if a much larger value of $C_T$ is allowed, this expansion is
not accurately reproduced.

At the $x$ (radial) boundaries, we implement shearing-periodic
boundary conditions \citep{1995ApJ...440..742H}, in which the azimuthal
(angular) velocity term 
$v_y\equiv R_0 (\dot \phi - \Omega_0)=(R_0/R) v_\phi -V_c$ 
is incremented or decremented by $\pm L_x \Omega_0$ in
mapping from the right$\rightarrow$ left or left$\rightarrow$ right
boundary, consistent with the equilibrium velocity field.  
In the $z$-direction, we adopt periodic boundary
conditions for the hydrodynamic variables, such that the total mass in
the domain is conserved.\footnote{We have found that except
for mass loss, the overall evolution is similar when we apply outflow
boundary conditions in the vertical direction.  Adopting periodic boundary
conditions for hydrodynamic variables makes it possible to maintain the
gas surface density $\Sigma$ at a constant value without devoting significant
computational resources to following the evolution of a tenuous corona
at large $|z|$.}  The gravitational potential solver applies open 
(i.e. vacuum) boundary conditions in $z$, as we next discuss.

\subsection{Poisson Solver}

We have developed a new method for obtaining the gravitational
potential of a disk in Cartesian geometry using Fast Fourier
Transforms (FFTs). Since the discrete Fourier Transformation allows only
periodic functions, a special approach is needed to solve for a disk
potential with vacuum boundary conditions outside the simulation domain.

Let us consider a simple case, consisting of an a uniform, isolated
gas sheet in the $z=0$ plane which has density
$\rho(z)=\sigma\delta(z)$. The corresponding gravitational potential
$\Phi(z)=2\pi G\sigma|z|$ is obtained by solving the Poisson equation,
\begin{equation}
\nabla^2 \Phi = 4 \pi G \rho, 
\end{equation}
with vacuum boundary conditions. If we
have a finite domain of size $L_z$ and suppose that the gas sheet lies
somewhere within the domain, then we only would require values of the potential
at locations within $\pm L_z$ of the sheet.  Thus, we may find the
potential within $z=(-L_z,L_z)$ in terms of discrete 
Fourier components as
\begin{eqnarray}
\hat{\Phi}_{\ell}&=&2\pi G\sigma C_{\ell},
\label{eq:Phi_ell1}
\end{eqnarray}
\begin{eqnarray}
C_{\ell}&=&
\int_{-L_z}^{L_z}|z|e^{2\pi i \ell \left(\frac{z}{2L_z}\right)}dz=
-\frac{2(1-e^{\pm i\pi \ell})}{\left(\frac{\pi \ell}{L_z}\right)^2}.
\label{eq:C_ell}
\end{eqnarray}
In Fourier space, the Poisson equation for one independent variable is 
\begin{equation}
-k^2 \hat \Phi_k = 4 \pi G \hat\rho_k.
\end{equation}
Thus, in terms of discrete Fourier components with 
$k_\ell=\ell (2\pi/2L_z)$, we have 
\begin{eqnarray}
\hat \Phi_\ell  = - \frac{4\pi G \hat{\rho}_{\ell}}
{\left(\frac{\pi \ell}{L_z}\right)^2}. 
\label{eq:Phi_ell2}
\end{eqnarray}
Equating the right-hand sides of equations (\ref{eq:Phi_ell1}) and
(\ref{eq:Phi_ell2}) and inserting the expression from equation 
(\ref{eq:C_ell}), this implies $\hat \rho_\ell = \sigma (1 - e^{\pm i \pi
  \ell})$ for the isolated sheet, so that the density 
in real space is obtained by taking an inverse transform:
\begin{eqnarray}
\rho(z)
&=&\frac{1}{2N_z}\sum_{\ell=0}^{2N_z-1}e^{-2\pi i \ell \frac{z}{2L_z}}
\hat{\rho}_{\ell},\\
&=&\frac{1}{2N_z}\sum_{\ell=0}^{2N_z-1}e^{-2\pi i \ell \frac{z}{2L_z}}
\sigma(1-e^{\pm i\pi \ell}),\\
&=&\frac{\sigma}{2N_z}\sum_{\ell=0}^{2N_z-1}e^{-2\pi i \ell \frac{z}{L_z}}
-\frac{\sigma}{2N_z}\sum_{\ell=0}^{2N_z-1}e^{-2\pi i \ell \frac{z\mp
L_z}{2L_z}},\\
&=&\sigma\delta(z)-\sigma\delta(z\mp L_z).
\end{eqnarray}
We see that the first term corresponds to the original density.
However, a second term has appeared as an image density with the
opposite sign from the real (physical) density, located a domain
length away.  This means that to obtain the correct solution for
$\Phi$ on the original domain $z=(-L_z/2,L_z/2)$, we need to
prepare twice as large a box in the vertical direction, and implement
the required image density within the augmented domain, at a distance
$\pm L_z$ from the physical slab.  Thus, a density slab at
$0\le z\le L_z/2$ 
would require an image slab at $z_{\rm image}=z-L_z$ in
$(-L_z,-L_z/2)$, and a
density slab at $-L_z/2\le z \le 0$ 
would require an image slab at $z_{\rm image}=z+L_z$
in $(L_z/2, L_z)$.  
Using a similar procedure, we have extended this idea to
the three dimensional case with an arbitrary density distribution
$\rho(x,y,z)$.  The details are described in the Appendix.  We note
that the numerical solution agrees with the solution obtained via
Green functions \citep{1987PThPh..78.1273M}, and is much faster to
compute because only FFTs (no direct sums) are needed.

\subsection{Cooling Function}\label{sec:cooling}

To allow for multiphase interstellar gas components, we must solve a
thermal energy equation that allows a wide range of conditions.  We
use a cooling function for the diffuse ISM derived by
\cite{2002ApJ...564L..97K}, 
which includes atomic gas cooling for the warm and
cold neutral medium (WNM, CNM), as well as cold molecular-phase
cooling (H$_2$, CO, and dust cooling).  We include a constant
volumetric heating rate to represent photoelectric heating by diffuse FUV.
This yields a standard 
(cf. \citealt{1969ApJ...155L.149F,1995ApJ...443..152W}) 
thermal equilibrium curve in which there is a
maximum density and pressure for the warm phase given by 
$1.0 ~ \cm^{-3}$ and $5.5\times 10^3 ~\kB~ \cm^{-3}\K$,
and a minimum density and pressure for the cold phase given by 
$8.7 ~ \cm^{-3}$ and $1.75 \times 10^{3} ~\kB~ \cm^{-3}\K$ 
(see Fig. \ref{fig:phase-1}).

HII regions in the real ISM
include photoionization of atoms and dissociation of molecules, and
radiative cooling of photoionized gas and warm molecular gas. These
effects depend on chemical fractions, as well as dust evaporation.
For this work, we are interested primarily in dynamics of the neutral
media, rather than the details of photoionized gas -- including the
complexities of ionization front propagation at sub-parsec scales.
The main requirement for capturing the large-scale 
dynamical effects of feedback
is thus to incorporate photoheating of gas in star-forming regions.
The simple but expedient approach we have chosen is to expose gas in
targeted regions to enhanced heating, while simply applying the same
cooling function we use for neutral gas. The enhanced heating we apply
yields thermal equilibrium for the ``photoheated'' gas at $\approx
10,000$K (see below), which is consistent with the temperatures that
would be attained if we had implemented realistic (but much more
computationally complex and expensive) photoprocesses.

Cooling and heating timescales often become much shorter than the hydrodynamic
time step (i.e. the flow or sound crossing time of a grid zone),
especially in HII regions, which have a high heating rate.  For
efficiency, we adopt implicit time integration for the heating and
cooling operators. In a given zone, the integral from the (j) to the
(j+1) time step is formally expressed as
\begin{eqnarray}
\int_{T_{j}}^{T_{j+1}}\frac{C_vdT}{\Gamma_{j}-n\Lambda(T)}
=\Delta t_{j},
\end{eqnarray}
where $C_v=\kB/(\gamma-1)$ is the heat capacity per particle.
This is a nonlinear equation with respect to $T_{j+1}$, with $T_j$ and
$\Delta t_j$ treated as parameters.  For this integral, we adopt
Simpson's rule and solve using the Newton-Raphson method.

\subsection{Thermal Conduction}

Thermal conduction determines the thickness of interfaces between
phases in the ISM, and proper incorporation of conduction is essential
in numerical simulations of a multiphase medium which is subject to
thermal instability
\citep{2004ApJ...601..905P, 2004ApJ...602L..25K,2008ApJ...681.1148K}.
The characteristic length scale set by conduction is
the Field length, 
\begin{equation}
\lambda_{\rm F}=2\pi\sqrt{\frac{KT}{n^2\Lambda(T)}},
\end{equation}
 \citep{1965ApJ...142..531F,1990ApJ...358..375B},
which corresponds to the critical wavelength of thermal instability.
For realistic values of the conductivity at $T\aplt 10^4 \K$,
$K=2.5\times 10^3\sqrt{T}$ erg cm$^{-1}$ s$^{-1}$ K$^{-1}$
\citep{1953ApJ...117..431P}, the 
Field length of 0.19 pc (at density $1 ~ \cm^{-3}$ and temperature 1,000 K) 
is much smaller than the size of interstellar clouds 
($\sim$ 1 -- 10 pc)
and we would 
need extremely high spatial resolution to resolve it -- and a
correspondingly high computational cost.
Instead, we adopt an approach somewhat analogous to the use of artificial
viscosity (far exceeding the true physical viscosity in magnitude) 
in resolving shocks on a numerical grid. Namely, we
adopt a sufficiently large numerical conduction coefficient 
to resolve the Field length on our chosen grid. We choose $K$ such that
for any simulation with resolution $\Delta x$, 
the Field number $N_{\rm F}\equiv\lambda_{\rm F}/\Delta x$ is equal to
1.7
at density and temperature typical of thermally unstable gas
(we use $n=1 \ \cm^{-3}$, $T=1,000$ K).
For example, the fiducial model Q11 has $\Delta x=1.5$ pc and 
the artificial conductivity gives $\lambda_{\rm F}=2.57$ pc for thermally 
unstable gas.

For low density gas, given our typical values of $K$ 
the thermal conduction term can become greater than
cooling/heating terms.
In order to limit the conduction in these regions, we adopt
\begin{eqnarray}
\tilde{K}&=&\frac{K}{1+n_{\rm crit}/n},
\end{eqnarray}
with $n_{\rm crit}=0.05 ~ \cm^{-3}$.

\subsection{Stellar Feedback Activity}

The primary focus of this work is to explore the dynamical effects of
strong, localized heating by OB stars in dense regions of the ISM.
Since this heating produces $T\sim 10^4$ K gas that is initially
overpressured by a factor $\sim 100$ or more compared to its
surroundings, HII regions expand rapidly.  This process is believed to
be an important source of turbulence both within self-gravitating GMCs
and in the surrounding diffuse ISM.  To study this process, ideally
one would implement (a) formation of OB stars from dense gas,
distributed throughout the space-time domain of the simulation; (b)
radiative transfer of ionizing photons from the OB stars through the
surrounding gas, with potentially multiple ionization sources
throughout the domain; (c) detailed ionization and heating of the gas
within HII regions.

In this first exploration, rather than attempting to model all of
these processes in an exact fashion, we instead adopt an idealized
approach, with the goal of gaining physical understanding.  First, we
apply certain criteria to determine when and where 
``star zones'' on the grid 
will be turned on.  Then, we apply strong heating to the gas in the
vicinity of each 
``star zone'' for the duration of its lifetime.
All 
``star zones'' 
have the same lifetime, $t_{\rm ms}$, which is set
to $3.7 \times 10^6$ yr, the typical lifetime of OB associations 
in clouds whose mass is $10^5\msun$ \citep{1997ApJ...476..144M}.
Within HII regions, we assume a constant gas heating rate, set via a
control parameter $G_{\rm HII}$.  Each 
``star zone'' is therefore essentially a control flag
for whether or not strong
photoheating is locally applied near that zone (which does not move).
Rather than solving a radiative transfer problem, we use a
simple criterion based on the gravitational potential to determine
whether gas is subject to strong heating.  
Because our goal is to
identify gas localized around star-forming regions, it is necessary to
subtract out the background gravitational potential and retain just
the potential component due to an individual self-gravitating cloud.

The background potential is the potential averaged over horizontal
planes.  In terms of Fourier components, the relative gravitational
potential $\Phi^{(1)}=\Phi - \Phi_{\rm background}$ is defined as
\begin{eqnarray}
\Phi^{(1)}({\bf x})=\sum_{k_z\ne 0} 
\hat{\Phi}_{k}\hat{f}_{\bf k}\exp(-i{\bf k\cdot
x}). \label{eq:relative potential}
\end{eqnarray}
Here, $\hat{f}_{\bf k}$ is the Fourier component of a smoothing window
function,
\begin{eqnarray}
f({\bf x})=\frac{3}{4\pi r_0^3}
\frac{1}{1+\exp\left[\frac{10}{\Delta x}(\sqrt{x^2+z^2}-r_0)\right]},
\end{eqnarray}
where $r_0$ is a smoothing length.  This window function 
smooths the HII region within a radius $\approx r_0$.
Convolution of the relative potential with a smoothing function (or,
equivalently, multiplication in Fourier space as above) is desirable
so that any heating that is applied is resolved on the grid.  We have
adopted $r_0=3\Delta x$ as providing adequate resolution.

HII photoheated regions are identified as regions where the relative
potential, $\Phi^{(1)}$, falls below some specified level:
$\Phi^{(1)}<\Phi_{\rm HII}$.  We also employ the relative potential
$\Phi^{(1)}$ for setting one of the criteria for turning on feedback:
$\Phi^{(1)}<\Phi_{\rm SF}$ and $\rho>\rho_{\rm SF}$ must both be met
in a given zone for a ``star zone'' to be created at that
location.  Thus, our recipe ensures that feedback will only occur in
dense and self-gravitating regions, consistent with the fact that OB
stars are observed to form only under these conditions.

For the feedback prescription we have adopted, there are 
five control parameters: $\rho_{\rm SF}$, $\Phi_{\rm SF}$, $t_{\rm ms}$, 
$\Phi_{\rm HII}$, $G_{\rm HII}$. 
The detailed estimation of those parameters is described in the
remainder of this section.

\subsubsection{OB Star Formation Criterion}

We choose a density threshold for star formation as
\begin{eqnarray}
\rho_{\rm SF}&=& 10^3 \mu\, \cm^{-3}.
\end{eqnarray}
This density is comparable to that of clumps of gas within GMCs.
The free fall time at this density, 
1.4 Myr, is
sufficiently small compared to the orbital time that structures
satisfying this threshold evolve independently of the global
environment.  Note that the local Jeans length
\begin{eqnarray}
\lambda_{\rm J}=\left(\frac{\pi c_s^2}{G\rho_{\rm SF}}\right)^{1/2} 
=4.1\,{\rm pc}\left(\frac{c_s}{0.9 ~\kms}\right)
\end{eqnarray}
must be resolved by a few zones in order to prevent fragmentation
occurring as a consequence of numerical artifacts
\citep{1997ApJ...489L.179T}.

To obtain an estimate for the potential threshold for star formation, 
we consider a cloud
with uniform number density $\bar{n}$ and radius $R_{\rm cl}$.  For a
spherical cloud, the radius is related to the cloud mass using 
\begin{eqnarray}
R_{\rm cl}&=&\left(\frac{3M_{\rm cl}}{4\pi\bar\rho}\right)^{1/3}
=19 ~  {\rm pc} \times 
\bar{n}_{2}^{-1/3}M_{\rm cl,5}^{2/3}, 
\end{eqnarray}
where the fiducial value of 
$\bar{n}_2\equiv \bar{n}/(10^2~\cm^{-3})$ is chosen using a
typical mean density within GMCs, 
and $M_{\rm cl,5}\equiv M_{\rm cl}/(10^5 ~ \msun)$.

Since our grid is
two-dimensional, the control parameter 
$\Phi_{\rm SF}$ must be based on a
cylindrical regions of a given density.  For a cylinder of radius
$R_{\rm cl}$, the potential difference between the center and a
distance $r$ ($>R_{\rm cl}$) is $\Phi(0)=\Phi(r) - 2 \pi G \bar\rho
R_{\rm cl}^2 [\ln(r/R_{\rm cl}) + 1/2]$.  
The logarithmic term corresponds to the potential difference between
$r$ and $R_{\rm cl}$, and the $1/2$ is the contribution between the
cloud's surface and its center.
If the cloud is created out
of all of the mass originally within a disk of surface density
$\Sigma$ within a range $2r\equiv L_{\rm eqv}$ of the cloud center,
then $\pi R_{\rm cl}^2 \bar\rho = \Sigma L_{\rm eqv}$, and a fiducial
distance for defining the effective zero of the potential is
\begin{eqnarray}
L_{\rm eqv}&=&\frac{\pi\bar\rho R_{\rm cl}^2}{\Sigma},\\
&=&393~ {\rm pc} \times \bar{n}_{2}M_{\rm cl,5}^{2/3}
\left(\frac{\Sigma}{10\msun\pc^{-2}}\right)^{-1}.
\end{eqnarray}
Here, the
radius is expressed 
in terms of that of an equivalent spherical cloud with a given 
mass.
If we set the potential at $r=L_{\rm eqv}/2$ to zero, then the
potential at the center of the cloud will be 
\begin{equation}
\Phi(0)= -\Phi_{\rm SF}\equiv - 2 \pi G \bar\rho R_{\rm cl}^2 
\left[\ln\left(\frac{L_{\rm eqv}}{2R_{\rm cl}}\right)
  + \frac{1}{2}\right].
\label{eq:Phi_0}
\end{equation}
For  
$\bar n = 100~ \cm^{-3}$ and  $R_{\rm cl}=19$ pc, the value of 
$\Phi(0)$ is $3.4 \times 10^{11} \cm^2 ~{\rm s}^{-2}$
times an order-unity
factor that varies only logarithmically in the ratio of cloud surface
density to mean ISM surface density.
We choose to adopt a potential threshold for 
star formation 
$\Phi_{\rm SF} = 3.4 \times 10^{11} \cm^2 ~{\rm s}^{-2}
=(5.8 {\rm km} ~{\rm s}^{-1})^2$; we have
tested sensitivity to the value of $\Phi_{\rm SF}$, and found that
results are insensitive to the exact choice, except as described below.

\subsubsection{Definition of Photoheated Regions}

First, consider an HII region centered on the origin of a uniform cloud
with spherical crossection.
If we assume the radius of the HII region is $R_{\rm HII}$ ($<R_{\rm
  cl}$), then if the center of the cloud has potential 
$\Phi(0)$,
the potential at the ionization front for a cylindrical cloud with
uniform density 
is $\Phi_{\rm HII}=\Phi(0) + \pi G\bar\rho R_{\rm HII}^2$.  (For a
spherical cloud, the second term is smaller by a factor $2/3$.)
Taking the difference with $\Phi(L_{\rm eqv}/2)$ in order to represent
a relative potential, using equation (\ref{eq:Phi_0}), and
substituting $\Phi(0) \rightarrow
-\Phi_{\rm SF}$ (since the criterion for star formation must be
satisfied if feedback has turned on) 
this implies that the relative potential at the location of the HII
region would be
\begin{equation}
\Phi_{\rm HII}=-\Phi_{\rm SF}\left[
1-\frac{\frac{1}{2}}
{\ln\left(\frac{L_{\rm eqv}}{2R_{\rm cl}}\right)+\frac{1}{2}}
\left(\frac{R_{\rm HII}}{R_{\rm cl}}\right)^2
\right].
\label{eq:R_HII_init}
\end{equation}
$R_{\rm HII}$ is given, for example,  by the Str\"omgren radius in a
uniform medium:
\begin{eqnarray}
R_{\rm HII}&=&\left(\frac{3S}{4\pi \bar{n}^2\alpha^{(2)}}\right)^{1/3}
= 2.97~{\rm pc}~
\bar{n}_{2}^{-2/3}
S_{49}^{1/3},
\end{eqnarray}
where $S$ is the number ionizing photons per 
unit time and $S_{49}\equiv S/(10^{49}{\rm s}^{-1})$, and 
$\alpha^{(2)}=3.09\times 10^{-13} ~ \cm^3 {\rm s}^{-1}$
is the case B hydrogen recombination coefficient at 
$T=8,000$ K \citep{1978ppim.book.....S}.   
$S_{49}$ is equal to unity
for the ionizing luminosity in a 
typical $10^5 \msun$ GMC \citep{1997ApJ...476..144M}. From 
equation (\ref{eq:R_HII_init}), when the density is comparable to the
mean density within a GMC, $R_{\rm HII}/R_{\rm cl}\ll 1$, and the HII
region is buried deep within the GMC at 
$||\Phi_{\rm HII}|-\Phi_{\rm SF}|/\Phi_{\rm SF}\ll 1$. 

HII regions are initially highly overpressured, however, and 
will expand rapidly until breaking out of the surrounding GMC, creating
a blister HII region.  
For the purposes of considering the momentum input to the system, the 
$R_{\rm HII}\rightarrow R_{\rm cl}$ limit is most appropriate for
defining the photoheated region.  
Thus, we suppose that the HII region has expanded, leaving a very low
density interior and a shell of radius $R_{\rm HII}<R_{\rm cl}$ 
in which most of the mass has piled up.  The potential in the interior
of the shell is then given by 
\begin{equation}
\Phi_{\rm HII}= -\Phi_{\rm SF}
\left[1- \frac{\frac{R_{\rm HII}^2}{R_{\rm cl}^2-R_{\rm HII}^2}
    \ln(\frac{ R_{\rm cl}}{R_{\rm HII}})}
{\ln\left(\frac{L_{\rm eqv}}{2R_{\rm cl}}\right)+\frac{1}{2}}
\right].
\label{eq:R_HII_shell}
\end{equation}
When $R_{\rm HII}\rightarrow R_{\rm cl}$, 
this expression is of course the same as if we had taken 
$R_{\rm HII}/R_{\rm cl}\rightarrow 1$ in equation 
(\ref{eq:R_HII_init}), that is, the potential near the surface
of the initially-uniform cloud.
For convenience, we 
introduce a dimensionless parameter $\epsilon$:
\begin{equation}
\Phi_{\rm HII} \equiv -\Phi_{\rm SF}(1-\epsilon),
\label{eq:HII_eps}
\end{equation}
where
$\epsilon=(1/2)
[\ln\left(\frac{L_{\rm eqv}}{2R_{\rm cl}}\right)+\frac{1}{2}]^{-1}$
when $R_{\rm HII} \rightarrow R_{\rm cl}$.
For the fiducial parameter values discussed above, $\epsilon$ is equal
to 0.18.  
We therefore
adopt $\epsilon =0.2$ as our ``standard'' value, although we have
tested how the results differ for much smaller values.

\subsubsection{Heating Rate in HII Regions}

During the period that ``star zones'' are alive, UV photons
enhance the heating rate within HII regions, defined as described
above.  For the heating rate in any zone due to UV photons, we adopt
the on-the-spot approximation given by:
\begin{eqnarray}
\Gamma &=& 
2.16\times 10^{-26}G_{\rm FUV} ~ {\rm erg} ~ \cm^{-3} {\rm s}^{-1},\\
G_{\rm FUV}&=&
\left\{
\begin{array}{ll}
 G_{\rm HII} \quad & 
\mbox{($\Phi^{(1)}<\Phi_{\rm HII}$)},\\
G_0 & \mbox{(otherwise)},\\
\end{array}
\right.
\end{eqnarray}
where $G_0=1.0$ is the background FUV field in the Galaxy.  

We choose $G_{\rm HII}=10^3$ throughout this paper, 
although we have also tested
results with lower and higher $G_{\rm HII}$.
In practice, the exact value of $G_{\rm HII}$ is not important, since the
purpose of this added heating is simply to increase the maximum
density at which a warm phase is present.  From our
Fig. \ref{fig:phase-1} (see also e.g. Fig. 7 of
\citealt{1995ApJ...443..152W}), a value of $G_{\rm HII}=10^3$ boosts
this to $n \sim 10^3 ~ \cm^{-3}$.

\section{Model Parameters}

The models in this paper are characterized by three main parameters:
the total gas surface density $\Sigma$, the orbital angular velocity
$\Omega$ in the center of the domain, and the stellar density at the 
midplane $\rho_{\ast}$.  The first parameter defines the amount of 
raw material
available for distribution between the dense and diffuse ISM phases,
while the second two parameters define the galactic environment in
which the gas evolves in response to the galactic tidal, rotational,
and shear stresses.  The effectiveness of self-gravity in forming
massive clouds depends on all of these parameters, as well as on the
turbulent state that develops as a consequence of star formation
feedback.

The simulation domain we model is a radial-vertical slice through a
disk. For the vertical direction, we require a domain $L_z$ that
encloses most of the mass of the ISM, which is confined by a
combination of stellar gravity and gas self-gravity.  The largest
scale height (excluding the hot ISM, absent in these models) 
is that of the WNM, and an upper
limit is obtained by neglecting self-gravity, which yields a Gaussian
distribution with scale height:
\begin{eqnarray}
H_{w}&=&\frac{c_{w}}{\sqrt{4\pi G\rho_{\ast}}}
=95 ~\pc
\left(\frac{c_{w}}{7 ~ \kms}\right)
\left(\frac{\rho_{\ast}}{0.1 ~ \msun \pc^{-3} }\right)^{-1/2}.
\end{eqnarray}
Here, $c_{w}=(k T_{w}/\mu)^{1/2}$ is the thermal speed of the warm
medium, which typically has $T_w=8,000-10,000$ K.
We require $L_z > 2H_w$ for our numerical models.

If the ISM consisted only of WNM, then with $n_{w,0}=\rho_{w,0}/\mu$ 
the central volume
density, the total surface density would be 
\begin{eqnarray}
\Sigma_w=\sqrt{2\pi} \rho_{w,0} H_w 
=1.6 ~ \msun \pc^{-2}
\left(\frac{n_w}{0.2~ \cm^{-3}}\right)
\left(\frac{c_w}{7 ~ \kms}\right)
\left(\frac{\rho_{\ast}}{0.1 ~\msun \pc^{-3} }\right)^{-1/2}.
\end{eqnarray}
The maximum density for which the warm phase is possible (when
$G_0=1$) is 
$n_{w, {\rm max}}=P_{w, {\rm max}}/(\mu c_w^2)\approx 1 ~{\rm cm}^{-3}$; 
this implies a maximum
possible total surface density of warm gas
$\Sigma_{w, {\rm max}}=(2G\rho_{\ast})^{-1/2} P_{w, {\rm max}}/c_w$.  
In practice, the midplane
density of the warm medium is closer to the value $0.23 ~ \cm^{-3}$, 
which represents
the warm medium density that is in pressure equilibrium with cold gas
at $P_{c, {\rm min}}/\kB =1.75 \times 10^3 \cm^{-3}\K$.  
We are interested in multiphase disks
similar to those observed in the Solar neighborhood and at smaller
radii in normal spirals; hence we choose surface density of at least
several $\msun \pc^{-2}$ such that the pressure is high enough
to permit a cold phase to form, i.e. 
$\Sigma > (2G\rho_{\ast})^{-1/2} P_{c, {\rm min}}/c_w$.  

The radial domain should be sufficient to capture the
largest-scale gravitational instabilities, which in a disk galaxy are 
limited by angular momentum.  
The Toomre length \citep{1964ApJ...139.1217T} is the maximum scale for
axisymmetric modes in a thin, cold disk; 
\begin{eqnarray}
\lambda_{T}&=&
\frac{4\pi^2 G\Sigma}{\kappa^2}=\frac{4\pi^2G\Sigma}{(4-2q)\Omega^2},\\
&=&1.36 ~ {\rm kpc}
\left(\frac{\Omega}{25 ~{\kms ~ {\rm kpc}^{-1}}  }\right)^{-2}
\left(\frac{\Sigma}{10 ~ \msun/\pc^{2} }\right).
\end{eqnarray}
We require $L_x>\lambda_{T}$ for our numerical models, 
typically by a factor 1.3.

The parameters of our models are summarized in Table \ref{tbl:models}.
In order to cover the three-dimensional parameter space and explore
environmental dependences systematically, we consider four 
series of models: Q, K, S, and R.
For each Series, we hold two quantities fixed and vary a third
quantity, as follows:
\begin{itemize}
\item Series Q: $\kappa/\Sigma$ and
      $\sqrt{\rho_{\ast}}/\Sigma$ are constant while $\Sigma$ varies;
\item Series K: $\kappa$ and $\sqrt{\rho_{\ast}}/\Sigma$ are constant
  while $\Sigma$ varies;
\item Series R: $\kappa/\Sigma$ and $\rho_{\ast}$ are constant while 
$\Sigma$ varies;
\item Series S: $\Sigma$ and $\rho_{\ast}$ are constant while $\kappa$
  (and $\Omega$) varies.
\end{itemize}

The value of the Toomre $Q$ parameter for 
the gas component, for a radial velocity dispersion
$\sigma_R$, is defined as
\begin{eqnarray}
Q&\equiv &\frac{\kappa\sigma_R}{\pi G\Sigma}\\
&=&1.82
\left(\frac{\Omega}{25 {\rm km~s^{-1}~kpc^{-1}}}\right)
\left(\frac{\sigma_R}{7\,{\rm km~s^{-1}}}\right)
\left(\frac{\Sigma}{10 \msun {\rm pc}^{-2}}\right)^{-1}.
\label{eq:ToomreQdef}
\end{eqnarray}
Since Toomre's parameter is proportional to $\kappa/\Sigma$, Series Q
and R would have constant thermal $Q=\kappa c_s/(\pi G \Sigma)$ for
the gas if the
sound speed $c_s$ were constant (which is true for warm gas).  The Q
and R series correspond to values of $Q=2.1 (\sigma_R/7 ~ \kms)$.  
Assuming a
constant stellar velocity dispersion, $\Sigma_{\ast} \propto
\sqrt{\rho_{\ast}}$, so the stellar Toomre parameter 
($Q_{\ast}\propto \kappa/\Sigma_{\ast}$) 
would also have the same value for all members of Series Q.  
As a fiducial model, we choose $\Sigma=10.6 ~ \msun/\pc^2$,
$\Omega=31.2 ~ \kms ~ {\rm kpc}^{-1}$, 
and $\rho_{\ast}=0.07 ~ \msun /\pc^3$, similar to conditions
in the Solar neighborhood; this is denoted as the
Q11 model in Table \ref{tbl:models}.

Relative to conditions similar to those in the Solar neighborhood, we
may think of the members of Series Q as representing conditions
ranging from slightly larger radii down to radii of a few kpc, for a
disk that has constant $Q$ and $Q_{\ast}$ -- i.e. larger gas and stellar
densities at small radii.  We may think of Series R as models spanning
a similar range of radii, except for a disk that has stellar density
(and the corresponding vertical gravity) independent of radius, while
the gas surface density increases inward (such that the gas
self-gravity can become dominant). We may think of Series S as
relocation of the Solar neighborhood conditions of gas and stellar
density to either further in or further out in the galaxy's potential
well, where rotation and shear are stronger or weaker, respectively.
We may think of Series K as choosing a fixed location in the galactic
potential well (dominated by dark matter), and then varying the gas
and stellar surface densities in tandem.

To initialize our models, we set the density to a uniform value (given
by $\bar n$ in Table \ref{tbl:models})
throughout the domain, and set the pressure to 
$P/\kB=4,860 ~ \cm^{-3}\K$ which is in the thermally unstable regime.  
The value of the initial pressure in fact does not matter, since the
gas rapidly separates into WNM and CNM due to thermal instability.
We also impose on the initial conditions isobaric density
perturbations 
(with a flat spectrum
at wavenumbers smaller than $kL_z/2\pi = 32$, and 10 \% total amplitude).
The results are also not sensitive to the amplitude or power spectrum
of the initial perturbation spectrum; this is simply a convenience to
seed the initial evolution into thermal instability and then Jeans
fragmentation of cold gas.  
In order to reach a quasi-steady state with repeated feedback cycles, 
we run our models for two orbital periods in Series Q, R and K 
and up to $t_{\rm final}=5.57\times 10^8$ year for Series S.

\subsection{Hydrostatic Models}

Because an important focus of this work is to assess the effects of
turbulence, it is important to ascertain how our dynamical models
differ from the situation in which there are no motions other than
background sheared rotation.  For these comparison models, we
calculate the one-dimensional hydrostatic equilibria in the vertical
direction.  These models include heating and cooling, but no feedback
from star formation.  We consider two series, HSP and HSC which have
stellar volume density $\rho_*$ 
either {\it proportional} (P) to the square of
the gas
surface density $\Sigma$ or {\it constant} (C), respectively.  Note that
Series HSP corresponds to the dynamical Series Q and K, while Series
HSC corresponds to the dynamical Series R. Details of these model
parameters are listed in Table \ref{tbl:HS}.  Uniform 1024 grids are
used for all hydrostatic models.  Note that the hydrostatic models
only allow one-dimensional vertical structure, and require higher
spatial resolution than the dynamic models because the scale height in
the cold layer near the midplane becomes very small when there is no feedback.

\section{Evolution of the Fiducial Model}

This section describes the overall evolution of the fiducial model.
After the initial thermal instability, the cold gas collapses into the
midplane, due to the vertical gravitational field.  The cold,
high-density midplane gas then fragments rapidly, due to self-gravity.
The time scale for this 
fragmentation is characterized by the Jeans time for a thin disk,
$t_{\rm J}=c_s/G\Sigma\approx 6 ~{\rm Myr}(c_s/0.3~{\rm km~s^{-1}})$.
The dense fragments collapse, with some of them coagulating, until the
feedback criteria are met, and heating is turned on to create
local HII regions.  These HII regions expand, causing the dense gas to
be swept outward, forming shells that then break up into filaments.
New overdense regions continue to form, collapse, and 
be dispersed by feedback.  

\begin{figure*}
\epsscale{1.0}
\plotone{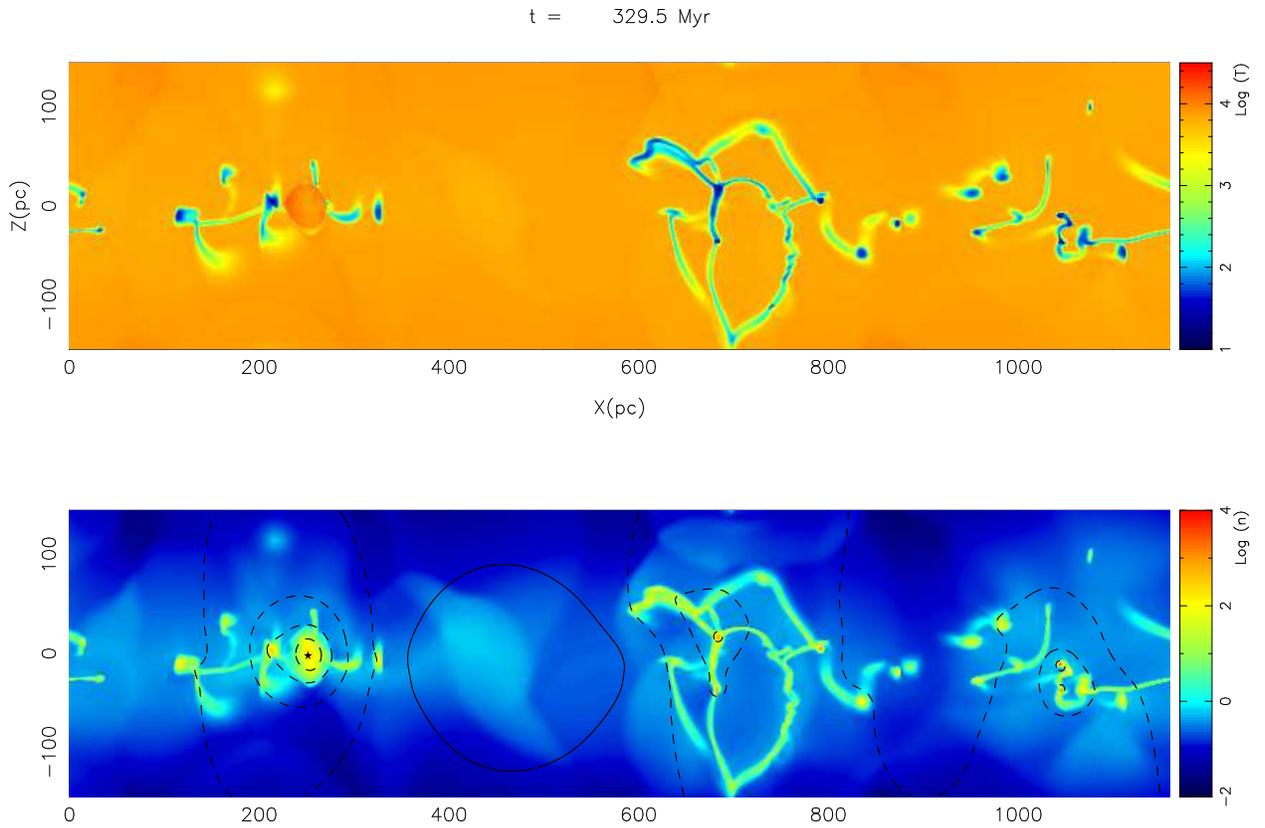}
\caption{A snapshot of Model Q11, with $\Sigma=10.6 \msun \pc^{-2}$: 
{\it Upper} panel shows temperature, and
{\it Lower} panel shows density; both have
logarithmic color scales as shown. 
Contours in the lower panel are loci of constant 
 $\Phi^{(1)}$, the relative potential, with solid and dashed curves for 
 $\Phi^{(1)}>0$ and $\Phi^{(1)}<0$, respectively. The black asterisk
 at the potential minimum in the left part of the panel indicates the
 location of a ``star zone'' feedback center.  
The cold, dense gas is highly clumpy and
 filamentary, and is concentrated toward the midplane by self- and
 external gravity.
}
\label{fig:snap-1}
\end{figure*}

Figure \ref{fig:snap-1} shows a snapshot from the fiducial model
(Model Q11) at a point after the system has reached a quasi-steady
state, in terms of the statistical distributions of density,
temperature, and velocity.  The two panels show the temperature and
density throughout the domain.  The contours in the lower panel denote
relative potential $\Phi^{(1)}$: solid and dashed lines show positive
and negative values, respectively.  At the time of this snapshot,
there are three large clouds consisting of collections of dense filaments
that create minima in the gravitational potential
(dashed lines in the lower panel).  Most of the dense filaments and
clumps in the lower panel correspond to cold gas in the upper panel.
In the upper panel, the orange circle associated with the cloud
near $x=250$ pc shows an active HII region, in which the gas is both
warm and dense and hence is expanding rapidly.  Expansion of shells
slows at later times (after the pressure drops to ambient values and
the enhanced heating turns off).  Since most of the time for any given
shell is spent near the maximum expansion, the widely-expanded
structure of the middle cloud is typical, in terms of the
time-averaged state.

\begin{figure*}
\epsscale{1.0}
\plotone{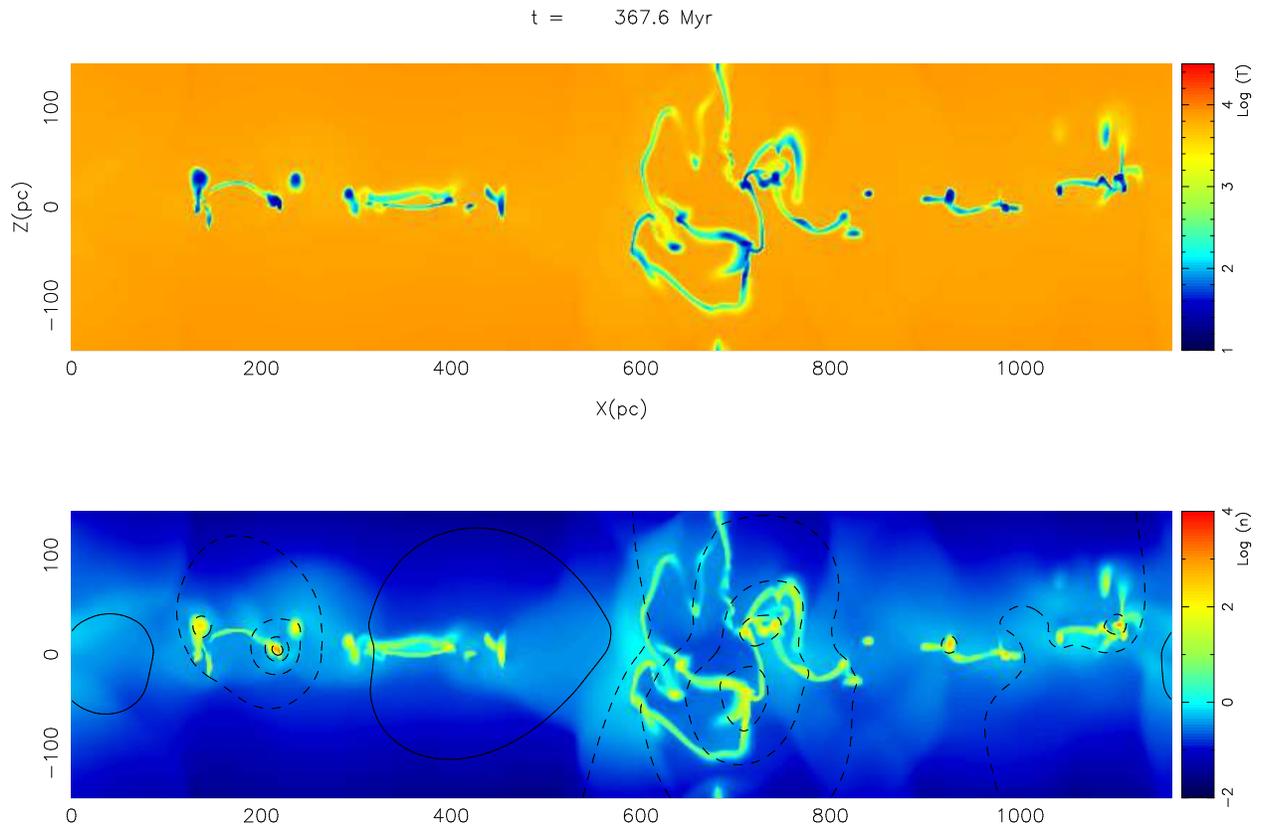}
\caption{A later-time ($\Delta t$= 38 Myr) 
snapshot of temperature (top) and density (bottom) from
Model Q11, showing that individual structures have dynamically
evolved.  
See Figure \ref{fig:snap-1} legend for details.}
\label{fig:snap-2}
\end{figure*}

\begin{figure}[p]
\plottwo{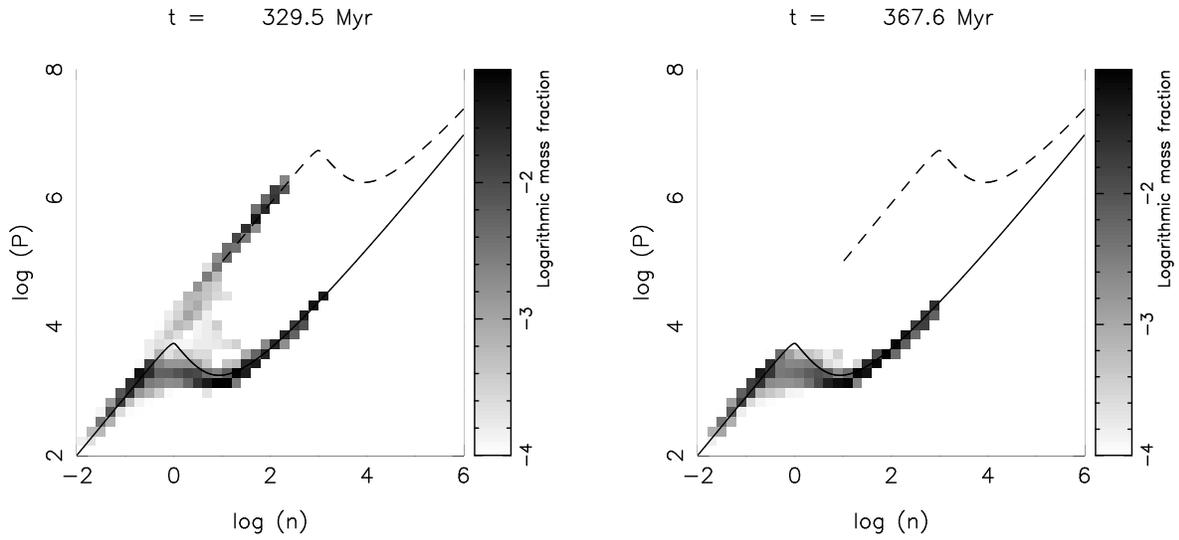}{Q13e2_1125phase.ps}
\caption{Phase diagram for Model Q11  snapshots shown in Figures
 \ref{fig:snap-1} and \ref{fig:snap-2}. The gray scale shows the relative 
fraction of the total mass
 at each point in the phase plane.  
The solid line denotes the thermal
 equilibrium curve at the background heating rate 
($G_{\rm FUV}=1$); the 
dashed curve shows thermal equilibrium in HII regions ($G_{\rm
 FUV}=10^3$). The snapshot from Figure \ref{fig:snap-1} is taken while
 an HII region is very young, and there is considerable gas on the
 hot, high-density branch; there is very little gas on the HII region
 branch from the second snapshot.
Overall, gas is preferentially -- but not exclusively -- 
found near the thermal equilibrium curves.
}
\label{fig:phase-1}
\end{figure}

Figure \ref{fig:snap-2} shows a snapshot of the density and
temperature in the same model Q11 at a time 38 Myr later.  Overall,
the structure is qualitatively similar, although details change
because the state is highly dynamic.  There are still three main
collections of filaments; the middle cloud has a large shell while the
left- and right-side clouds have contracted onto the midplane and have
nearly reached the point at which new HII regions will be born.

Three large ``clouds'' within the 1.16 kpc horizontal length of the
domain corresponds to mean separations of 390 pc.  
One might expect the number of cloud entities to be related to the
properties of star formation feedback, and for our adopted
prescription to the parameter $\epsilon$, which effectively determines
the maximum volume of an HII region: large $\epsilon$ corresponds to
large HII regions, whereas small $\epsilon$ corresponds to HII regions
only in the immediate vicinity of a potential minimum defined by a
high-density clump of gas.  In a situation with multiple local minima
in the gravitational potential, if $\epsilon$ is large then a single
HII region could engulf what would be multiple HII regions in the case
of small $\epsilon$.  Expansion and shell collision of many small HII
regions would produce more (but smaller) clouds than expansion and
collision of a few large HII regions.  In fact, when we run the same
model but set $\epsilon=0.02$, we find that there are typically 4-5
clouds in the same domain.  We conclude that the number of large
clouds is not {\it very} sensitive to $\epsilon$, but since this
control parameter can only approximately model the effects of real HII
regions, the current study cannot provide an exact prediction for the
size or mass of GMCs.  We note that the typical separation is,
however, in the same range as the two-dimensional Jeans length at the
typical velocity dispersion, $\lambda_{\rm J,2D}\equiv \sigma^2/(G\Sigma)=
340 \pc (\sigma/4 ~{\rm km~ s^{-1}})^2$ for this model.

Figure \ref{fig:phase-1} shows the phase diagram ($\rho-P$ plane) for
the same snapshot in Figure \ref{fig:snap-1} and \ref{fig:snap-2}.  The gray scale is
proportional to the fraction of the total mass in the domain that is
found at a given point in the phase plane.  We overlay the thermal
equilibrium curves for both the cases of ``normal'' heating (solid
curve) and the enhanced heating in HII regions (dashed curve).
Clearly, most of the gas resides near thermal equilibrium, due to the 
short cooling times compared to the longer hydrodynamical times.

\begin{figure}[p]
\plotone{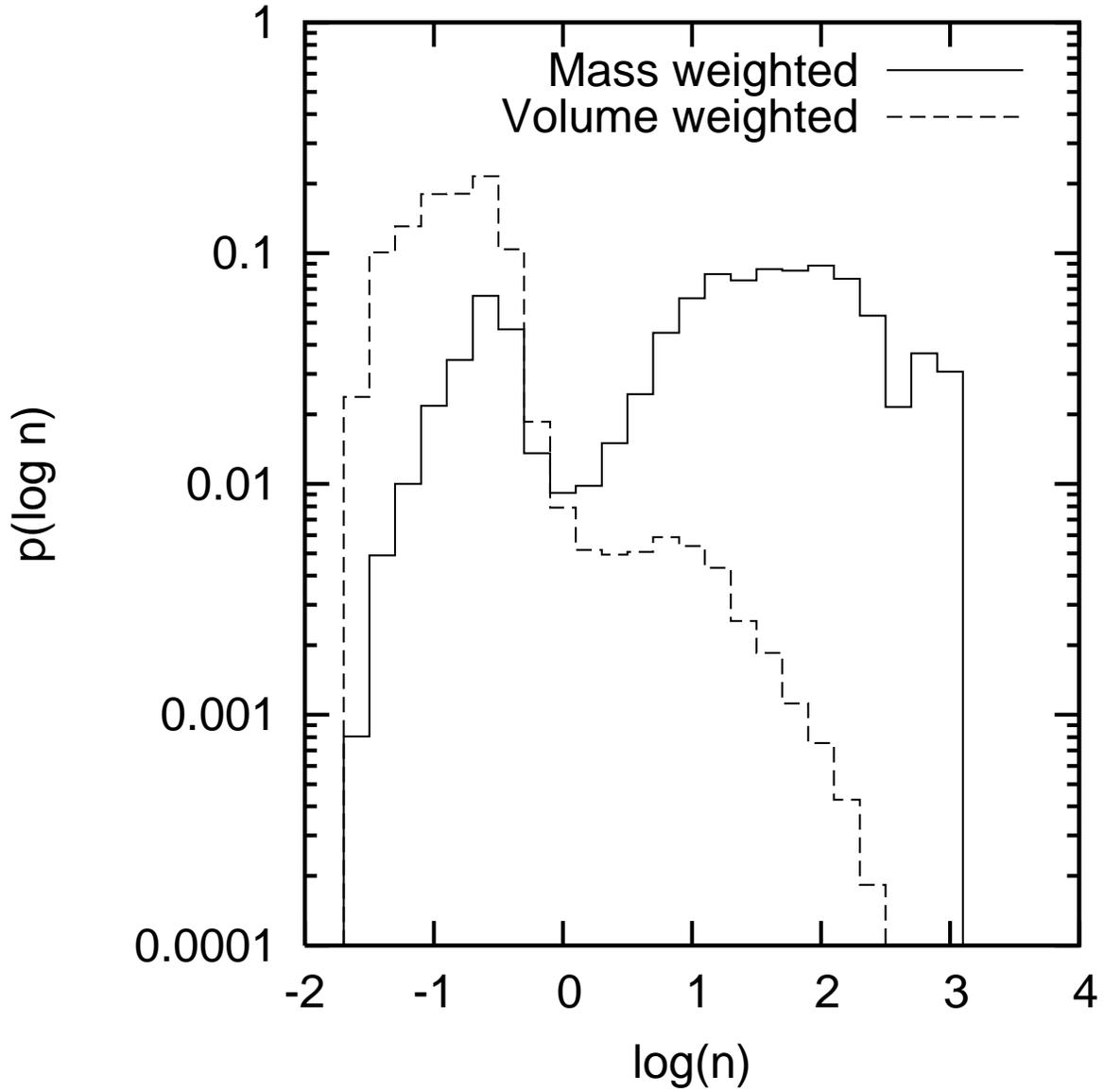}
\caption{Density distributions for Model Q11 snapshot shown in Figure
 \ref{fig:snap-1}.
Solid and dashed lines denote mass- and
volume- weighted density PDFs.  The bimodal distribution of mass
is characteristic of the modified two-phase warm/cold medium that
 develops for a bistable cooling function with a range of pressures
 (due to vertical stratification, self-gravity, and turbulent dynamics).}
\label{fig:PDF-1}
\end{figure}

The range of properties of the gas can also be seen in the Figure
\ref{fig:PDF-1}, which shows the probability distribution functions (PDFs)
 of gas density.  Solid and dashed lines show mass and volume weighted 
probabilities, respectively.  The volume PDF shows that 
the volume is mainly occupied by warm and diffuse gas (WNM) at
densities of a few $\times 0.1~ \cm^{-3}$.  
In terms of the mass PDF, there are two peaks: one corresponds to the WNM,
and the other to a cold component at density above $10~ \cm^{-3}$.

\begin{figure}[p]
\plotone{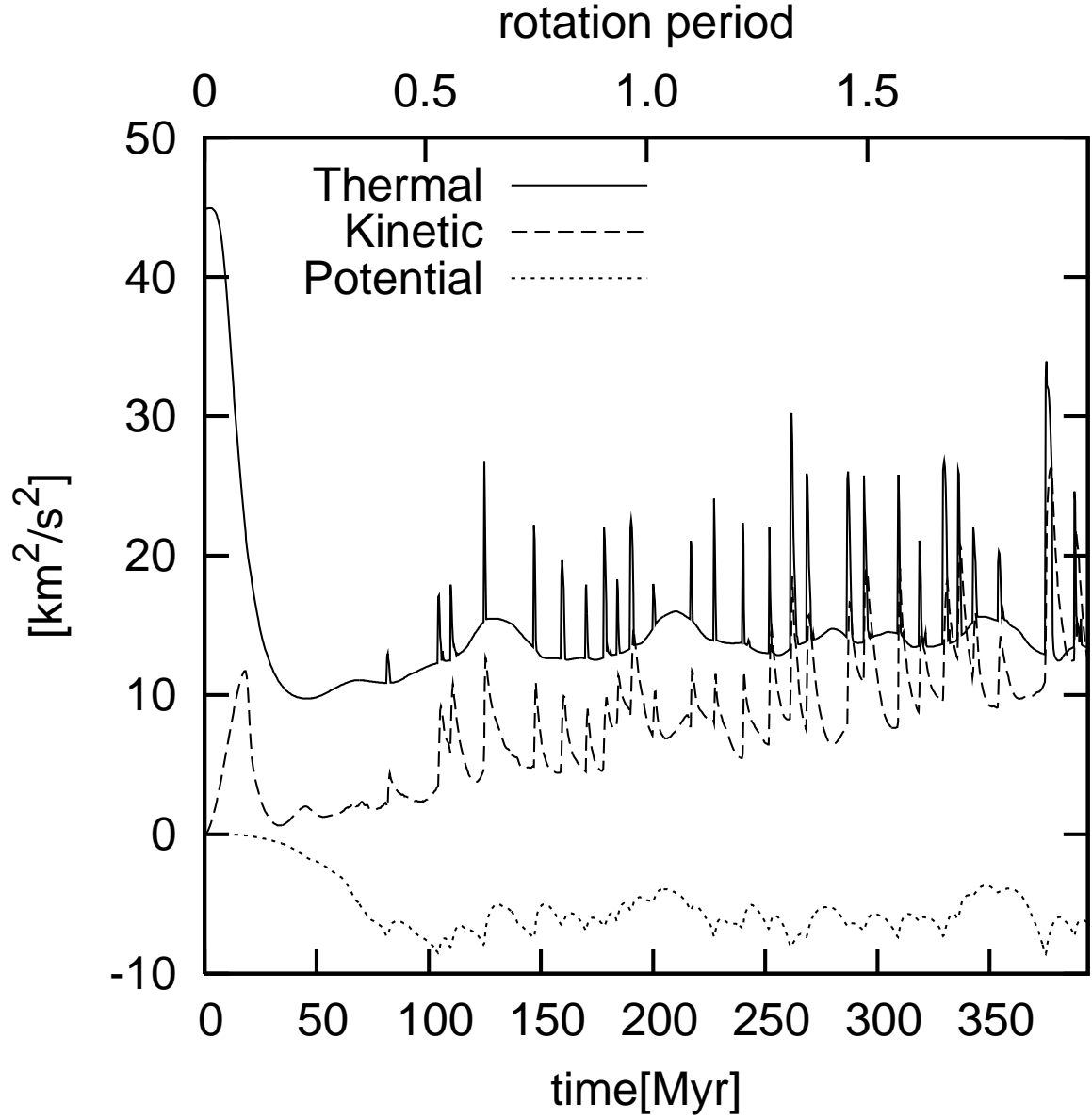}
\caption{Time evolution of component energies in Model Q11: Solid, Long-dashed,
 and Short-dashed lines denote the thermal, kinetic, and potential
 energies, respectively, integrated over the domain.
 For the potential energy, we use the relative potential $\Phi^{(1)}$ 
(see eqn. \ref{eq:relative potential}). }
\label{fig:time}
\end{figure}

Figure \ref{fig:time} shows the time evolution of thermal, kinetic and
potential energies averaged over the domain, for Model Q11.  For the
potential energy, the background disk potential is subtracted out;
i.e. we use $\Phi^{(1)}$ (see eq. \ref{eq:relative potential}).  There
are many sharp spikes in both thermal and kinetic energies, which
correspond to times when new HII regions are born and then rapidly
expand.  The number of spikes corresponds to the number of stellar
generations in the model; note that this number must be proportional
to the domain size.  In the second rotational period (i.e., $1 \le t
\le 2$ rotation), there are 18 generations per 1.16 kpc, or 6
generations per massive cloud per rotation period 
(i.e. an interval of $3.3 \times 10^7$ yr) 
if the mean number of clouds is 3 in this fiducial model.

\begin{figure*}[p]
\plotone{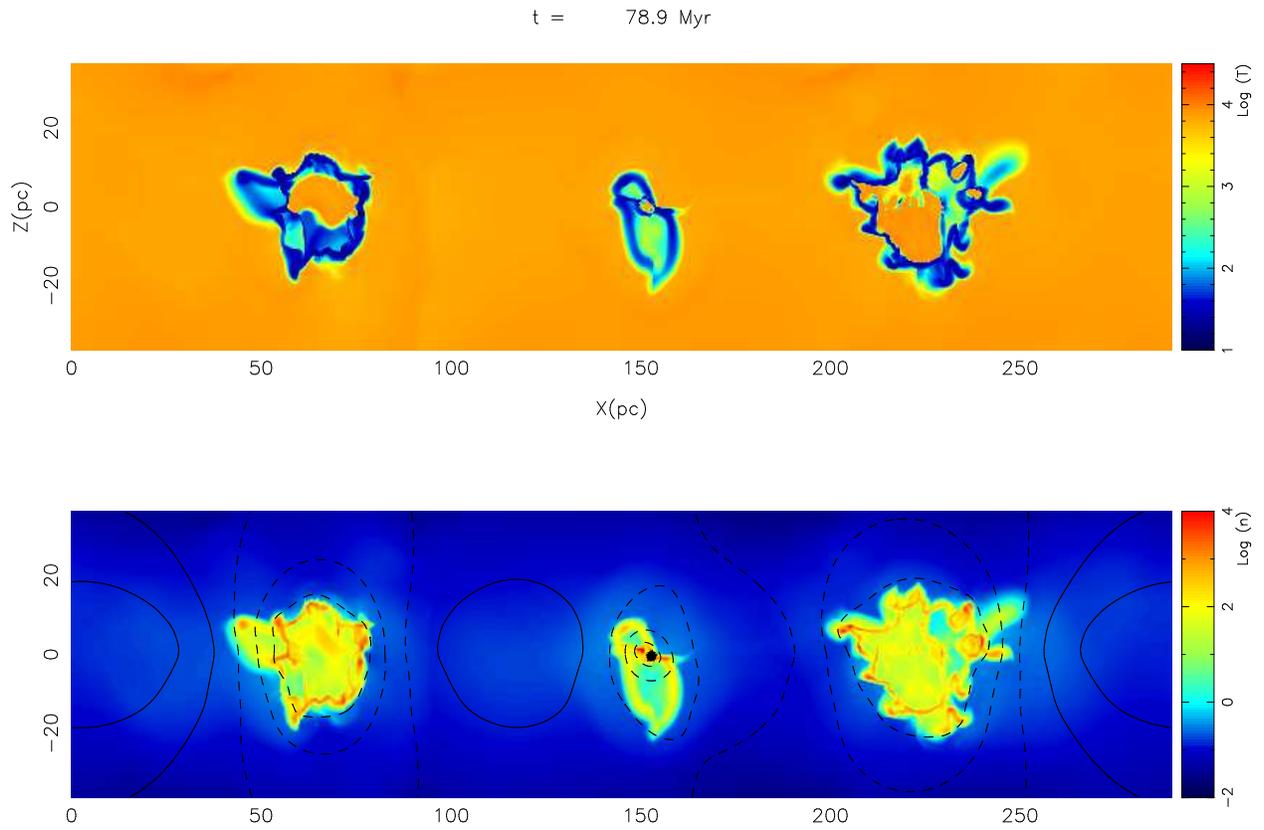}
\caption{A snapshot of temperature ({\it top}) and
 density ({\it bottom}) in high surface density Model Q42. See 
Figure \ref{fig:snap-1} legend for details.}
\label{fig:snap-3}
\end{figure*}

Other models show similar overall behavior in terms of the evolution
of physical structure (consisting of clumps and filaments that
disperse and re-collect), as well as the distribution of mass in the
phase plane.  As environmental parameters vary along a sequence,
however, there are some characteristic changes in structure.  The most
notable difference is that for high gaseous surface density cases,
clouds are often more physically concentrated 
(i.e., more compact and dense) because of the higher
stellar and gaseous gravity.  
For example, Figure \ref{fig:snap-3} shows a snapshot
from Model Q42, which has $\Sigma$ and 
$\rho_{\ast}$ four and 16 times larger, respectively, than the values
in the
fiducial model Q11. The three clouds that are seen in the figure are
more compact than in the lower-$\rho_{\ast}$ case.  For a given $\Phi_{\rm
  SF}$ and $\epsilon$, increasing the mass of a cloud implies that 
HII region cannot break out as easily.

\section{Parameter Dependence of Statistical Properties}

All of our models show a turbulent, multiphase ISM with several
generations of feedback from photoheating.  In this section, we
analyze how the statistical properties of those models depend on the
environmental parameters, both along a given series and from one
series to another.  The statistical properties that we study are based
on averages of the fluid variables over space and time.  First we
describe how these averages are defined in general, and then we turn
to the particular statistics.

\subsection{Space and Time Averaging Procedure}

For a variable of $A$ that is averaged over both space and time,
we use mass-weighted averages defined by:
\begin{eqnarray}
\langle A(a)\rangle &\equiv&
\frac{\int_{a}  A\,dm}
{M(a)},\\
M(a)&\equiv& \int_{a} \,dm,
\end{eqnarray}
where the argument `$a$' denotes a given phase or component of the gas
(such as WNM).
The time averaging is then defined via:
\begin{eqnarray}
\overline{\langle A(a)\rangle}&\equiv&
\int_{t_1}^{t_2} \langle A(a)\rangle \,dt \Big/ S,
\label{eq:space-time-av}
\\
S&\equiv&
\int_{t_1}^{t_2} \theta ( M(a) )dt,
\end{eqnarray}
where the step function $\theta$ gives $1$ 
if $M(a)>0$ 
and $\theta=0$ otherwise, 
so that only intervals in which the component is present are included
in the averaging.
Note that $S=t_2-t_1$ if the component `$a$' is present somewhere in
the domain at every interval during the simulation.
Thus, we denote spatial and temporal averages using angle brackets and 
overlines, respectively.
To avoid the initial transients at the beginning of the simulation,
we adopt the interval between $t_1=t_{\rm final}/2$ and $t_2=t_{\rm final}$
for our time average.
For the purpose of averaging, our sampling rate is 
$\Delta t=0.1 t_{\rm ms}$, where $t_{\rm ms}=3.7 \times 10^6$ yr 
is the adopted lifetime
of ``star zone'' flags that control photoheating feedback.

\subsection{Mass Fractions}

The models of this paper focus on dynamics rather than chemistry, so
rather than dividing the gas into distinct phases, we simply bin it
according to density.  The neutral gas (i.e. the gas that is not
within the limits defining HII regions) is separated into four bins.
The first bin ($n<1 ~ \cm^{-3}$) 
corresponds approximately to the WNM, with densities 
below the maximum for which a warm phase is possible in thermal
equilibrium.  The second bin  ($1~ \cm^{-3} <n \le 100 ~ \cm^{-3}$) 
extends up to the maximum density that is in pressure equilibrium with
WNM gas, and corresponds approximately to the CNM (phase diagrams
show that the thermally-unstable regime is not highly populated for
our models; see e.g. Figs \ref{fig:phase-1} and \ref{fig:PDF-1}).
The dense medium (hereafter DM) is all the gas at 
$n \ge 100 ~ \cm^{-3}$, 
which in thermal equilibrium is above the maximum pressure
for the warm phase and therefore only exists in regions that are
internally stratified due to gravity.  The DM gas corresponds
approximately to the molecular component of the ISM; we divide it into two 
bins, DM2 ($100<n \le 10^3$) and DM3 ($10^3<n $).  Gas that is within
the limits defined for enhanced heating is labeled as ionized gas
(hereafter HII).  So that the mass fractions $f(a)$ 
of all components add to unity,  $\sum_a f(a)=1$, 
we use a slightly different definition from that of equation
(\ref{eq:space-time-av}).  The 
mass fraction of each component $a$ in (WNM, CNM, DM2, DM3 or HII) is
defined as 
\begin{eqnarray}
f(a)=\frac{\displaystyle\int_{t_1}^{t_2}\frac{M(a)}{M_{\rm tot}}\,dt}{t_2-t_1}.
\label{eq:fraction}
\end{eqnarray}

\begin{figure}[p]
\epsscale{1.1}
\plottwo{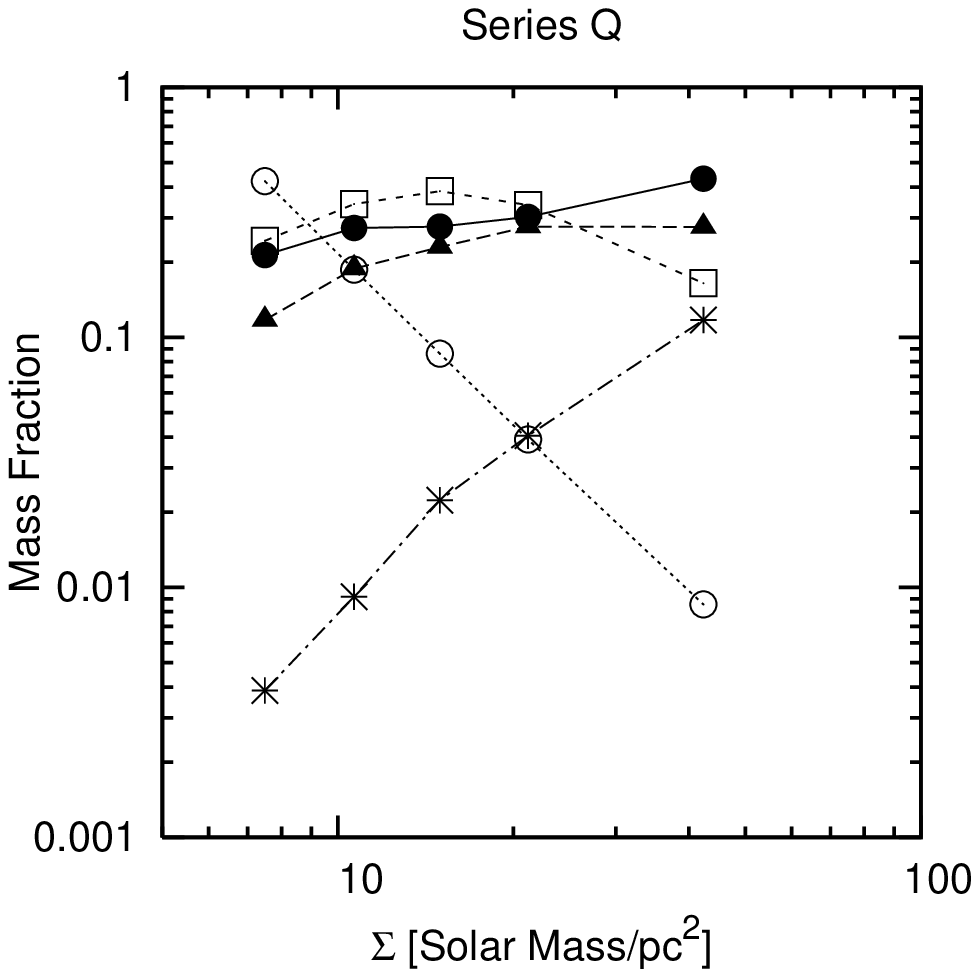}{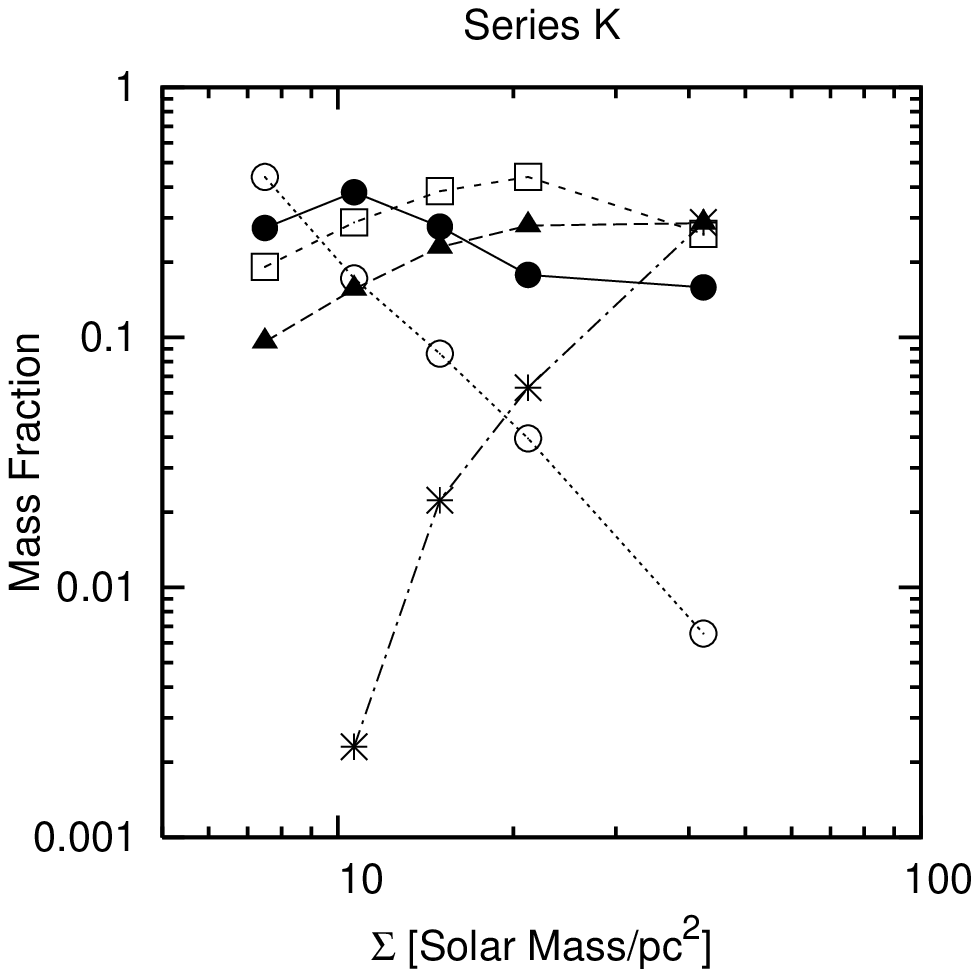}
\plottwo{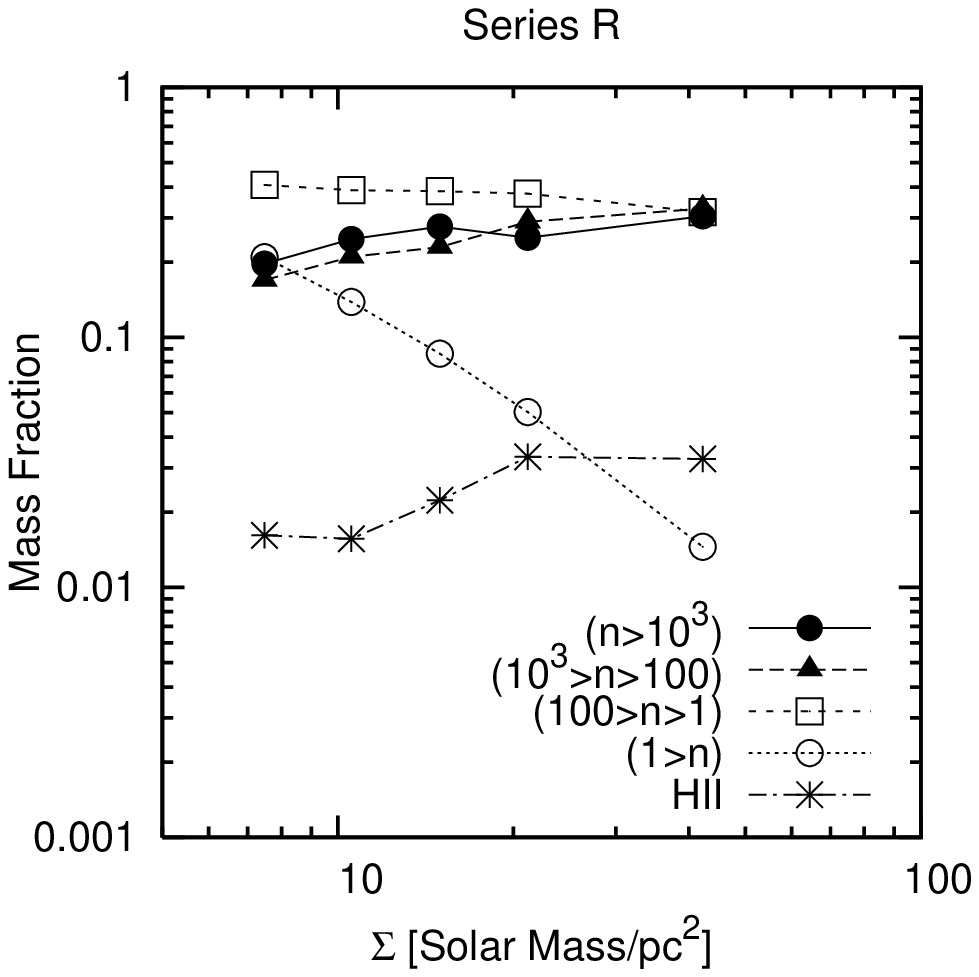}{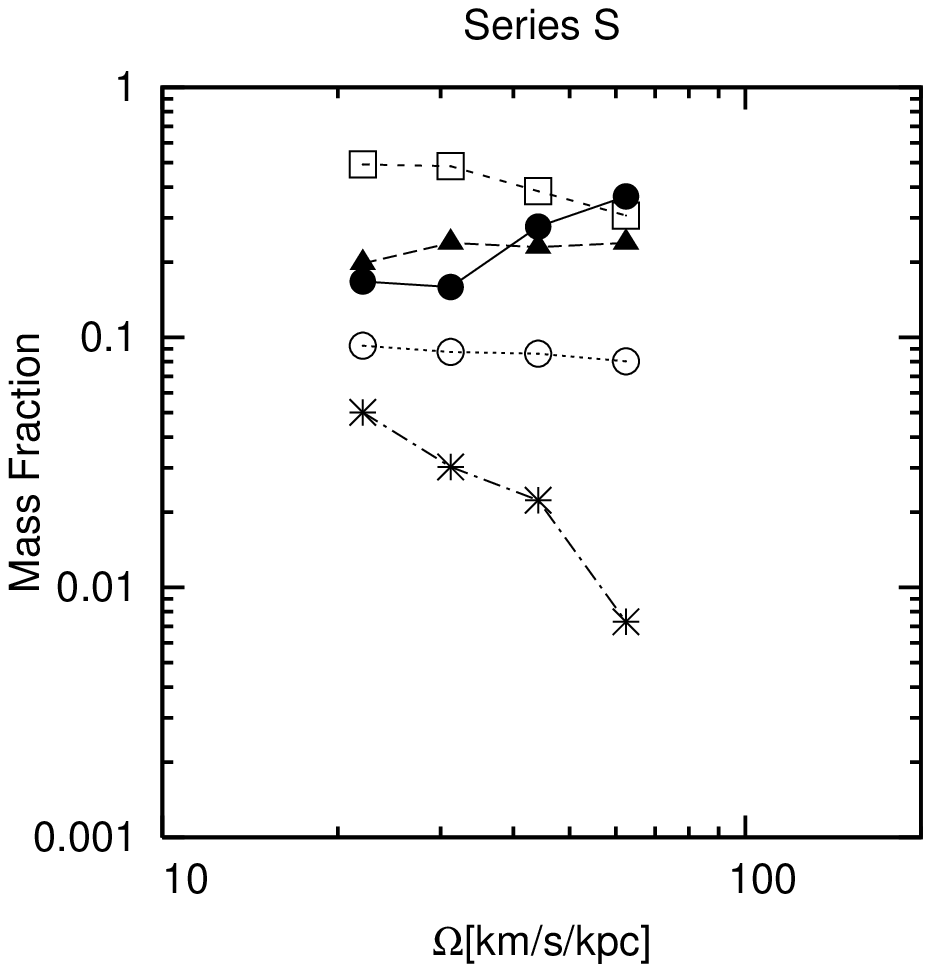}
 \caption{Mass fractions of gas components in all model series, 
as a function of surface density (Series Q, K, R) or 
angular velocity (Series S). Open symbols denote neutral 
gas at densities in the primarily-atomic regime; 
filled symbols denote neutral gas at densities in the
primarily-molecular regime; asterisks denote photoheated gas
corresponding to HII regions.}
\label{fig:mass}
\end{figure}

Figure \ref{fig:mass} shows the mass fraction of the various
components either as a function of surface density (Series Q, R and K)
or angular velocity (Series S). For Series Q and R (which most closely
correspond to the radial variations found within normal spiral
galaxies), at low $\Sigma$ the diffuse (WNM+CNM) components dominate,
while at high $\Sigma$ the dense (gravitationally-confined) components
(DM2+DM3) dominate. The behavior is somewhat different in Series K,
which is highly gravitationally unstable and thus extremely active
when $\Sigma$ is large (since $\kappa$ is constant), leading to larger
HII and CNM mass fractions at high $\Sigma$.  At low $\Sigma$, the
behavior in Series K is similar to that in Series Q and R. For all
series, the HII mass fraction increases at higher $\Sigma$ or lower
$\Omega$, corresponding to lower Toomre $Q$ (see Figure
\ref{fig:Qvalue}) and hence higher rates of stellar feedback
activity. The mass fraction of the WNM component secularly declines
with increasing $\Sigma$ in model series Q, K, and R.
Even though the models of Series S are most gravitationally unstable at
low $\Omega$, they remain dominated by diffuse gas (CNM) rather than
dense gas, because the total surface density is relatively modest for
this series ($\Sigma=15 ~ \msun \pc^{-2}$).

\subsection{Surface Density}

\begin{figure}[p]
\epsscale{1.1}
\plottwo{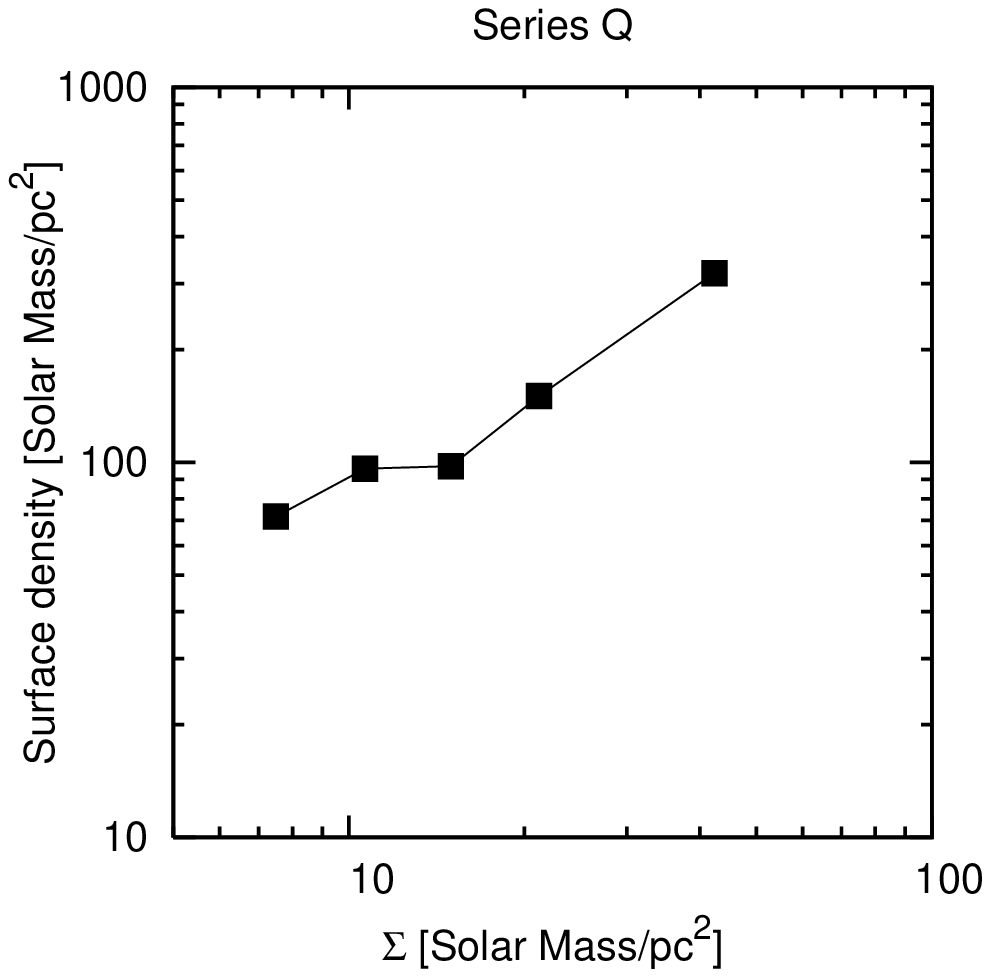}{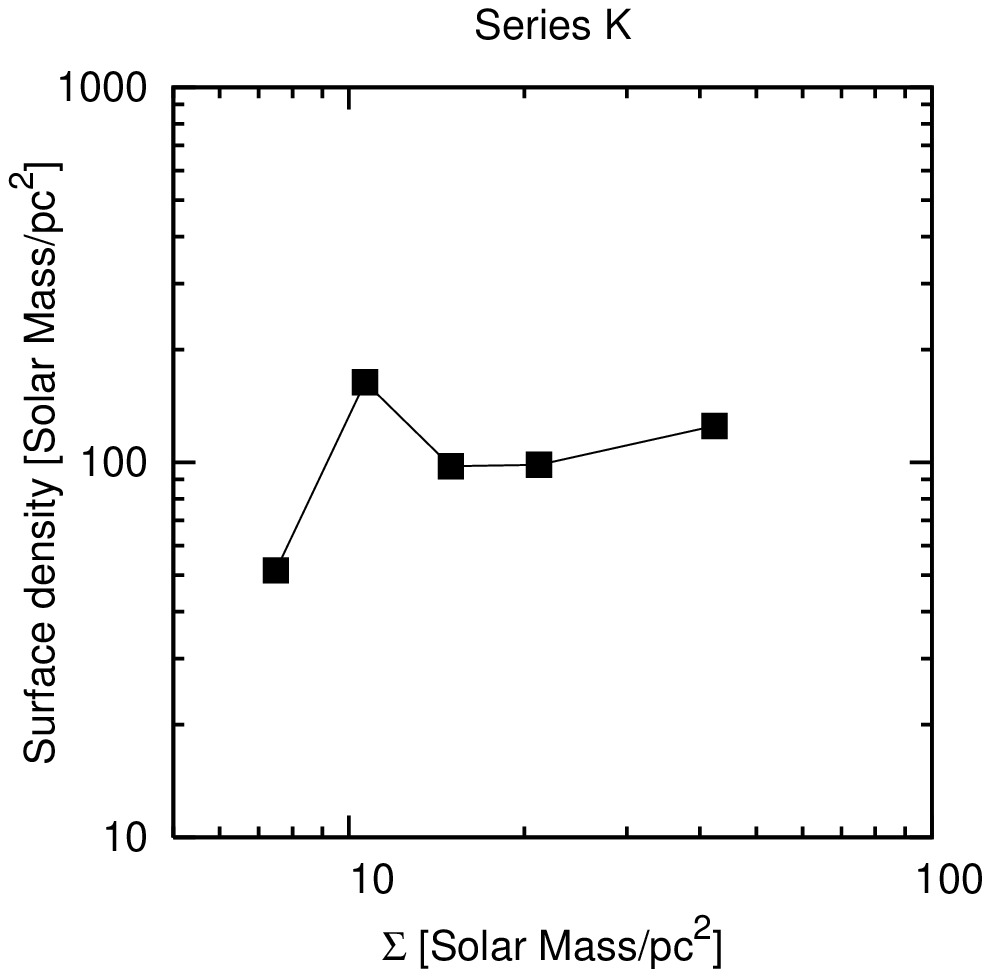}
\plottwo{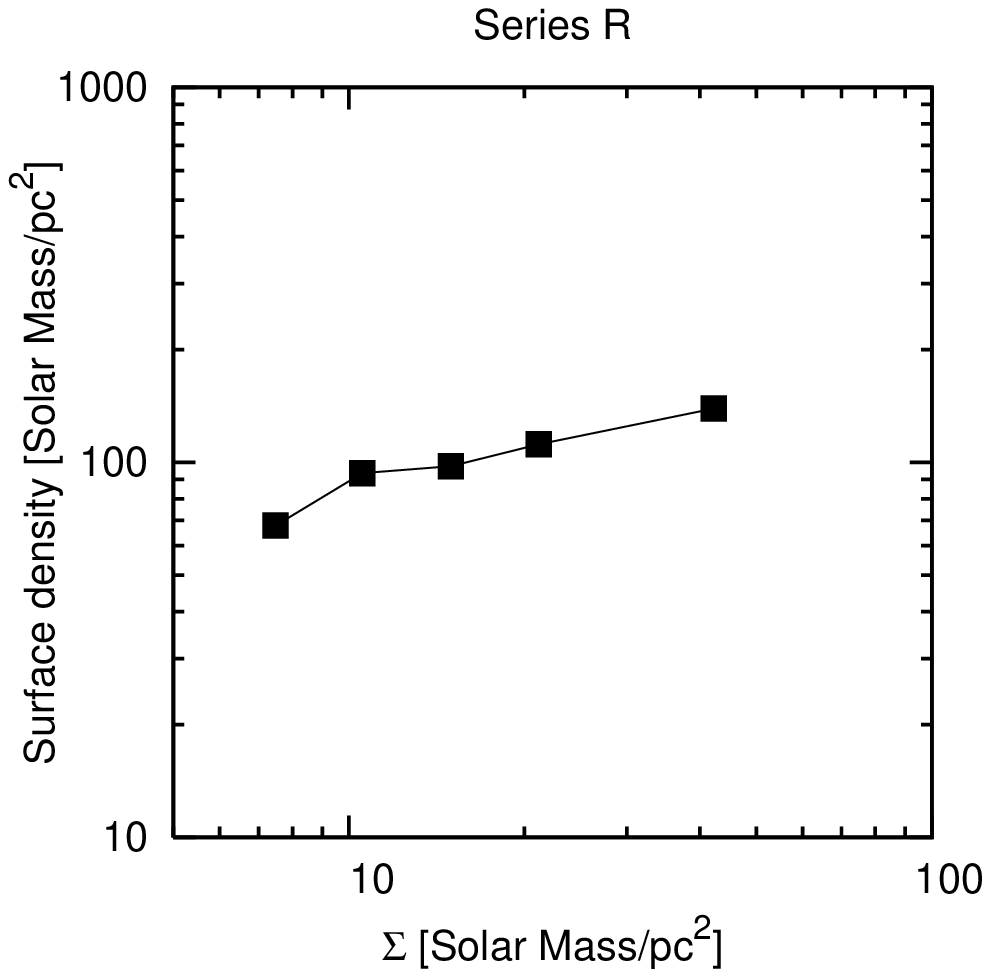}{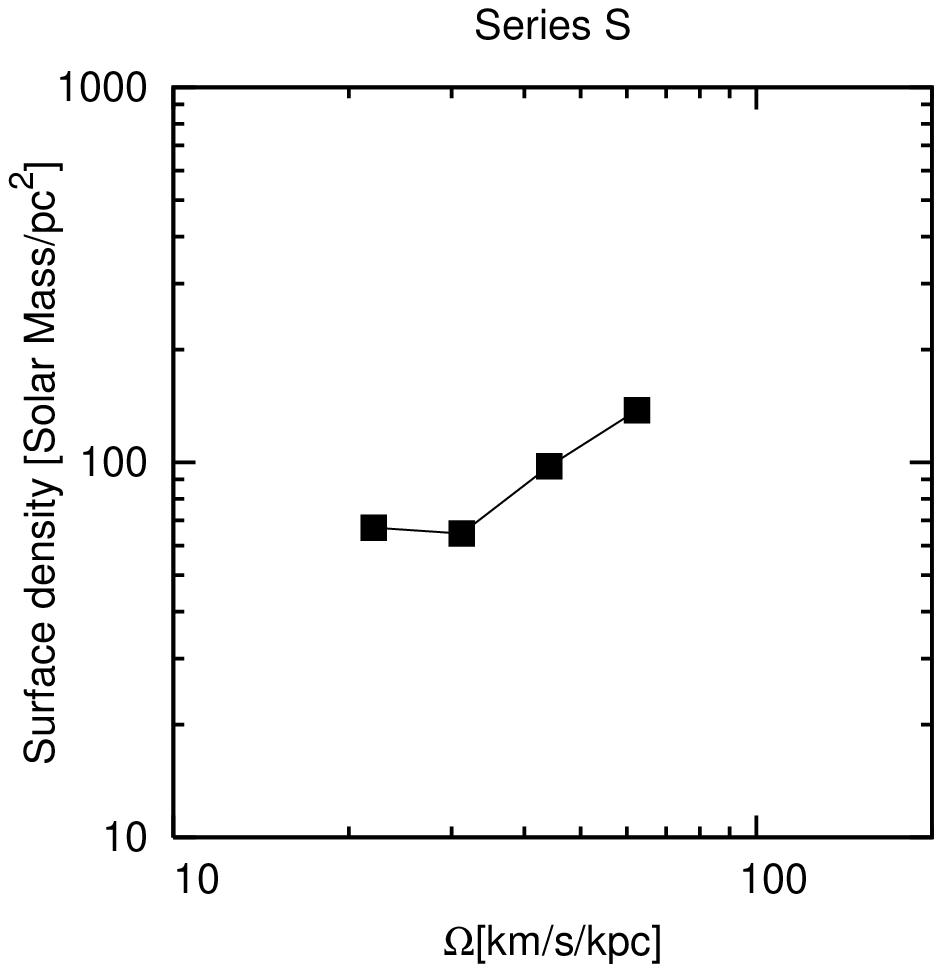}
 \caption{The mass-weighted surface density of gas ($\Sigma_{\rm
   cloud}$; see
   eq. \ref{eq:S}) as a function of area-weighted
   surface density $\Sigma$.  Values of $\Sigma_{\rm cloud}$ are
   similar to observed GMC surface densities, and relatively
   independent of environmental conditions.}
\label{fig:S}
\end{figure}

The simulation domain for our two-dimensional models represents a
radial-vertical ($x-z$) slice through a galactic disk, such that if we
viewed the corresponding galaxy face-on, the surface density as
a function of radius would be given by $\Sigma(x)=\int dz \rho(x,z) $.  The
area-weighted
mean surface density in any model is equal to 
$\langle\Sigma\rangle_A \equiv \int
dx dz\, \rho/L_x$; this is a conserved quantity for any simulation, and
is one of the basic model parameters (see column 2 of Table
\ref{tbl:models}; in general we omit the angle brackets and ``A'' subscript).
The value of the surface density weighted by mass rather than by area 
better represents the ``typical'' surface density of clouds found in the disk.
This is defined as
\begin{eqnarray}
\Sigma_{\rm cloud} = 
\frac{\int \Sigma(x)^2\,dx}{\int \Sigma(x)\,dx}.
\label{eq:S}
\end{eqnarray}
Figure \ref{fig:S} shows $\Sigma_{\rm cloud}$ as a function of
$\Sigma$ for all model series.  Interestingly, we find that
$\Sigma_{\rm cloud}$ does not strongly depend on parameters (either
$\Sigma$ or $\Omega$) throughout the four model series.
The largest value of $\Sigma_{\rm cloud}$ is $300 \msun \pc^{-2}$ and
the smallest is $50 \msun \pc^{-2}$, although for most cases the range
is even smaller: $\Sigma_{\rm cloud}=70-150\msun \pc^{-2}$.  This
factor-of-two range for $\Sigma_{\rm cloud}$ is significantly smaller
than the factor-of-six range of mean surface densities, $\Sigma = 7.5
- 42\msun \pc^{-2}$.  The range of $\Sigma_{\rm cloud}$ is also 
similar to the typical observed surface densities of giant molecular clouds
(see discussion in \S \ref{Summary}).

This weak variation of $\Sigma_{\rm cloud}$ among the various series
suggests that it is star formation feedback, rather than the
feedback-independent parameters, that determines the typical surface
density of clouds.
In particular, other tests we have performed suggest that it is 
 the gravitational potential threshold for star formation 
$\Phi_{\rm SF}$
that most
influences $\Sigma_{\rm cloud}$.  
A model in which $\Phi_{\rm SF} \to 0$ (with any $\epsilon$) would
have photoheating events independent of the local ISM properties; 
the consequent expansion of HII regions would thoroughly mix gas so
that $\Sigma_{\rm cloud} \to \Sigma$.  This is indeed what we find
when we run models with $|\Phi_{\rm SF}|$ a factor 10 below our adopted
value. On the other hand, larger values of $|\Phi_{\rm SF}|$ require
more massive and compact clouds in order to have star formation, which
would raise $\Sigma_{\rm cloud}$.  Tests with $|\Phi_{\rm SF}|$ a
factor 10 above our adopted value indeed result in larger $\Sigma_{\rm
  cloud}$ (although only by a factor $\sim 2$).
The dependence on $\epsilon$ is much weaker than the dependence on 
$\Phi_{\rm SF}$; reducing $\epsilon$ by a factor 10 changes 
$\Sigma_{\rm cloud}$ by only tens of percent, at our standard $\Phi_{\rm SF}$.

The comparison between Series Q and Series R is also interesting, in
this regard.  The difference between these two series is that the
stellar density $\rho_{\ast}$ increases with $\Sigma$ in Series Q, while
$\rho_{\ast}$ is constant throughout Series R.
Based on the larger
$\Sigma_{\rm cloud}$ value for the largest $\Sigma$ in Series Q
compared to Series R, when the stellar density is increased, the
surface density required in order to form clouds also increases.

\subsection{Temperatures}

\begin{figure}[p]
\epsscale{1.1}
\plottwo{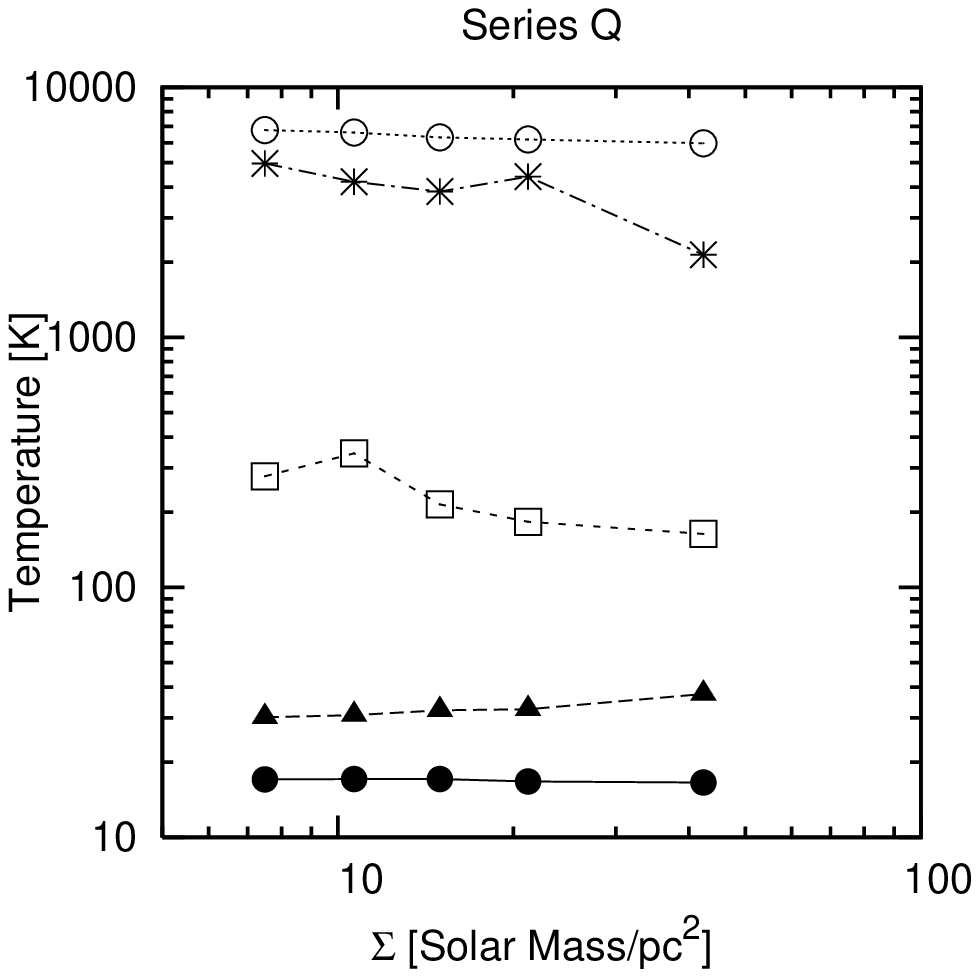}{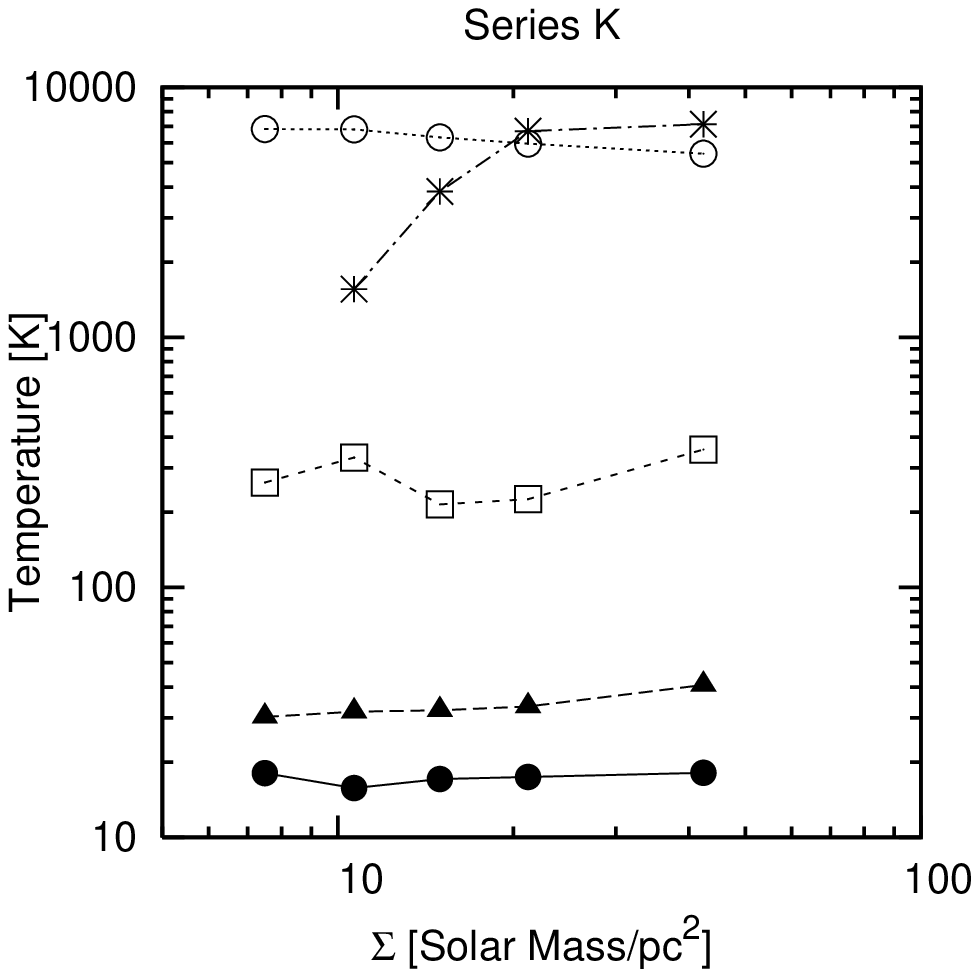}
\plottwo{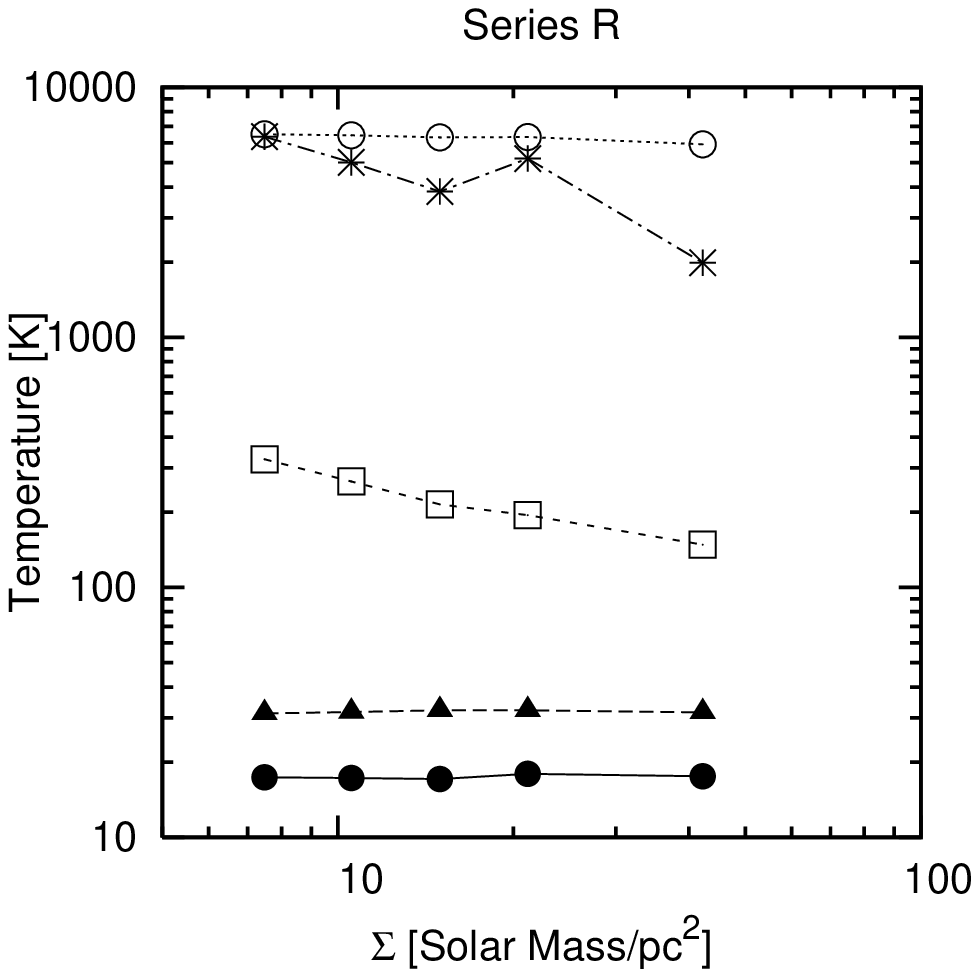}{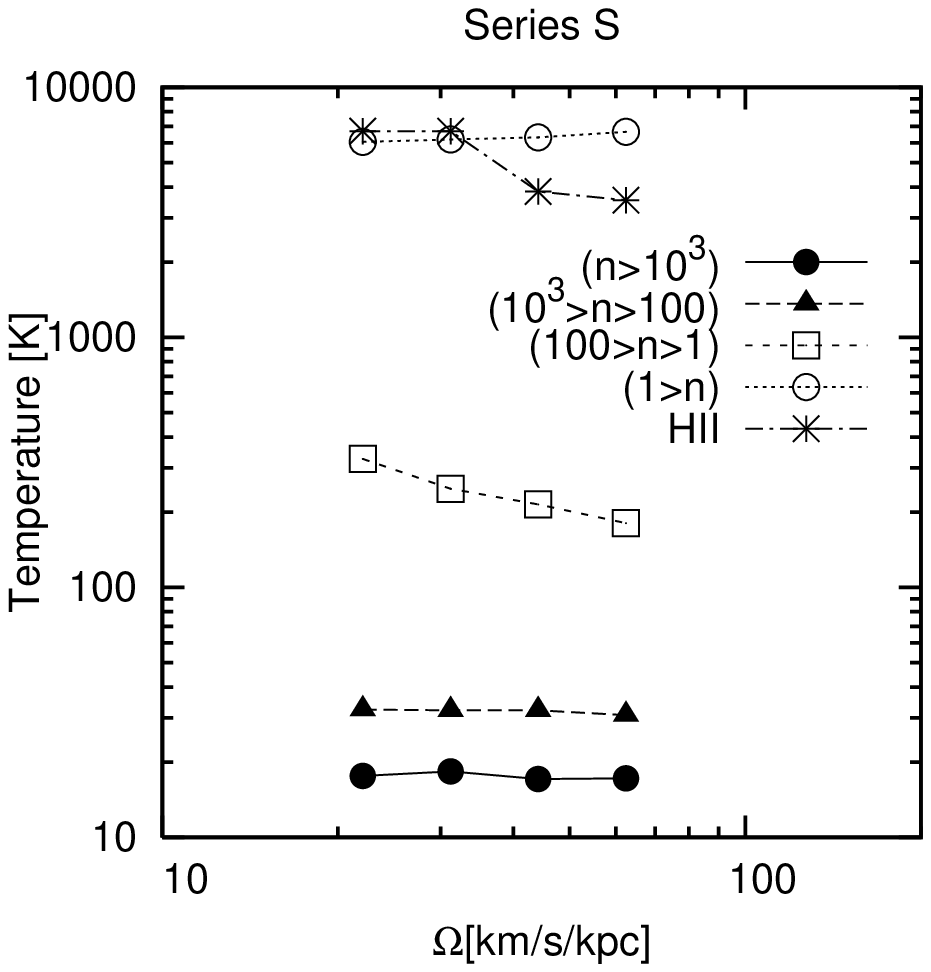}
 \caption{Mean temperature in varying density ranges.
Diffuse (WNM: open circles, CNM: open squares), dense 
(DM2: filled triangles, DM3: filled circles), and photoheated (HII:
 asterisks) components are defined as in Fig.
\ref{fig:mass}.}
\label{fig:temp}
\end{figure}

Figure \ref{fig:temp} shows the space-time averages of temperature 
for the components we have defined via density bins, 
$\overline{\langle T(a)\rangle}$, 
where the argument `$a$' denotes WNM, CNM, DM2, DM3 and HII.
Throughout the model series, the temperatures for the most dense and
most diffuse components are fairly constant; we find 
$T=6,000-7,000$ K for WNM,  $T=20-40$ K for DM2, and $T=10-20$ K
for DM3.  For the CNM component (which in fact includes
thermally-unstable gas when it exists), the range is somewhat larger, 
$T=100-400$ K, reflecting the larger range of conditions for this gas.
The gas which is subject to enhanced heating 
has mean temperatures for most models of 4,000 -- 8,000 K.

The link between density and temperature in our models implies that
the components we have defined via density bins also approximately
correspond to natural ISM phases.  This is because much of the gas
mass is close to thermal equilibrium (see Figure \ref{fig:phase-1}),
and we have chosen the bin edges so as to match up to points in the
phase plane with physical significance.  Temperature PDFs that we have
constructed show a bimodal distribution, as is expected based on the
cooling function.

\subsection{Turbulent Velocities}

Expansion of HII regions feeds kinetic energy into the ISM.
This kinetic energy is not imparted solely to expanding HII bubbles
and shells surrounding them, but is shared throughout the ISM as turbulence.
Our models provide a first look at the results of this form of
turbulent driving.  It is interesting to examine how the  
turbulent amplitudes vary from one component to another in a given
model, and how the overall levels vary between models
with different feedback rates as a consequence of different system parameters.

Figure \ref{fig:velocity} shows the turbulent velocity dispersions for
all series, defined for each component as:
\begin{eqnarray}
v(a)&\equiv&\sqrt{\overline{
\langle v_x(a)^2+\tilde{v}_y(a)^2+v_z(a)^2\rangle}},
\end{eqnarray}
where the argument `$a$' denotes WNM, CNM, DM2, DM3 and HII, and 
$\tilde{v}_{y}=v_y + \Omega x$ in order to subtract out the
velocity of unperturbed (sheared) rotation about the
galactic center.
The azimuthal velocities are excited by Coriolis forces 
so that
the relation for epicyclic motions 
\begin{equation}
\frac{{\tilde v}_{y}}{{v}_x}\simeq
\frac{\kappa}{2\Omega}=\frac{1}{\sqrt{2}}
\end{equation}
should apply \citep{1987gady.book.....B},
and we have checked that this is in fact satisfied.
We note that the velocity dispersion for each component is computed
by summing over the whole domain.  Thus, the measured velocity
dispersions are larger than they would be within smaller-scale clouds 
in the system.  However, we have found that there
are not large contributions to the velocity dispersion from
velocity differences of widely-separated regions; this is because
turbulence is driven by the expanding HII regions, such that the
maximum correlation scale is comparable to the effective thickness of
the disk.  For example, if we
divide the domain in Fig. \ref{fig:snap-1} 
horizontally into eight equal parts, the mean 
velocity dispersion of all gas 
within these sub-domains is $\sim 98\%$ of that of
the domain as a whole.  Considering just the dense gas, the velocity
dispersion for subdomains is $\sim 75\%$ of that in dense gas 
for the whole domain.  For the three large clouds seen in 
Fig. \ref{fig:snap-3}, the mean internal velocity dispersions are an order
of magnitude larger than the dispersion in mean velocities.

In general, the densest component (DM3) has the lowest velocity
dispersion, with the next-densest (DM2) the next-lowest.  The value of
the velocity dispersions for the dense components are highly
supersonic, and are similar to (or slightly below) those that are
observed within real GMCs (see \S \ref{Summary}).  
The CNM component in our models typically
has higher turbulence levels than the WNM component, because the
former is in closer (space-time) contact with energy sources.  
Because turbulent motions in our models 
are driven by the pressure of photoheated gas, 
$P=\rho c_s^2 \sim \rho v^2$,
the turbulent velocities have an upper limit of the 
sound speed in gas heated to 8,000 K, $c_{\rm HII}\sim 7 ~ \kms$. 
Since the driving is intermittent, this upper limit is not
usually reached; mean values for the diffuse gas are closer to $\sim 5~\kms$.
The diffuse-gas velocity dispersions in our models are lower by about 50\%
compared to observed levels, indicating (consistent with expectations)
that other turbulence sources are important in the real diffuse ISM.

The model series Q and R, which have $\Sigma \propto \Omega$ (and thus
effectively constant gaseous Toomre $Q$ if the velocity dispersion is
constant) show velocity dispersions that are insensitive to the value
of $\Sigma$.  Series K, on the other hand, has much higher turbulence
levels for large $\Sigma$.  This is because, with constant $\Omega$
($=\kappa/\sqrt{2}$) the high-$\Sigma$ models are quite susceptible to
gravitational instability (in terms of $Q$); this leads to very active
feedback which then raises the velocity dispersion.  A similar
physical effect is seen in Series S: the velocity dispersion is
highest at low $\Omega$, since these are the most
gravitationally-susceptible models among the series.  We discuss
measurements of the effective Toomre $Q$ values that account for
turbulence, in the next subsection.

\subsection{Effective Toomre $Q$ Parameters
\label{sec:Toomre}}

For a rotating disk that contains only thermal pressure,
susceptibility to growth of self-gravitating perturbations depends on
the Toomre $Q$ parameter, defined by setting $\sigma_R$ equal to the
thermal sound speed in equation (\ref{eq:ToomreQdef}).  An
infinitesimally-thin gas disk is unstable to axisymmetric
perturbations if the value of this thermal $Q$-parameter is $<1$
\citep{1964ApJ...139.1217T}.  Nonaxisymmetry, magnetic fields, and the
presence of active stars enhance gravitational instability
\citep{1965MNRAS.130..125G,1984ApJ...276..114J,2001MNRAS.323..445R,
2001ApJ...559...70K,2002ApJ...581.1080K,2003ApJ...599.1157K,
2007ApJ...660.1232K,2005ApJ...620L..19L}
allowing growth at higher $Q$, while nonzero disk thickness suppresses
gravitational instability
\citep{1965MNRAS.130...97G,2002ApJ...581.1080K,2007ApJ...660.1232K},
lowering the the critical $Q$ value.  Allowing for all of these
effects, threshold levels measured from
simulations are $Q \approx 1.5$.

Turbulence at scales below the wavelength of gravitational instability
can also help to suppress the growth of large-scale density
perturbations, by contributing to the effective pressure.  Since the
original Toomre parameter is arrived at based on effects of radial
pressure gradients, only the radial component of the velocity
dispersion should be added to the thermal velocity dispersion in
defining an effective $Q$ (see eq. \ref{eq:ToomreQdef}).  It is
natural to expect galactic disks to self-regulate the values of the
effective $Q$: growth of self-gravitating instabilities subsequently leads to
star formation and energetic stellar feedback, which drives
turbulence, raises $Q$, and tends to suppress further GMC formation.
Indeed, the suggestion that galactic star formation is self-regulated through
turbulent feedback dates back to the earliest work on large-scale
instabilities in galactic gas disks \citep{1965MNRAS.130..125G}, 
with \citet{1972ApJ...176L...9Q} making the related suggestion that
galaxies deplete their gas until they reach marginal stability.  
The self-regulation processes are complex, but they have begun to be
studied in recent numerical simulations 
(e.g.\citealt{2002ApJ...577..197W,2008ApJ...673..810T}).

\begin{figure}[p]
\epsscale{1.1}
\plottwo{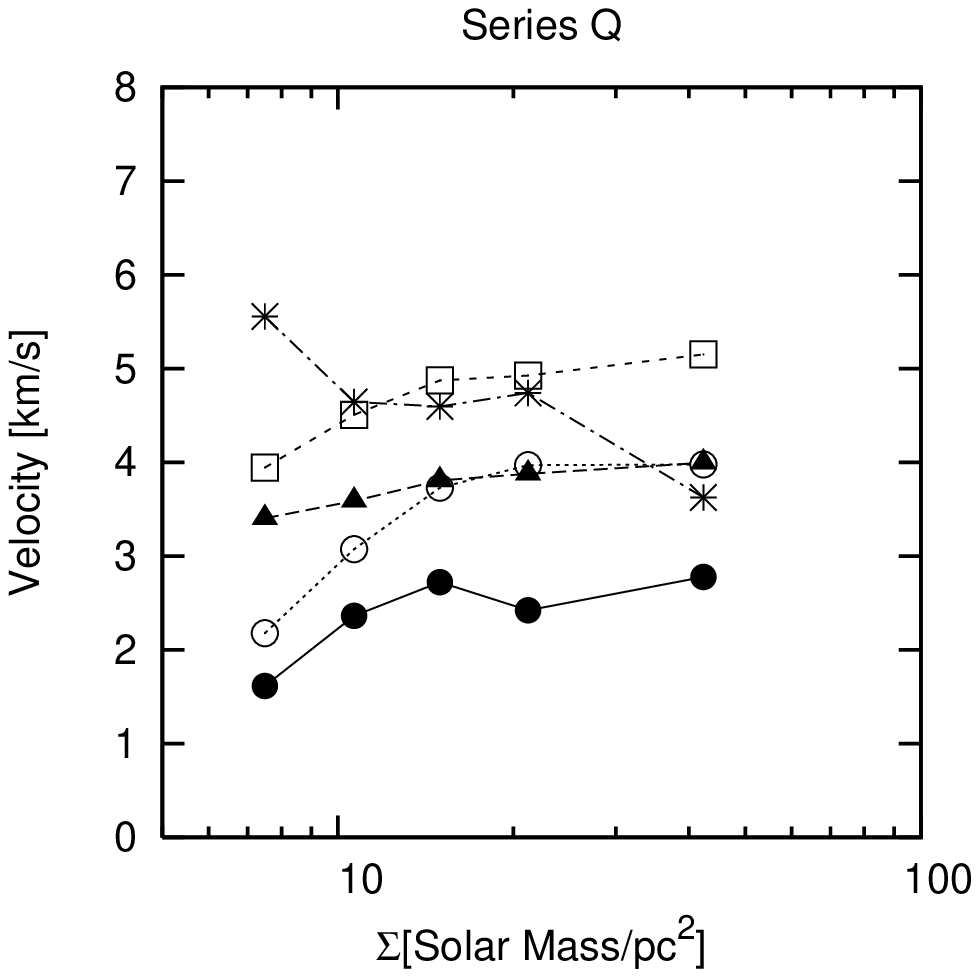}{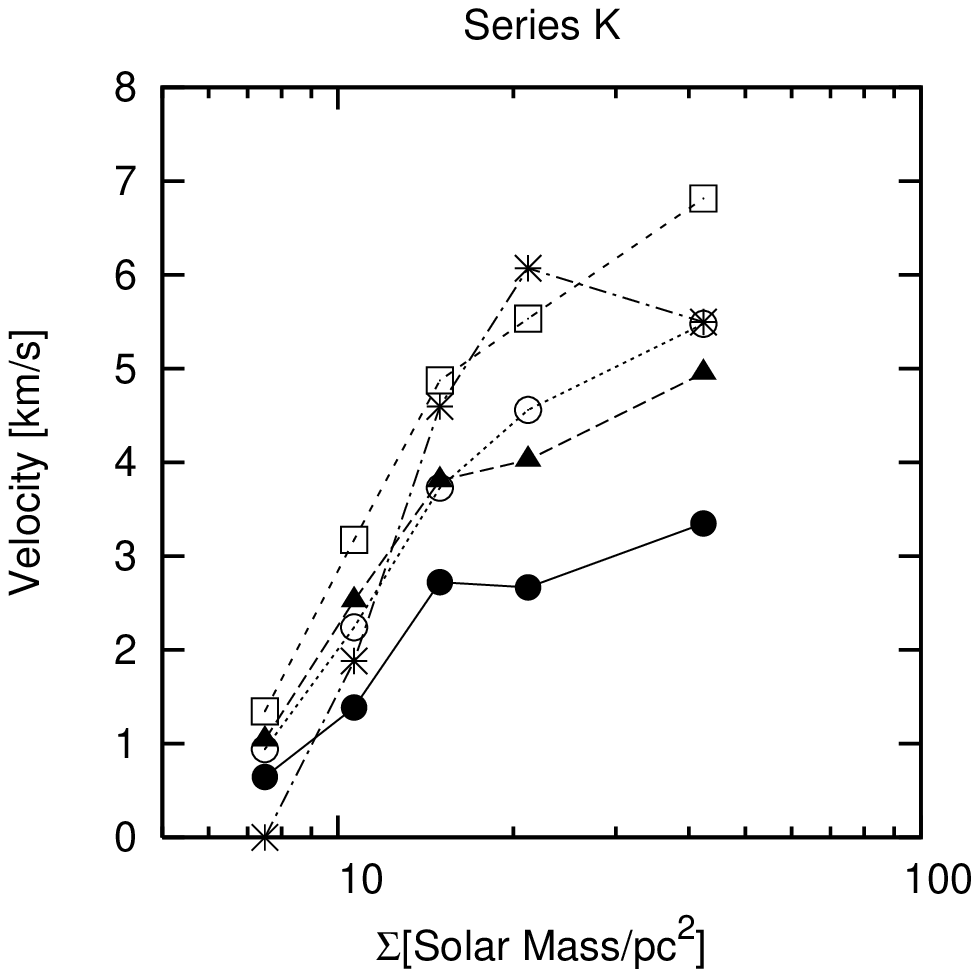}
\plottwo{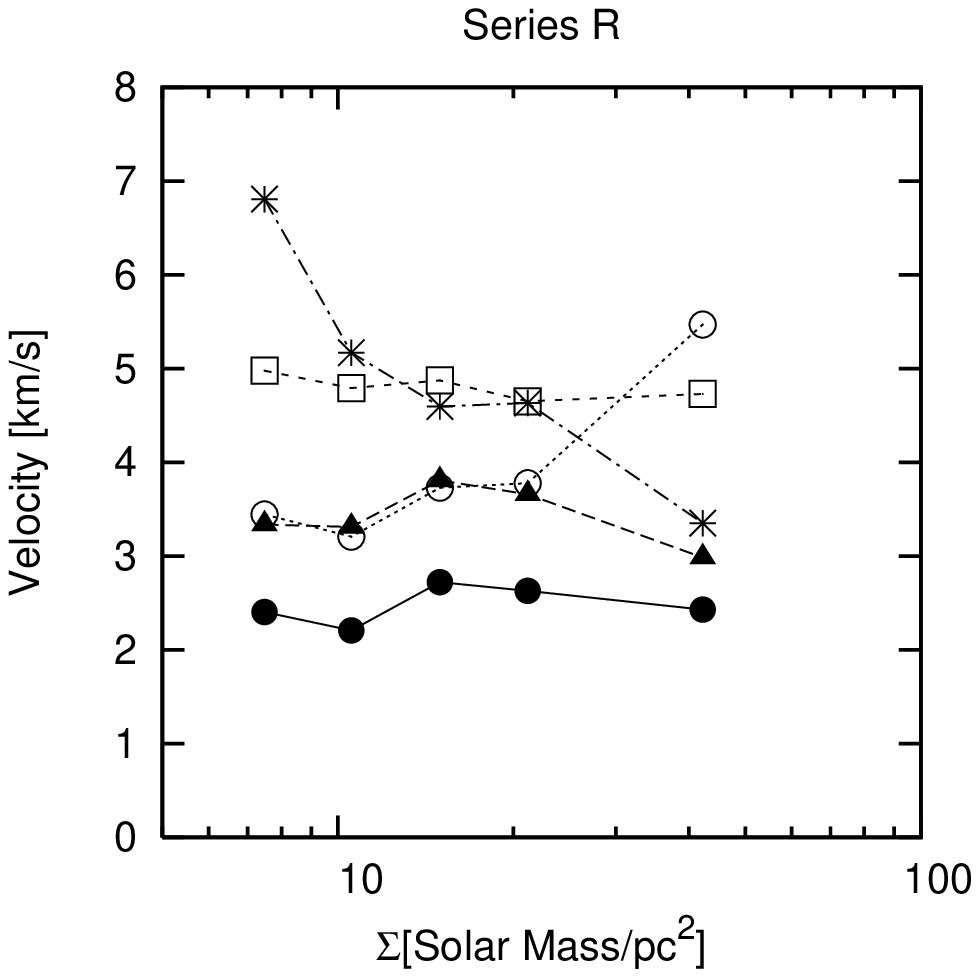}{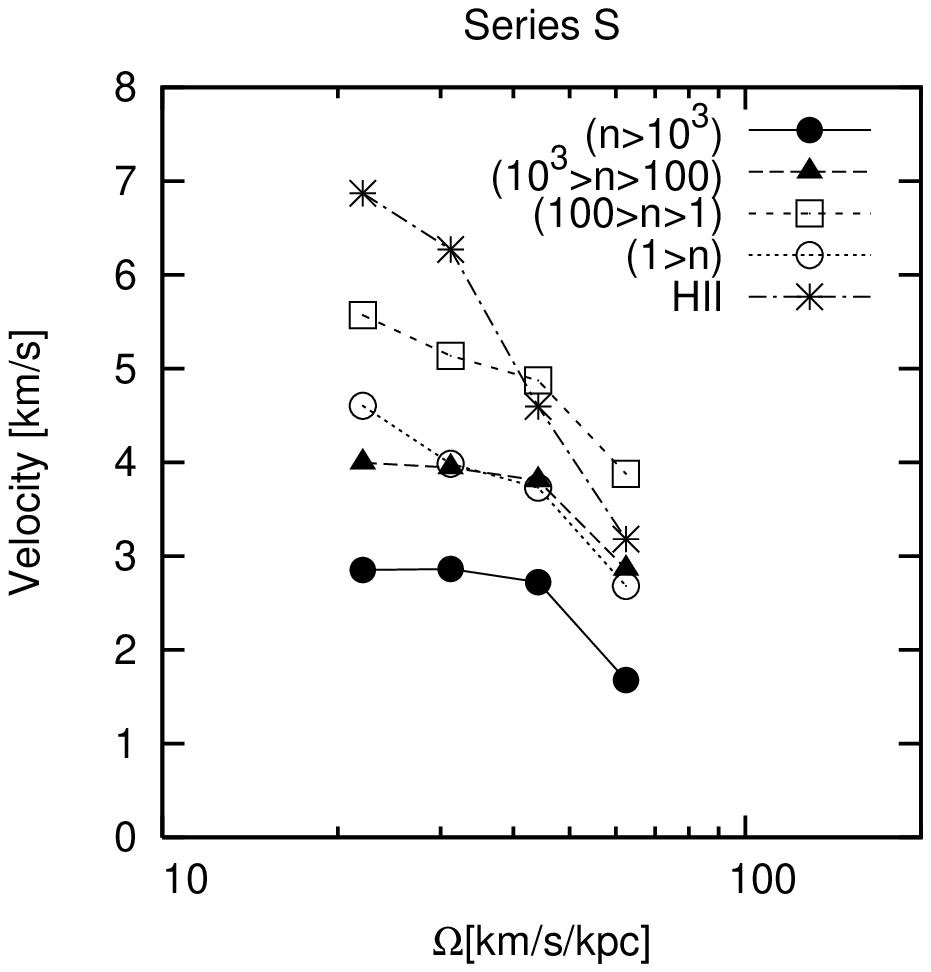}
 \caption{Mean turbulent 
velocity dispersion in each gas component.  Turbulence levels are
relatively independent of $\Sigma$ in Series Q and R which have
$\kappa$ and $\Omega \propto \Sigma$, but turbulence increases with $\Sigma$ in
Series K, which has $\kappa$ and $\Omega$ constant for all models and
is therefore
highly unstable at large $\Sigma$. Turbulence levels decrease with
increasing $\Omega$ in series S, because angular momentum stabilizes against
gravitational collapse and subsequent feedback.
}
\label{fig:velocity}
\end{figure}

\begin{figure}[p]
\epsscale{1.1}
\plottwo{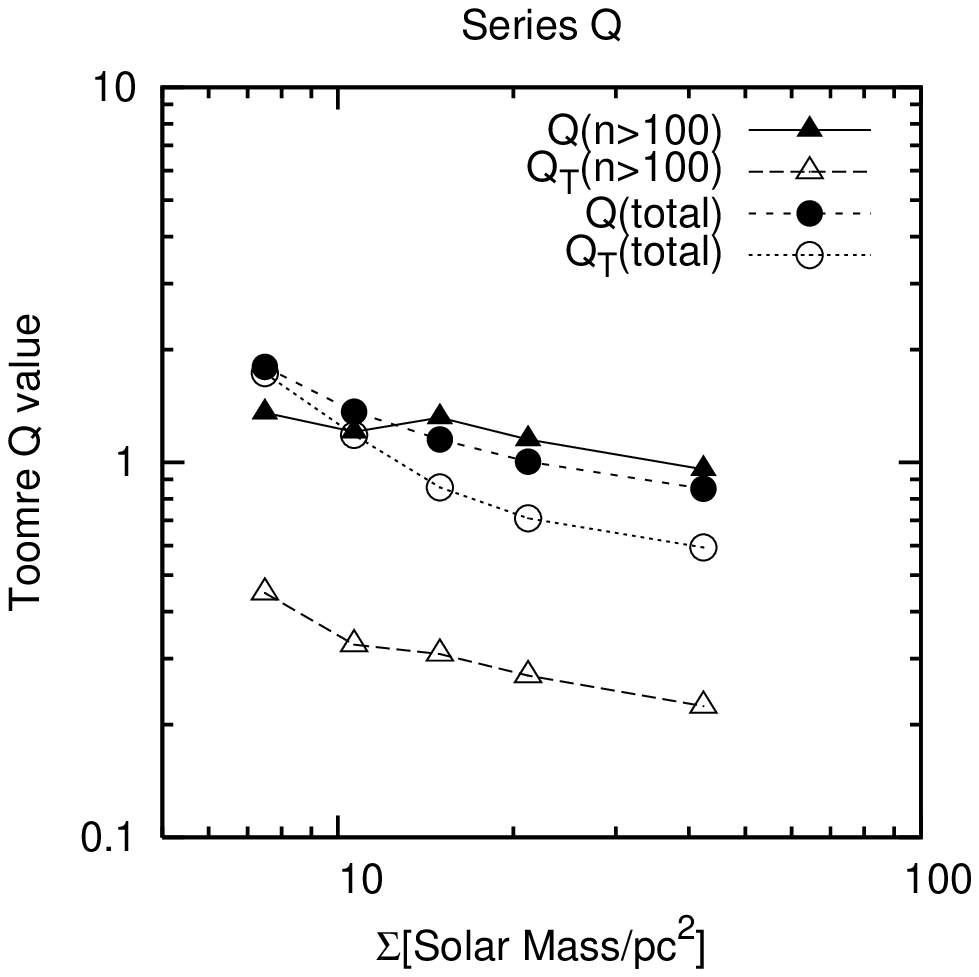}{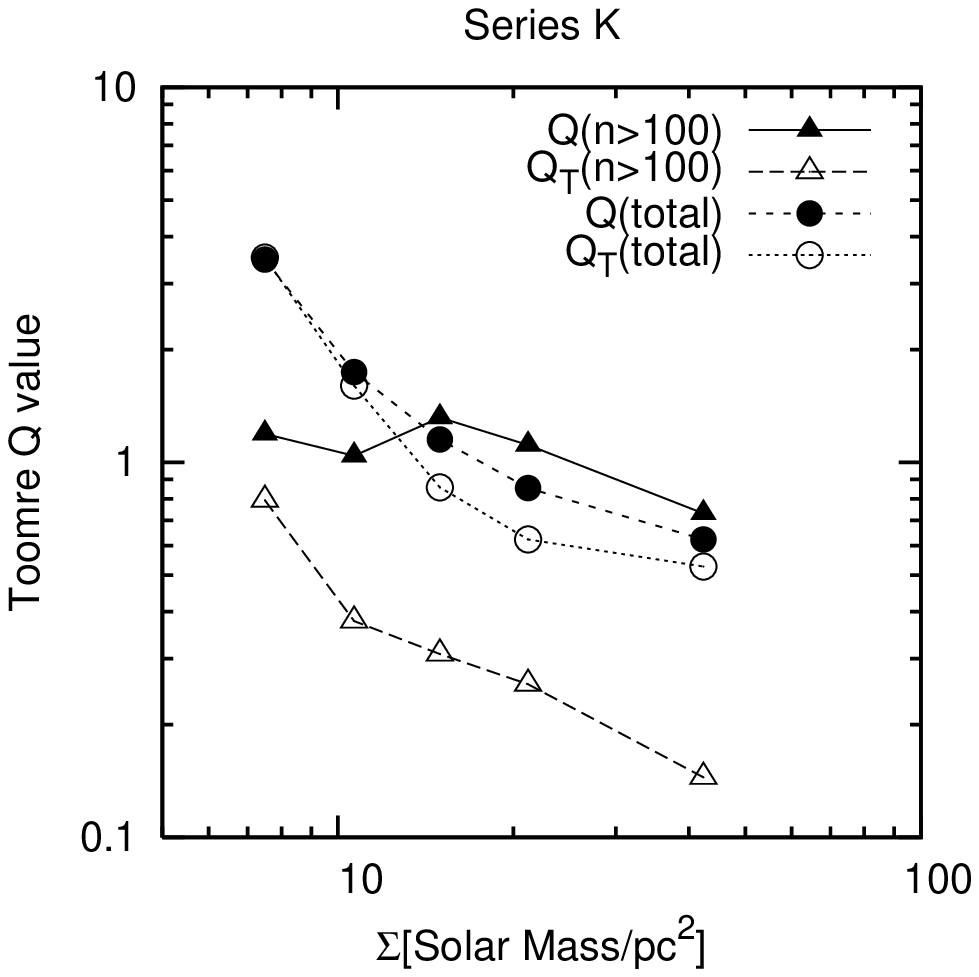}
\plottwo{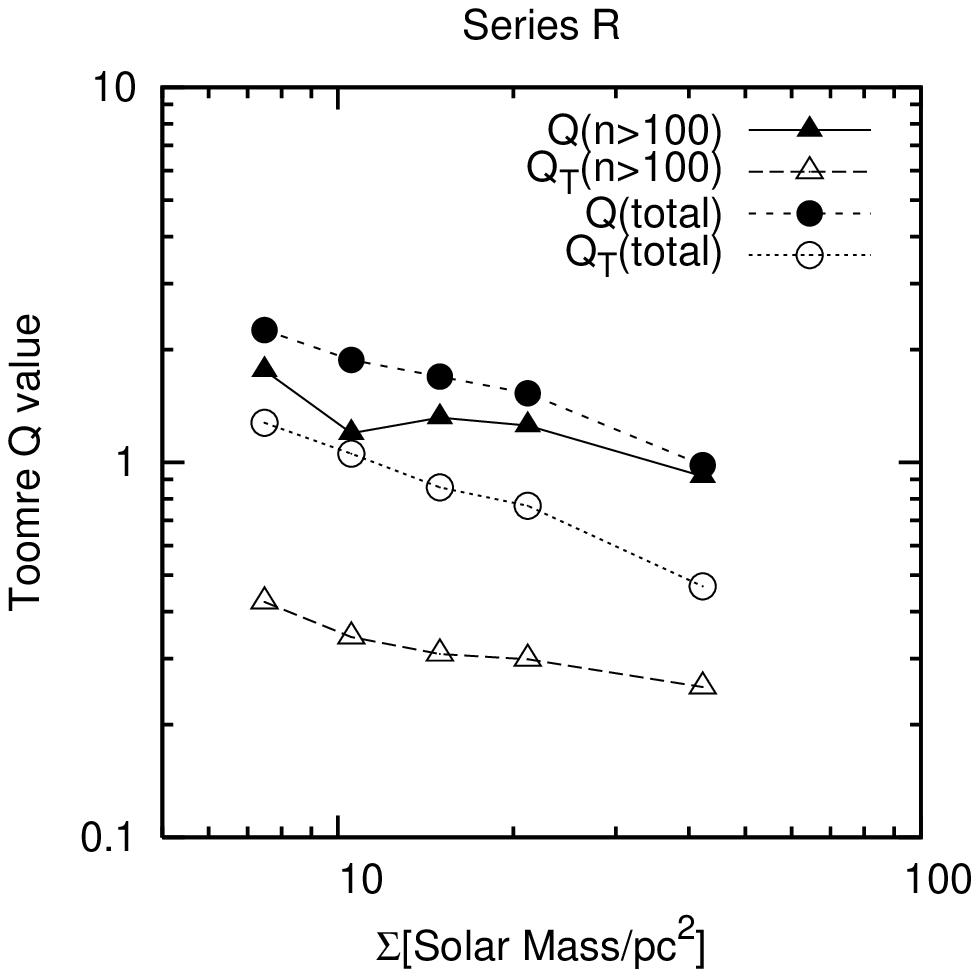}{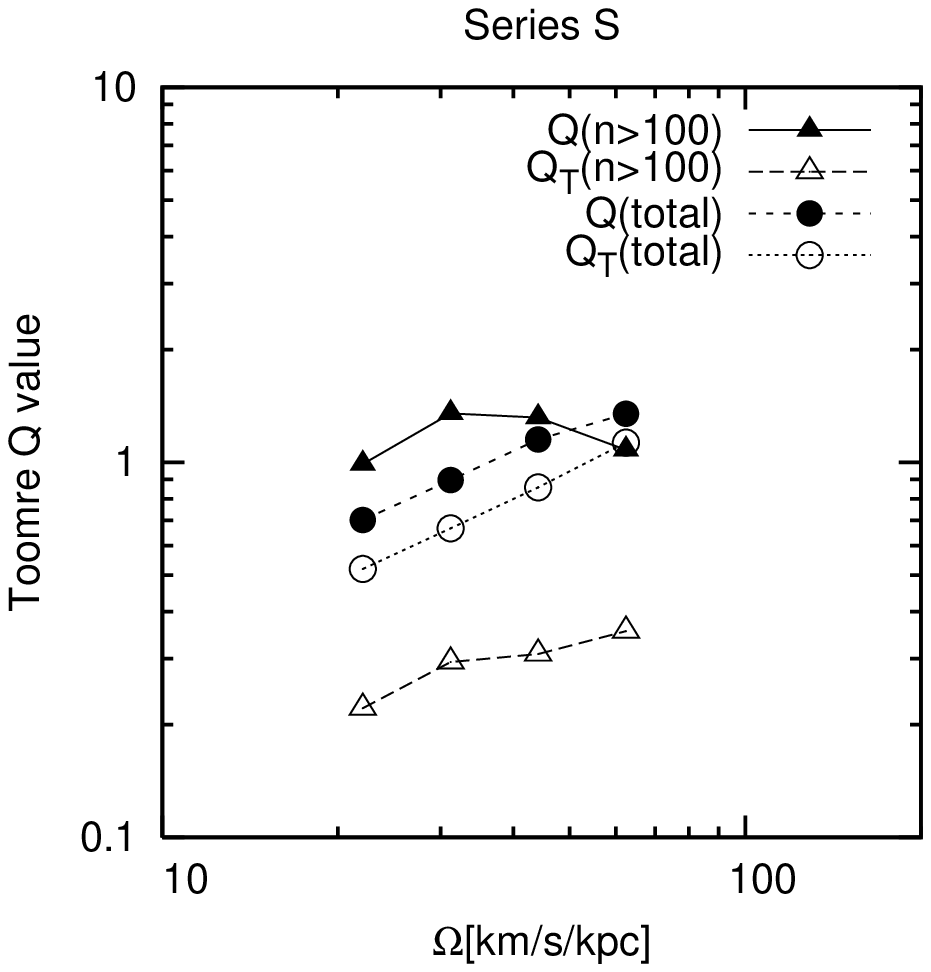}
 \caption{Values of the Toomre $Q$ parameter.
We separately compute $Q$ in four different ways:  including all gas
and either using the thermal  (subscript ``T'') or thermal+turbulent
 velocity dispersion to obtain $Q_T$(total) and  $Q$(total),
 respectively; and including just the (cold)
high-density gas to obtain (thermal) $Q_T(n>100)$ and 
(thermal+turbulent) $Q(n>100)$.  
Feedback drives turbulence and 
regulates the thermal+turbulent Toomre parameter to reach levels near unity,
especially for cold, dense gas.
See text for details.}
\label{fig:Qvalue}
\end{figure}

We use the results of our models to measure the values of the
effective Toomre parameter in the saturated state.  We compare four
different measurements of $Q$ in each model.  The first is closest to
Toomre's original definition for a gaseous medium
in that it is based on thermal velocity 
dispersion; since our medium has components at differing
temperatures, we use a mass-weighted thermal velocity dispersion:
\begin{equation}
Q_T({\rm total}) = \frac{\kappa 
}{\pi G\Sigma}\sqrt{\frac{\gamma \kB
}{\mu}
\overline{\langle T\rangle}}.
\end{equation}
The second measurement incorporates turbulence, again including all gas
and weighting by mass:
\begin{equation}
Q({\rm total}) = \frac{\kappa}{\pi G\Sigma}
\sqrt{\frac{\gamma \kB
}{\mu} \overline{\langle  T \rangle}
+\overline{\langle v_R^2\rangle}}.
\end{equation}
For the third and fourth measurements, 
we consider only the dense gas for both the
numerator and denominator:
\begin{equation}
Q_T(n>100) = \frac{\kappa}{\pi G\Sigma} \frac{\sqrt{ \frac{\gamma \kB
}{\mu} \overline{\langle T(n>100)\rangle}}} {f(n>100)}.
\end{equation}
and 
\begin{equation}
Q(n>100) = \frac{\kappa}{\pi G\Sigma} \frac{\sqrt{
\frac{\gamma \kB}{\mu} \overline{\langle T(n>100)\rangle} +\overline{\langle
v_R^2(n>100) \rangle}}} {f(n>100)}.
\end{equation}
Here, the mass fraction of dense gas $f(n>100)$ is given 
in eqn (\ref{eq:fraction}).  The turbulent velocities dominate
the dense gas when they are included, since the
thermal sound speed is $<0.5~ \kms$ whereas turbulent velocities
are several times larger; see Fig. \ref{fig:velocity}.
Note that $\kappa$ and $\Sigma$ are constant in time for any simulation.

Figure \ref{fig:Qvalue} shows the measured value of these four
quantities $Q_T({\rm total})$, $Q({\rm total})$, $Q_T(n>100)$, and $Q(n>100)$,
for all of our models.  In general, we find that the
saturated-state values when turbulence is included 
are near unity.  The only significantly larger
values are for the low-$\Sigma$ models in Series K, which have large
$\kappa$ and hence the thermal value $Q_T({\rm total})$ is large; when
turbulence is included this is raised even more.  Since Series Q and R
have $\kappa/\Sigma$ constant, the value of $Q_T({\rm total})$ is simply
proportional to the mass-weighted thermal velocity dispersion.  The
increased fraction of cold gas at high $\Sigma$ leads to a
corresponding decrease in $Q_T({\rm total})$.  Since the thermal and
turbulent velocity dispersions of the dense gas are small compared to
those of the diffuse components, the dense
components contribute to $Q_T({\rm total})$ and $Q({\rm total})$
mostly by lowering the mass fraction of the diffuse components, in the
numerator.  Because turbulent contributions are positive-definite,
$Q({\rm total})\ge Q_T({\rm total})$.

The strongest evidence of self-regulation by feedback-driven
turbulence is seen
in the saturated-state results for $Q(n>100)$.  With the low values of
the temperature in the dense component, the thermal-only values for
the dense gas are mostly $Q_T(n>100)<0.5$.
When turbulence is included, however, the saturated-state value of 
$Q(n>100)$ is between 1 and 2 for almost all models.  This is
consistent with expectations for marginal instability.  We note in
particular that velocity dispersions in Series K 
(see Fig. \ref{fig:velocity})
vary strongly with
$\Sigma$ (by a factor $\sim 5$), while $Q(n>100)$ varies weakly with
$\Sigma$ (by a factor $<2$); feedback evidently 
self-adjusts in these models so as to maintain a state of marginal
gravitational instability.

\subsection{Virial Ratios}

In a self-gravitating system that approaches a statistical steady
state, the Virial Theorem predicts that the specific kinetic and gravitational
energies $E$ and $W$ will be related by $2E+W=0$; this is modified
when magnetic terms are present
\citep{1953ApJ...118..116C,1956MNRAS.116..503M,1992ApJ...399..551M}.
Classically, the Virial Theorem has often been assumed to hold within
individual GMCs in order to obtain estimates of their masses, and
indeed this yields values that are consistent (within a factor $\sim
2$) with other measures of the mass
(e.g. \citealt{1987ApJ...319..730S}).  If individual GMCs are
short-lived, however, they may not satisfy the Virial Theorem because
the moment of inertia tensor is changing rapidly enough, and/or
surface terms are large enough, to be comparable to the kinetic and
gravitational energy integrated over the cloud volume
\citep{1999ApJ...515..286B, 2007ARA&A..45..565M,2007ApJ...661..262D}.
When averaged over an ensemble of clouds, there will be (partial)
cancellation of surface and time-dependent terms, as they appear with
opposite signs for forming and dispersing clouds.  An added
complication is that self-gravitating GMCs form out of diffuse gas, and when
they are destroyed (whether after a short or long time) they return to diffuse
gas; thus the different terms in the Virial Theorem may be observed in
different tracers depending on whether diffuse gas is primarily atomic
or molecular.

\begin{figure}[p]
\epsscale{1.1}
\plottwo{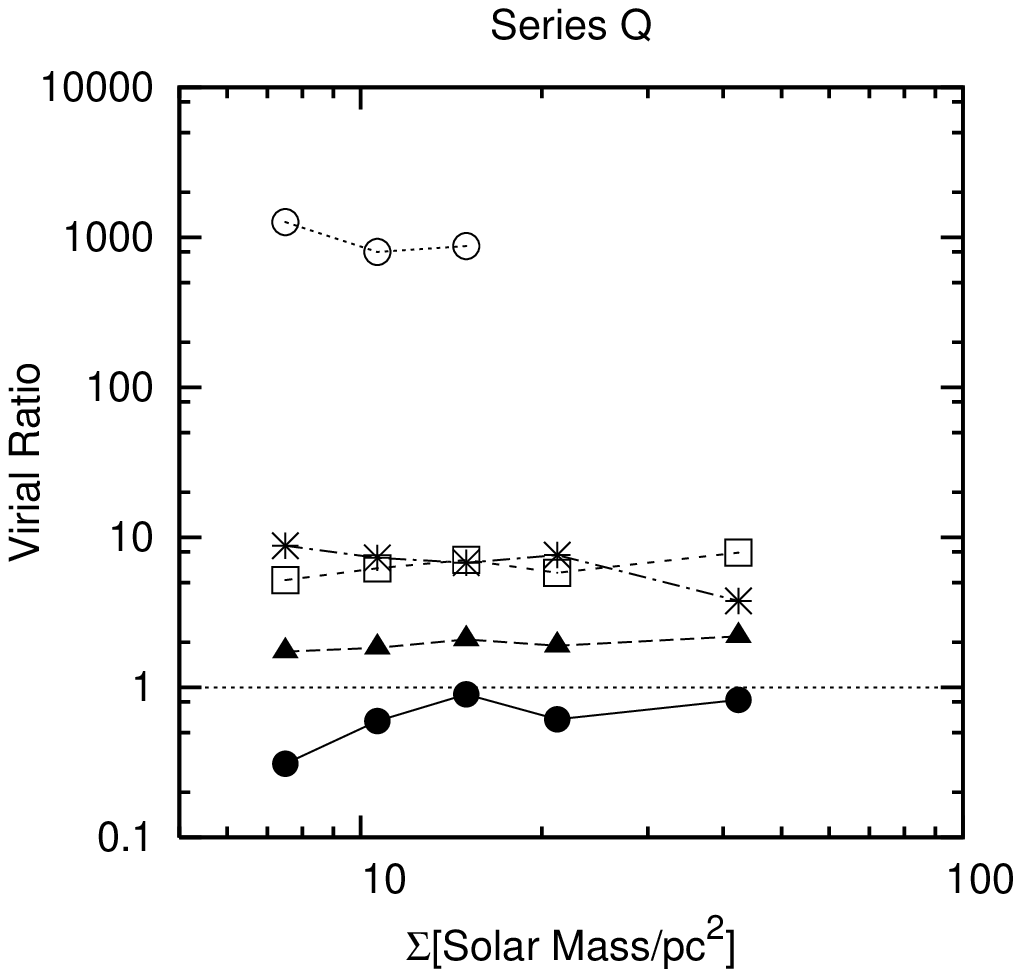}{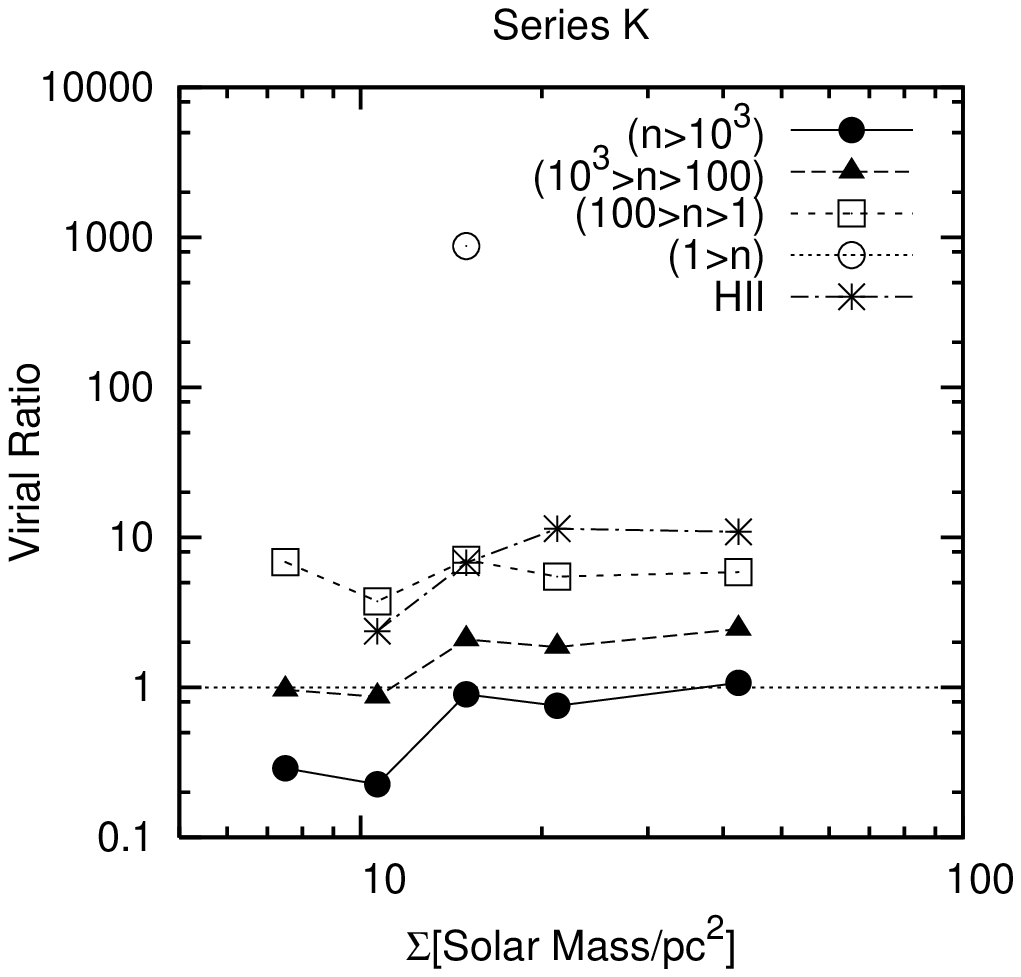}
\plottwo{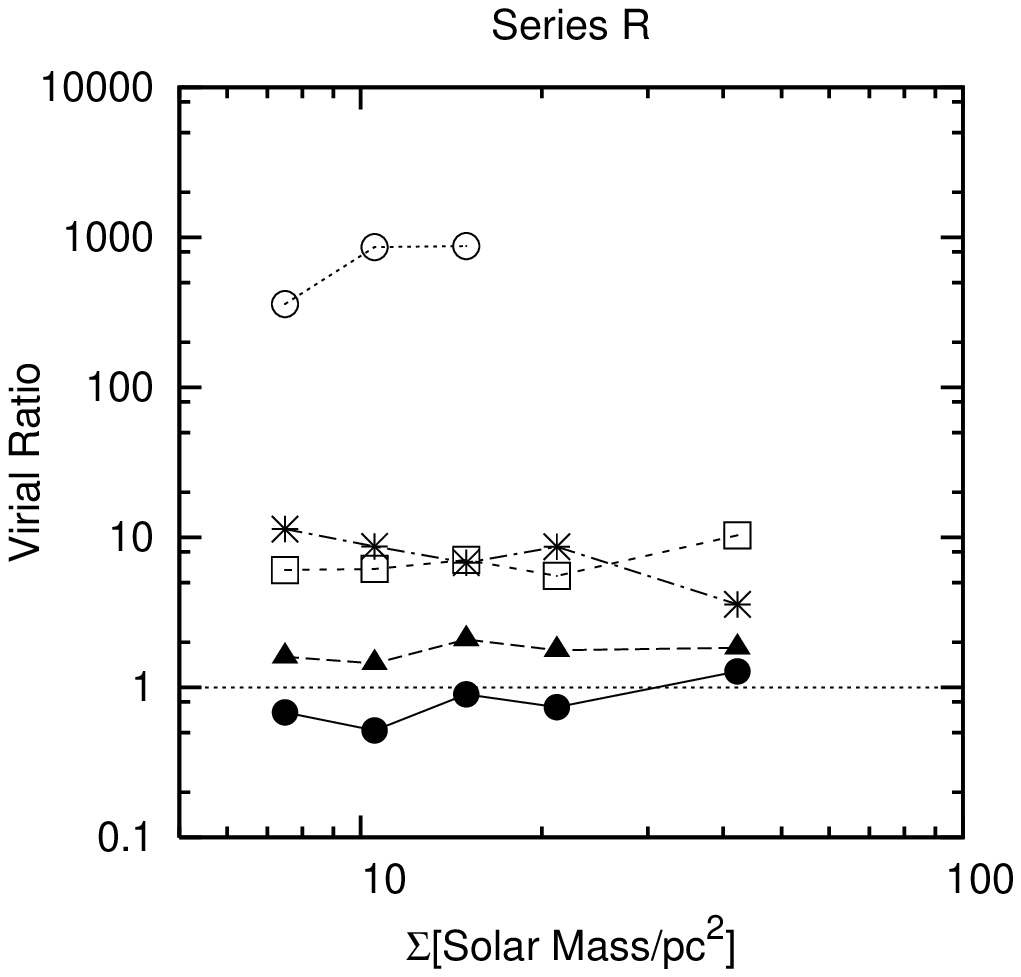}{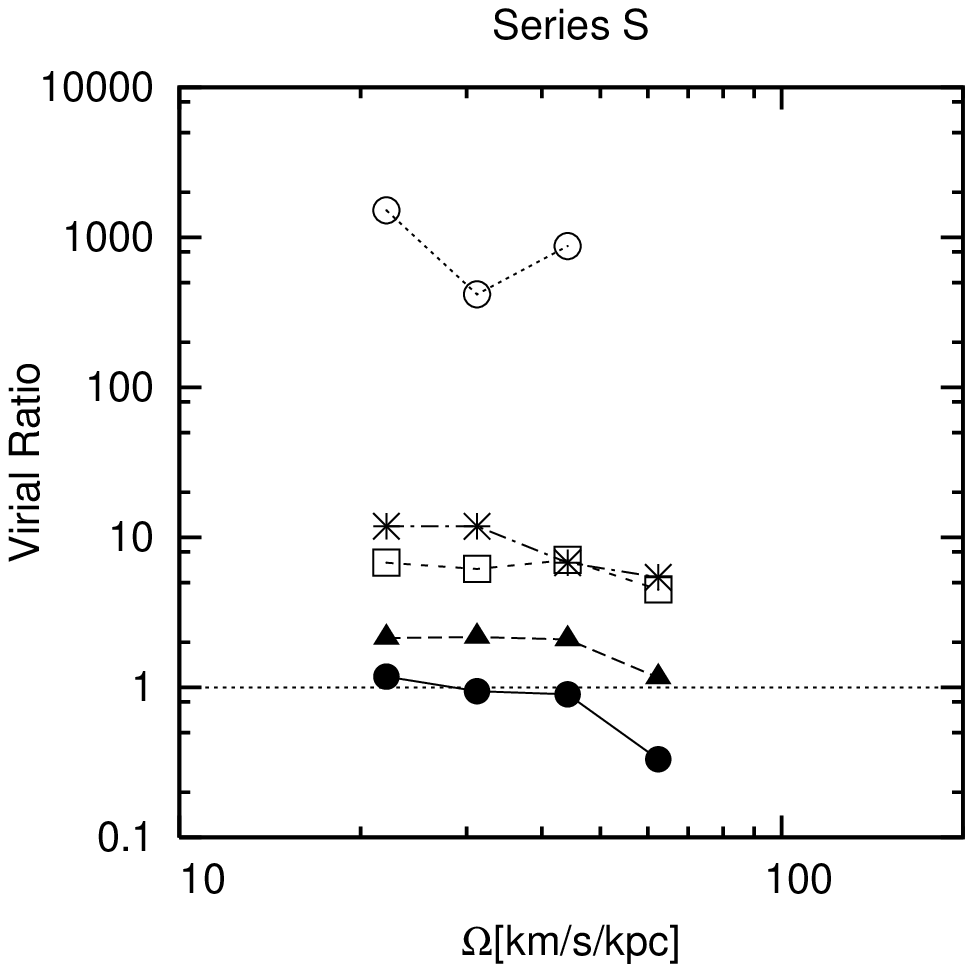}
 \caption{Virial ratio ${\cal R}\equiv 2E/|W|$ (see text), measured separately
 for each gas component. High density gas ($n>100\,\cm^{-3}$) is
 strongly gravitationally bound; low density gas is not.}
\label{fig:virial1}
\end{figure}

Here, we consider virial ratios 
\begin{eqnarray}
{\cal R}\equiv \frac{2E}{|W|}
\label{eq:vir}
\end{eqnarray}
separately for each component of the gas in our models.  
The term 
$E$ includes both thermal and bulk kinetic energy, computed via a
space-time average as:
\begin{equation}
E=\frac{1}{2}\overline{\langle v_x^2+\tilde{v}_y^2+v_z^2\rangle}+
\frac{1}{\gamma-1}\overline{\langle P/\rho \rangle},
\end{equation}
and for $W$ only the perturbed gravitational potential is used in
computing the space-time averaged value of the energy:
\begin{equation}
W=\frac{1}{2}\overline{\langle\Phi^{(1)}\rangle}.
\label{eq:Wdef}
\end{equation}
As for the other statistical properties we have considered, we measure
$\cal R$ separately for each component (separated into density bins) 
of the system.
Figure \ref{fig:virial1} shows the virial ratio of each component, for
all models in all series.  Note that ${\cal R}<2$ and ${\cal R}>2$
imply gravitationally bound and unbound states, respectively for any
component. As we do not separate the contributions to the potential
from the different density ranges, a given component may be bound
within a potential well that is created by more than one component.
Strictly speaking, the factor 1/2 in equation (\ref{eq:Wdef}) applies
only for self-potentials.

As expected, the lowest-density WNM component ($n<1~\cm^{-3}$) has $\cal R$
very large (above 100), and the intermediate-density CNM component
($1~\cm^{-3}<n<100 ~\cm^{-3}$) also is non-self-gravitating, with ${\cal R}$ in the range
$\sim 5-10$.  The HII (photoheated) component generally has values of
${\cal R}\sim 10$, similar to that of the CNM component.  
Although it has very large thermal energy (much greater than the CNM),
the photoheated gas by definition resides within deep parts of the 
gravitational potential well.
The two dense
components, DM2 ($10^2~\cm^{-3}<n<10^3~\cm^{-3}$) and 
DM3 ($10^3~\cm^{-3}<n$), on the other hand,
are consistent with being marginally or strongly gravitationally
bound, with ${\cal R} \aplt 2$ and ${\cal R} \aplt 1$, respectively.
For the majority of models, the value of ${\cal R}$ for the densest
component, DM3, is quite near unity, indicating consistency with
virial equilibrium for the component as a whole.  For a few models,
${\cal R}$ is as low as 0.3 for the DM3 component; this indicates that
the dense gas is transient, with dense regions being rapidly
dispersed into lower-density gas by the feedback process.  Overall, we
find no significant differences in the trends for ${\cal R}$ between
different series or different models within any series.  There is a
weak correlation between $Q$ and ${\cal R}$, with lower-$Q$ (more
unstable) models having slightly higher virial ratios.

\subsection{Vertically-Averaged Density and Free-Fall Time
\label{sec:rhoave}}

Although the ISM consists of many phases at different densities, all
of this gas resides within a common potential well which tends to
confine material near the galactic midplane.  The scale height of each phase
depends on the support provided by thermal and kinetic pressure (plus
support by magnetic stresses and cosmic rays, although these may be 
less significant).  In \citet{PaperII} we consider in detail
the vertical distribution of gas within our models, and show that
vertical equilibrium is a good approximation for the system as a
whole, provided that appropriate accounting is made for the differing
velocity dispersions of different components.  We also discuss dependence
of the mean scale height on model parameters.

For the purpose of assessing gravitational timescales of the
overall ISM system, it is useful to measure the density when averaged
over large scales (i.e. a volume at least comparable to the scale height).
To evaluate this volume average in our models, we first compute the
vertical scale height, defined using the following averaging:
\begin{equation}
H_{\rm ave}=\sqrt{
\frac{\sum_{\rm all\ zones}\rho z^2 }{\sum_{\rm all\ zones}\rho}
}
\end{equation}
where $z$ is the vertical coordinate relative to the midplane. 
We further average the values of $H_{\rm ave}$ over time.
For a Gaussian density profile, 
$\rho(z)=\rho_0 \exp(-z^2/2H^2)$, the midplane density is
related to the surface density and scale height by 
$\rho_0=\Sigma/(\sqrt{2\pi}H)$, and 
the mass-weighted mean value of the average density is given by
$\rho_0/\sqrt{2}$.  We therefore define an average density in our
models as:
\begin{equation}
\rho_{\rm ave}\equiv 
\frac{\Sigma}{2\sqrt{\pi}H_{\rm ave}}
\end{equation}
(see also Appendix in \cite{PaperII}).

\begin{figure}[p]
\epsscale{1.1}
\plottwo{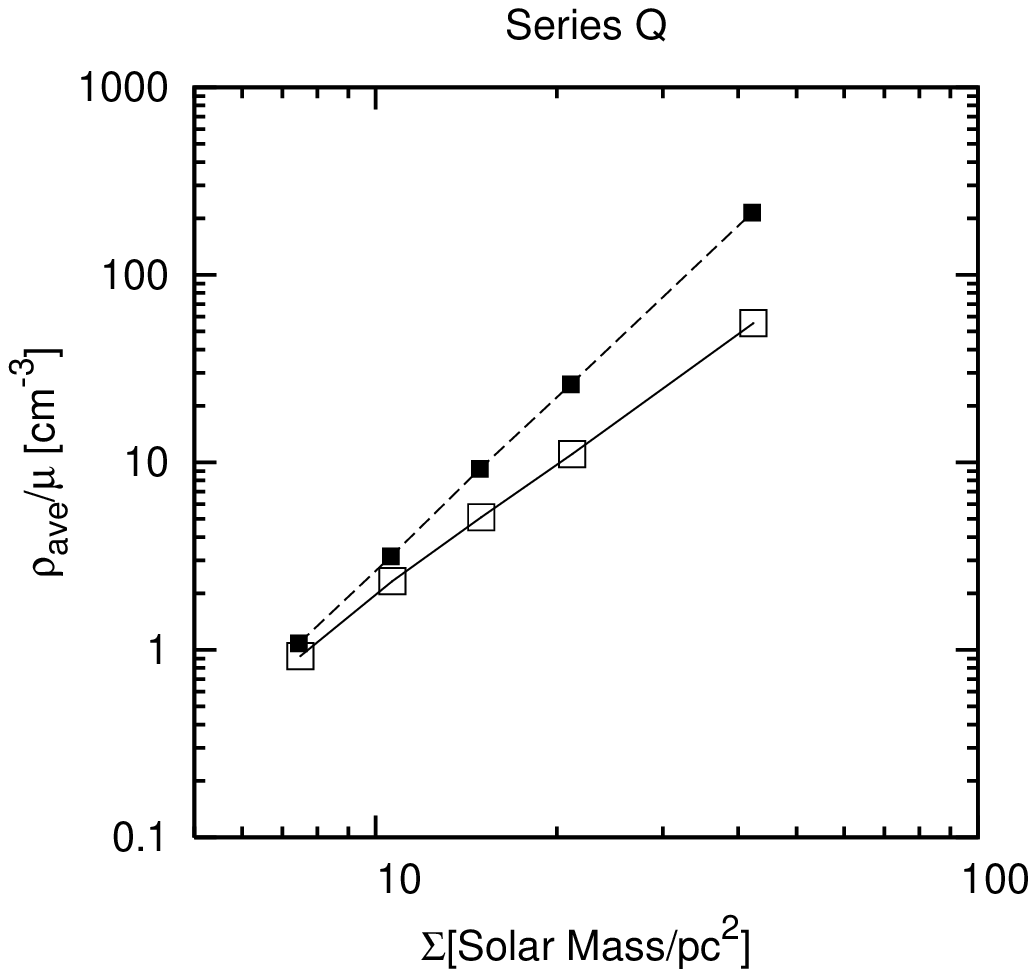}{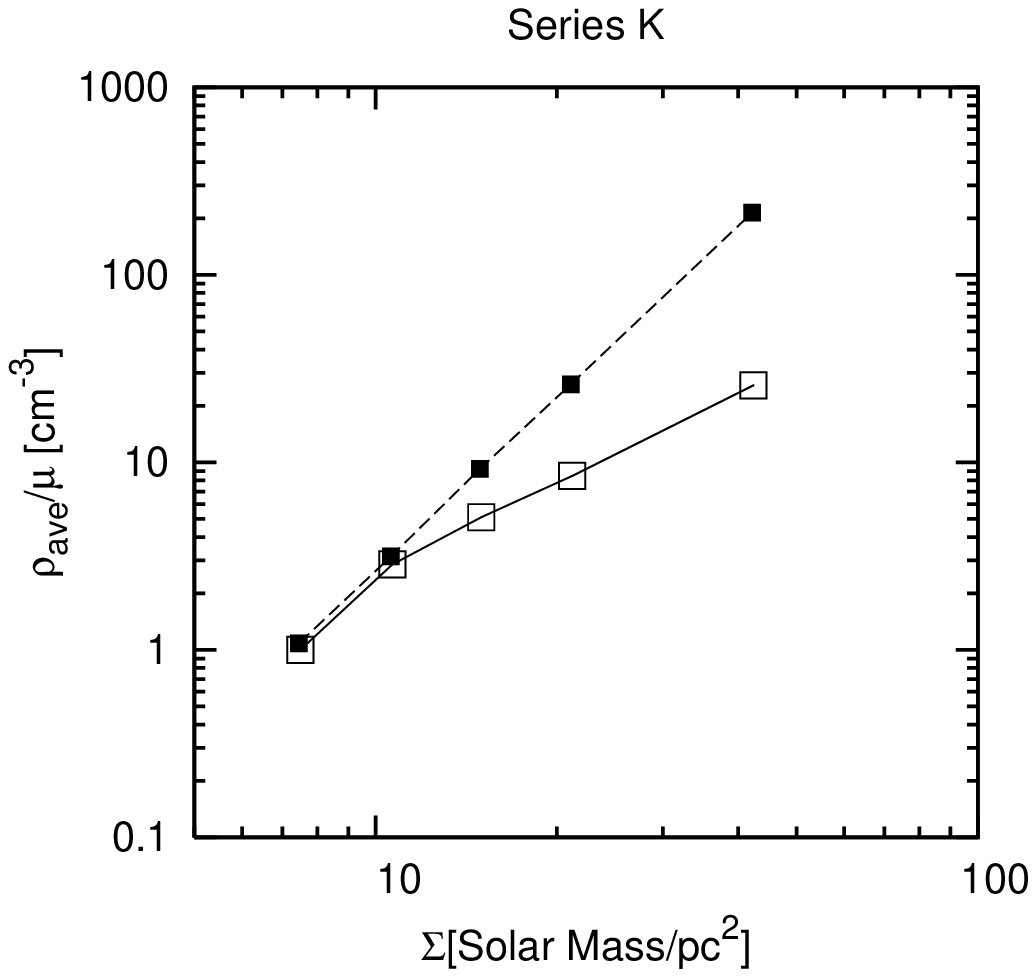}
\plottwo{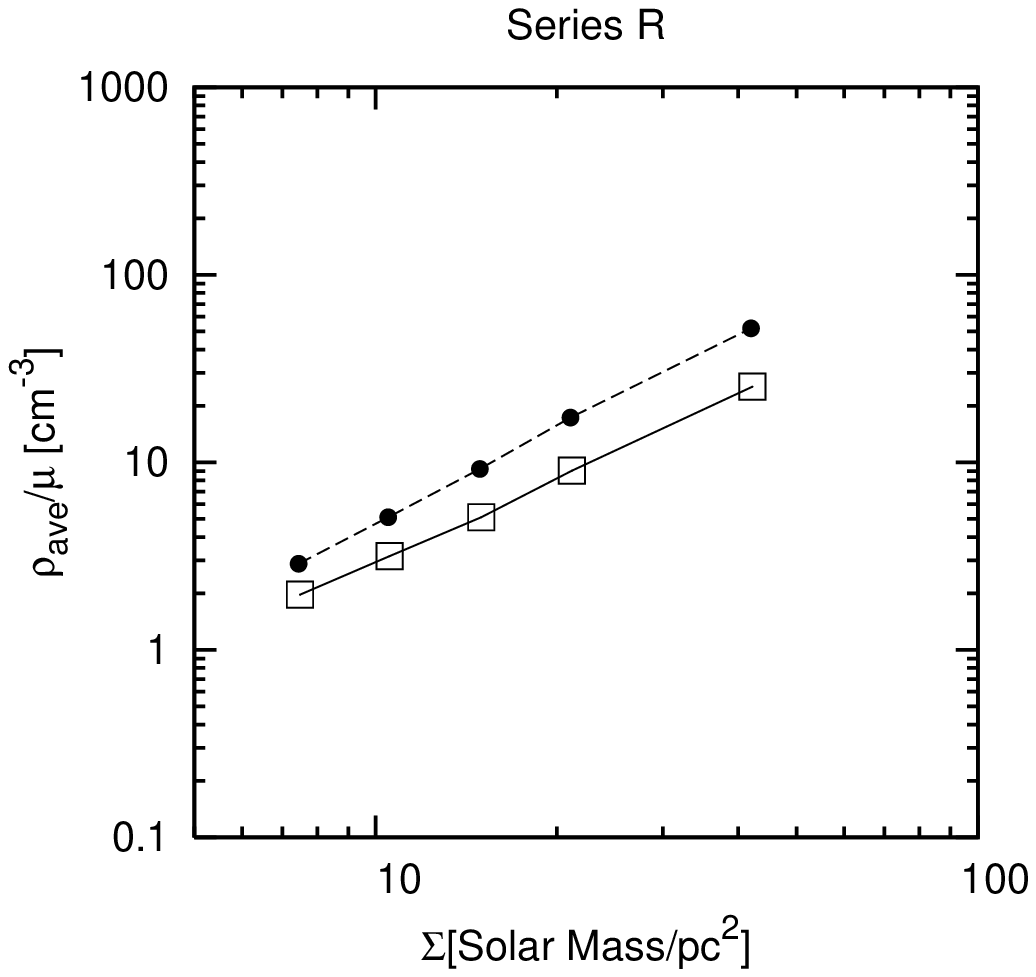}{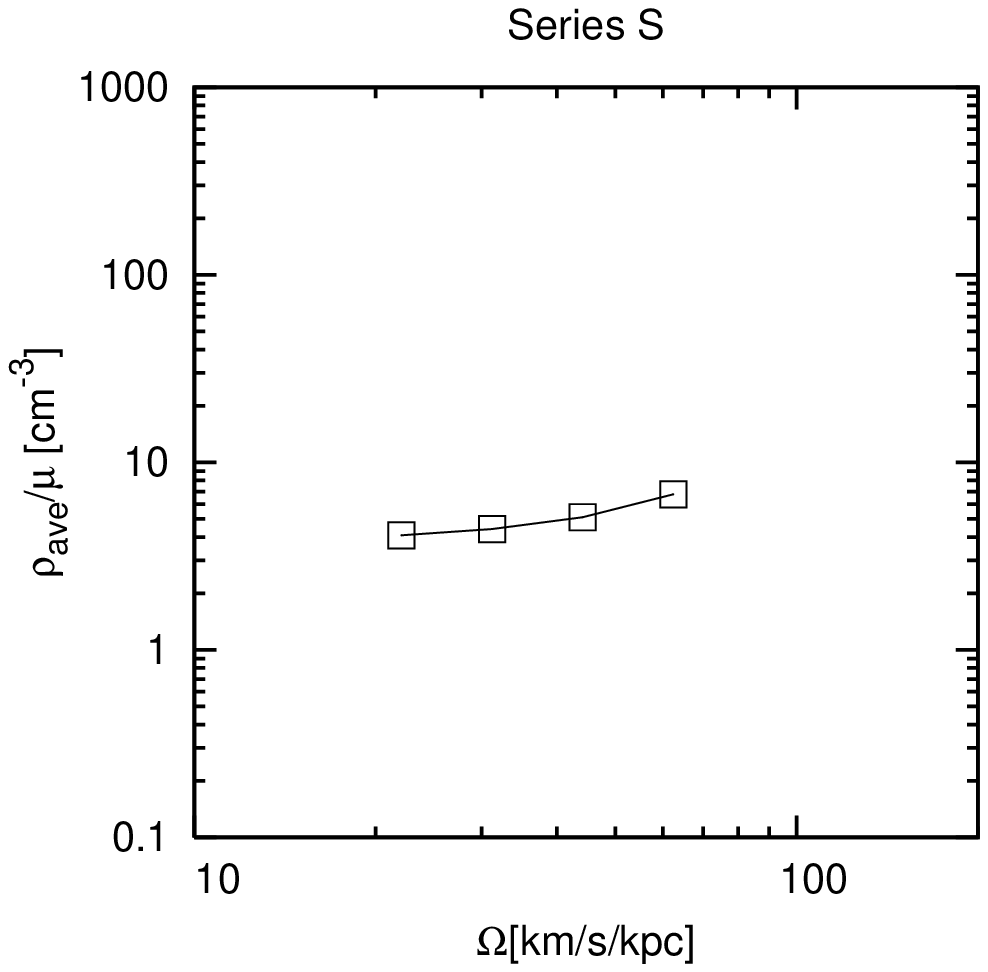}
 \caption{Vertically-averaged density for all dynamic (open squares) 
and hydrostatic 
model series. Filled squares and circles denote hydrostatic series
 HSP and HSC, respectively. Turbulence reduces the mean density
 (averaged over the disk scale height) relative to the hydrostatic
 case. Mean density increases with $\Sigma$ more rapidly in Series Q
 compared to Series R because the latter adopts the same stellar
 vertical gravity value for all models.}
\label{fig:rave}
\end{figure}

Figure \ref{fig:rave} shows the vertically-averaged density for all
models in all series.  In general, we find that the average density
increases with the total surface density of gas in the disk.
A slightly shallower increase of $\rho_{\rm ave}$ with $\Sigma$ 
is obtained in Series K compared to Series Q, 
which can be attributed to the large velocity
dispersions in strongly unstable (small $Q$) disk models.  
Series R also has a shallower slope than in Series Q, because the
stellar gravity does not increase at large $\Sigma$ in the former.
For reference, we also plot in Figure \ref{fig:rave} the values of the
vertically-averaged density from our comparison hydrostatic model
series.  The slope of Series HSC (lower-left panel) is shallower than that in
Series HSP (top panels), again because the stellar density does not
increasingly compress the gas at large $\Sigma$ in Series HSC.
The volume-averaged densities of the dynamic models are  
lower than those of the hydrostatic models by up to an order of
magnitude; the difference increases at large surface density where
turbulence plays an increasingly important role (see also 
\cite{PaperII}).

Using the mean density and the definition of the free-fall time,
\begin{eqnarray}
t_{\rm ff}(\rho)&=& \left(\frac{3\pi}{32G \rho}\right)^{1/2},
\label{eq:tff}
\end{eqnarray}
we can calculate the free-fall time for the
system as a whole, $t_{\rm ff}(\rho_{\rm ave})$. 
Since $\rho_{\rm ave}$ increases with $\Sigma$ in our models,
$t_{\rm ff}(\rho_{\rm ave})$ will decrease with increasing $\Sigma$.
Because star formation requires gas to become self-gravitating, a
widespread notion is that the star formation timescale, when averaged
over large scales in a galaxy, will be proportional to the large-scale
average of $t_{\rm ff}$, i.e. $t_{\rm ff}(\rho_{\rm ave})$.  
Since star formation takes place within GMCs
that have much higher density than the mean value in the ISM, the
conditions that control star formation where it actually takes place
are not those of the large-scale ISM.  Thus, implicit in the notion
that star formation times should be related to the large-scale mean
$t_{\rm ff}(\rho_{\rm ave})$
is the idea that the formation of GMCs (on timescales
closer to $t_{\rm ff}(\rho_{\rm ave})$) 
is the principal means of regulating star
formation.  If the star formation efficiency per GMC is constant, then the GMC
formation rate would control the star formation rate.  Alternatively,
the star formation rate might be related to the large-scale $t_{\rm
  ff}(\rho_{\rm ave})$
if the densities within GMCs are proportional to the
large-scale mean densities of the ISM, 
$\langle \rho_{\rm GMC}\rangle \propto \rho_{\rm ave}$.

\begin{figure}[p]
\epsscale{1.1}
\plottwo{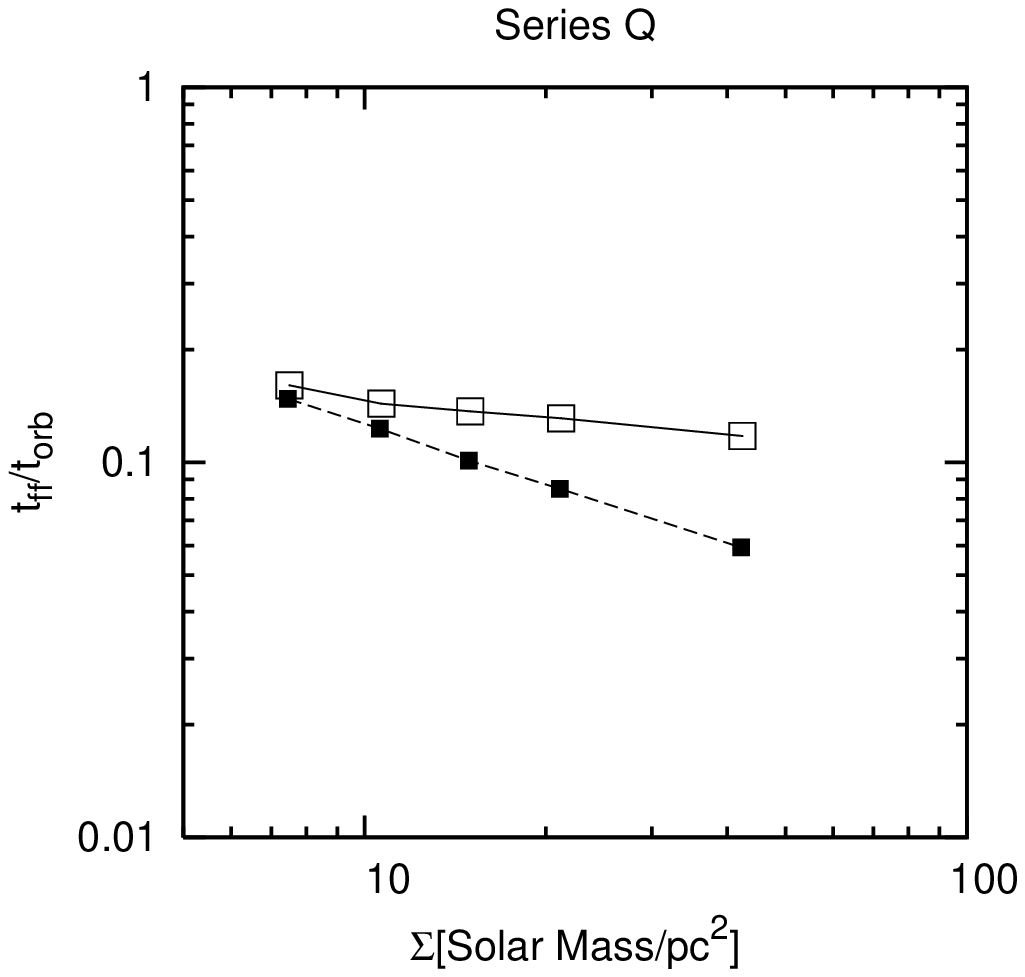}{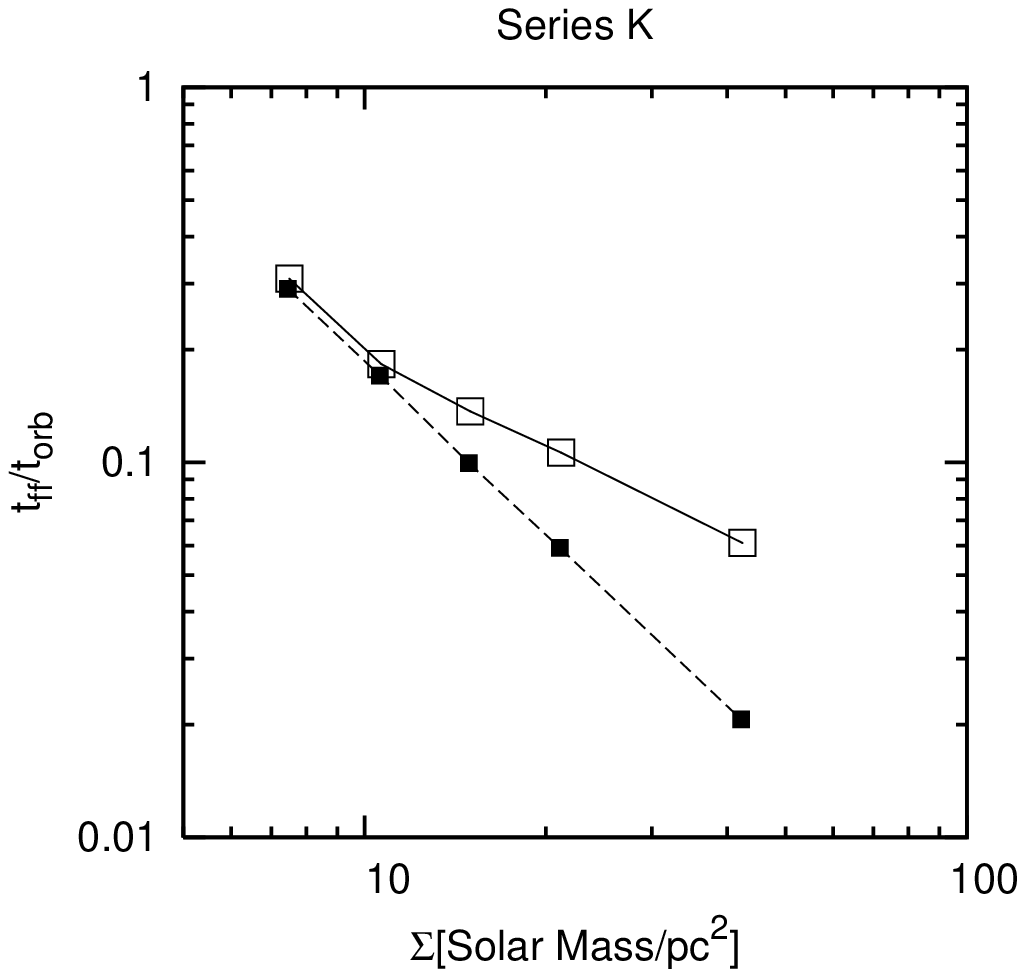}
\plottwo{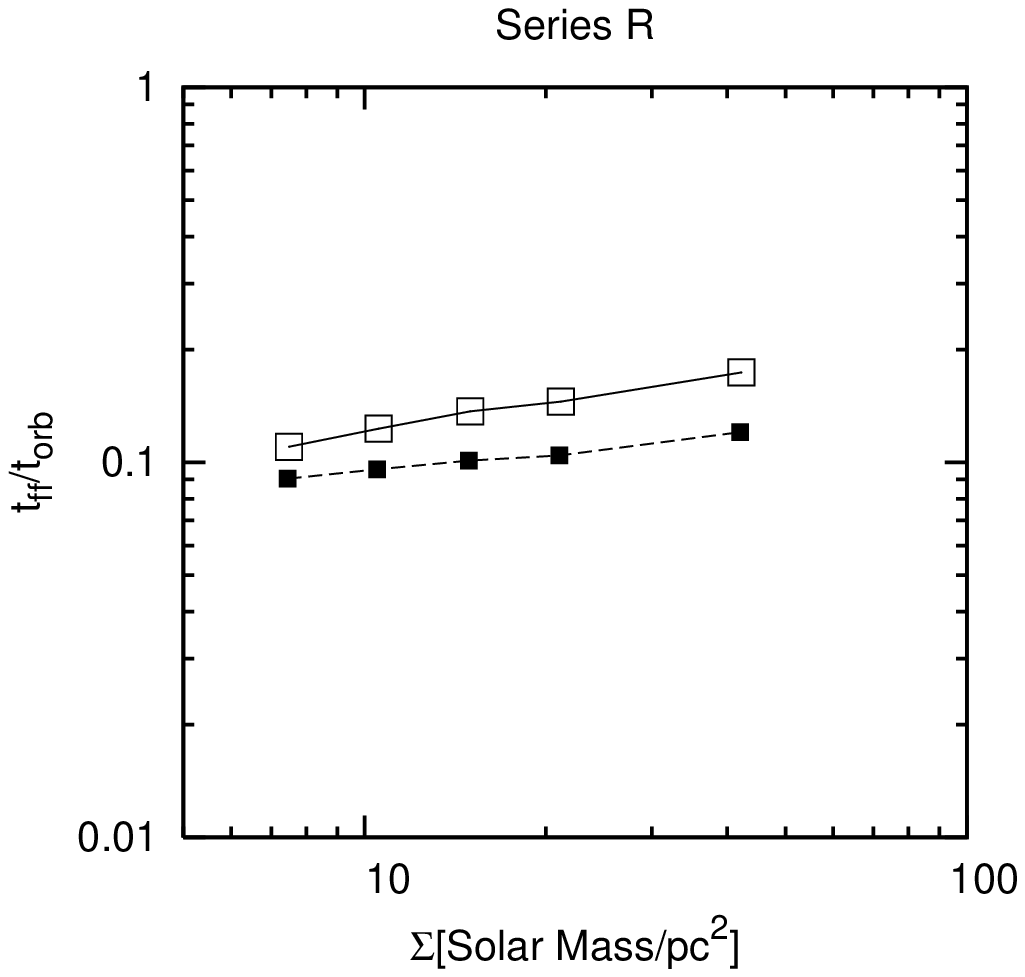}{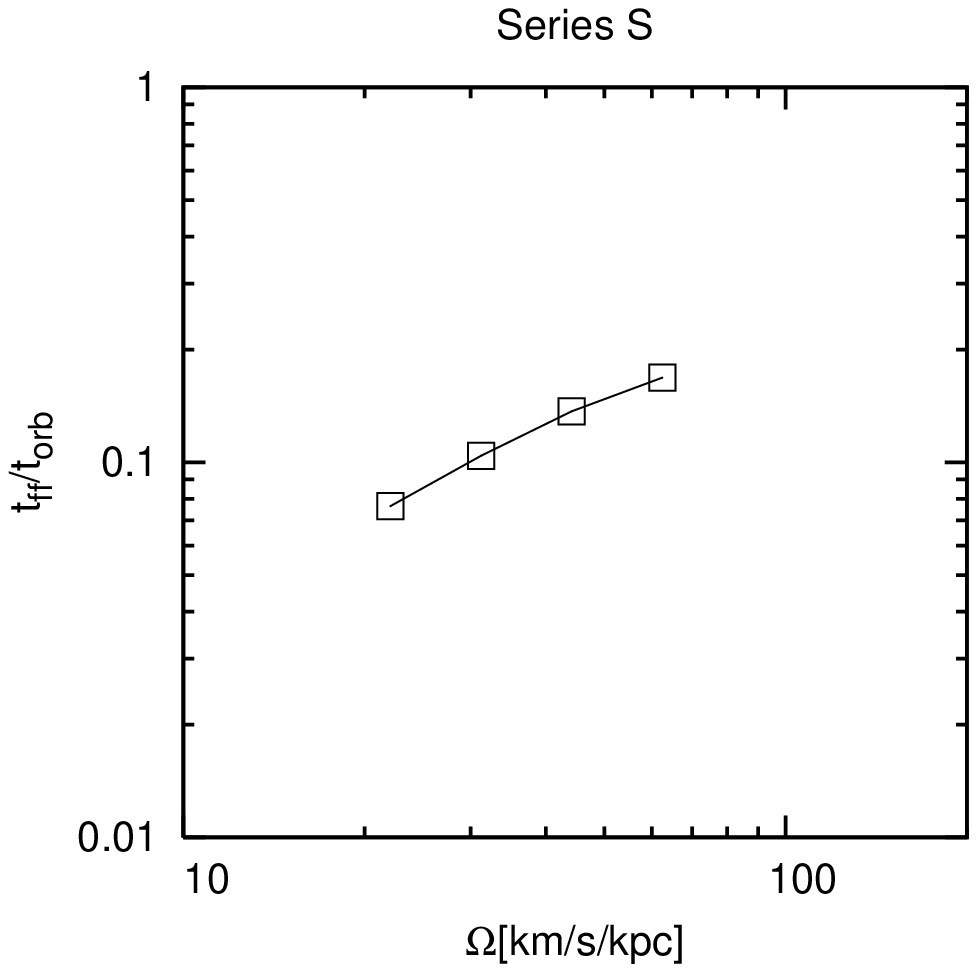}
 \caption{Ratio of mean free-fall to orbital time. 
Open squares denote hydrodynamic models, and filled squares
   denote the hydrostatic series (HSP and HSC).}
\label{fig:tfftorb}
\end{figure}

Another important dynamical timescale in disk galaxies is the orbital
time, $t_{\rm orb}=2 \pi/\Omega$.  Growth of large-scale
self-gravitating perturbations in disks in fact occurs at timescales
longer than $t_{\rm ff}(\rho_{\rm ave})$ (provided pressure limits small-scale
collapse), and more comparable to $t_{\rm orb}=2 \pi/\Omega$.
Observations \citep{1998ApJ...498..541K} 
show that empirically-measured star formation
timescales in disk galaxies tend to be correlated with the orbital
time, with $\sim 10\%$ of gas being converted to stars per galactic
orbit. It is useful to compare $t_{\rm ff}(\rho_{\rm ave})$ 
with $t_{\rm orb}$ in our
models.  Figure \ref{fig:tfftorb} shows the ratio of $t_{\rm
ff}(\rho_{\rm ave})/t_{\rm orb}$ 
for all hydrodynamic and hydrostatic series.  For the
hydrodynamic series, the typical ratio is 
$0.06-0.2$; for the
hydrostatic models, the densities are much higher at large $\Sigma$ 
so that 
$t_{\rm ff}(\rho_{\rm ave})/t_{\rm orb}\sim 0.02-0.2$.
For series Q and R, the ratio $t_{\rm ff}(\rho_{\rm ave})/t_{\rm orb}$ 
varies relatively weakly with $\Sigma$, and lies in the range $0.1-0.2$.
The comparison hydrostatic models for these series also show 
$t_{\rm ff}(\rho_{\rm ave})/t_{\rm orb}$ varying only modestly with $\Sigma$.  
For these series, $t_{\rm orb}\propto 1/\Sigma$.  Since turbulent velocity
dispersions do not depend strongly on $\Sigma$ for Series Q and R,
$\rho_{\rm ave}$ does not strongly depart from a scaling $\propto \Sigma^2$,
yielding behavior similar to $t_{\rm ff} \propto \Sigma \propto t_{\rm
  orb}$.
Interestingly, the K series, which has constant $t_{\rm
  orb}$, shows a smaller range of $\rho_{\rm ave}$ than
the Q series.  This is indicative of self-regulation: high
feedback activity in the highest-$\Sigma$ models of series K yield
high turbulent amplitudes, which lead to lower values of $\rho_{\rm ave}$.
As a consequence, the ratios of $t_{\rm ff}/t_{\rm orb}$ are more
modulated in the hydrodynamic models for Series K than in the corresponding
hydrostatic series.

\section{Implications for Star Formation}

In the present work, we do not explicitly follow star formation.  
Nevertheless, it is interesting to explore the 
consequences of our statistical results, within the context of recipes
that are commonly adopted for star formation in numerical models.
We compare estimates of the implied star formation timescale both to
observations and to various fiducial dynamical times.

\subsection{Star Formation Rates and Timescales 
\label{sec:SFR}}

A common practice in numerical simulations of galactic evolution is to
assume that the star formation rate per unit volume (in a
computational region) is proportional to the gas density per unit
volume divided by the free-fall time at that density.  When a minimum
density threshold for star formation is imposed, the total star
formation rate (SFR, in mass of new stars per unit time) takes the
form
\begin{eqnarray}
\dot{M}_{\ast}\equiv 
\frac{\epsilon_{\rm ff}(\rho_{\rm th})
M(\rho>\rho_{\rm th})}{t_{\rm ff}(\rho_{\rm th})}
\end{eqnarray}
provided that the density PDF decreases above the threshold, so that
most of the star forming activity is in gas near $\rho_{\rm th}$.  
Here, the star formation efficiency per free-fall time, 
$\epsilon_{\rm ff}(\rho_{\rm th})$, 
is an arbitrary constant parameter that is adopted, 
generally by comparing to observations. 
In practice, the parameter $\epsilon_{\rm ff}(\rho_{\rm th})$ in this
sort of 
recipe enfolds many different effects that limit star formation
compared to the fastest possible rate.
Within GMCs, turbulence and magnetic fields limit the rate of core formation
and collapse, and feedback from star formation limits GMC lifetimes;
at larger scales, dynamical processes in the diffuse ISM limit GMC formation.
Depending on the value of the threshold density, either more (low 
$\rho_{\rm th}$) or fewer (high $\rho_{\rm th}$) processes are implicitly 
packaged in the single efficiency parameter $\epsilon_{\rm ff}(\rho_{\rm th})$.

For a given star formation rate, the star formation timescale is
defined by dividing the total gas mass by the total SFR,
 $\dot{M}_{\ast}$:
\begin{eqnarray}
t_{\rm SF}&\equiv& \frac{M_{\rm tot}}{\dot{M}_{\ast}}
=\frac{M_{\rm tot}}{M(\rho>\rho_{\rm th})}
\frac{t_{\rm ff}(\rho_{\rm th})}{\epsilon_{\rm ff}(\rho_{\rm th})},\\
&=&\frac{1}{f(\rho>\rho_{\rm th})}
\frac{t_{\rm ff}(\rho_{\rm th})}{\epsilon_{\rm ff}(\rho_{\rm th})}
\equiv \frac{\tau_{\rm SF}}{\epsilon_{\rm ff}(\rho_{\rm th})}.
\end{eqnarray}
The latter expression uses the mass fraction $f$ as defined in
equation (\ref{eq:fraction}); $M_{\rm tot}$ is the total gas mass.
Because $f$ and $\epsilon_{\rm ff}$ are, by definition,
less than 1, the star formation time always exceeds the free-fall time at
the threshold density.
Since the efficiency per free-fall time is arbitrary 
(from the point of view of simulations), it is 
convenient to introduce 
$\tau_{\rm SF}\equiv \epsilon_{\rm ff}(\rho_{\rm th}) t_{\rm SF}$ such that
$\tau_{\rm SF}=t_{\rm ff}(\rho_{\rm th})/f(\rho>\rho_{\rm th})$.
This scaled 
star formation time then depends only on the choice of density threshold
and the fraction of the total gas mass above this threshold.

\begin{figure}[p]
\epsscale{1.1}
\plottwo{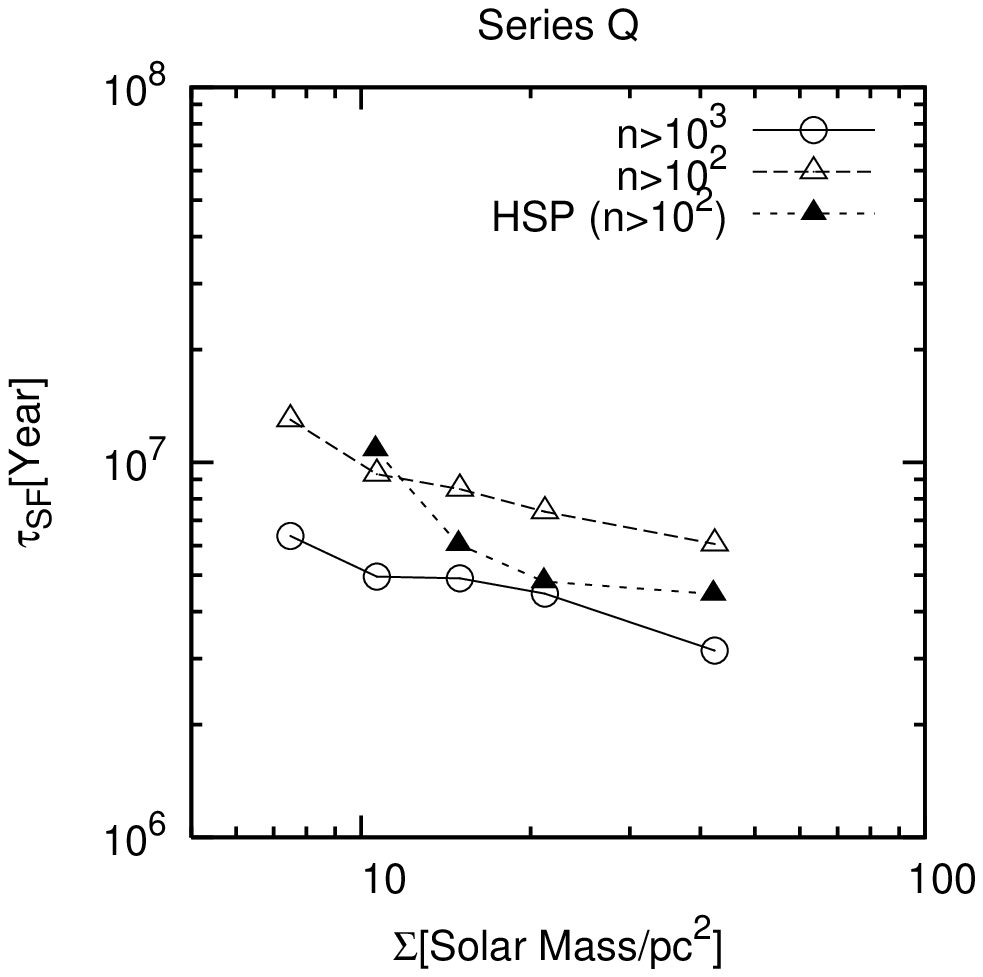}{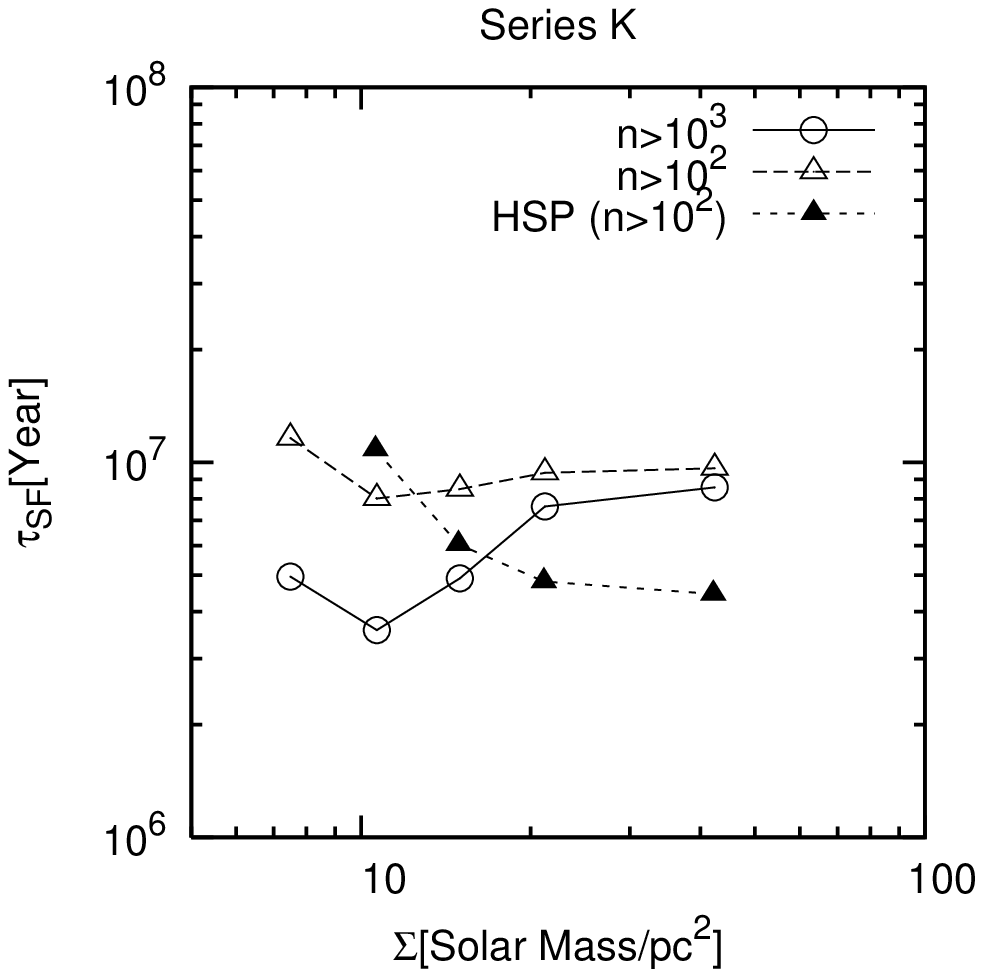}
\plottwo{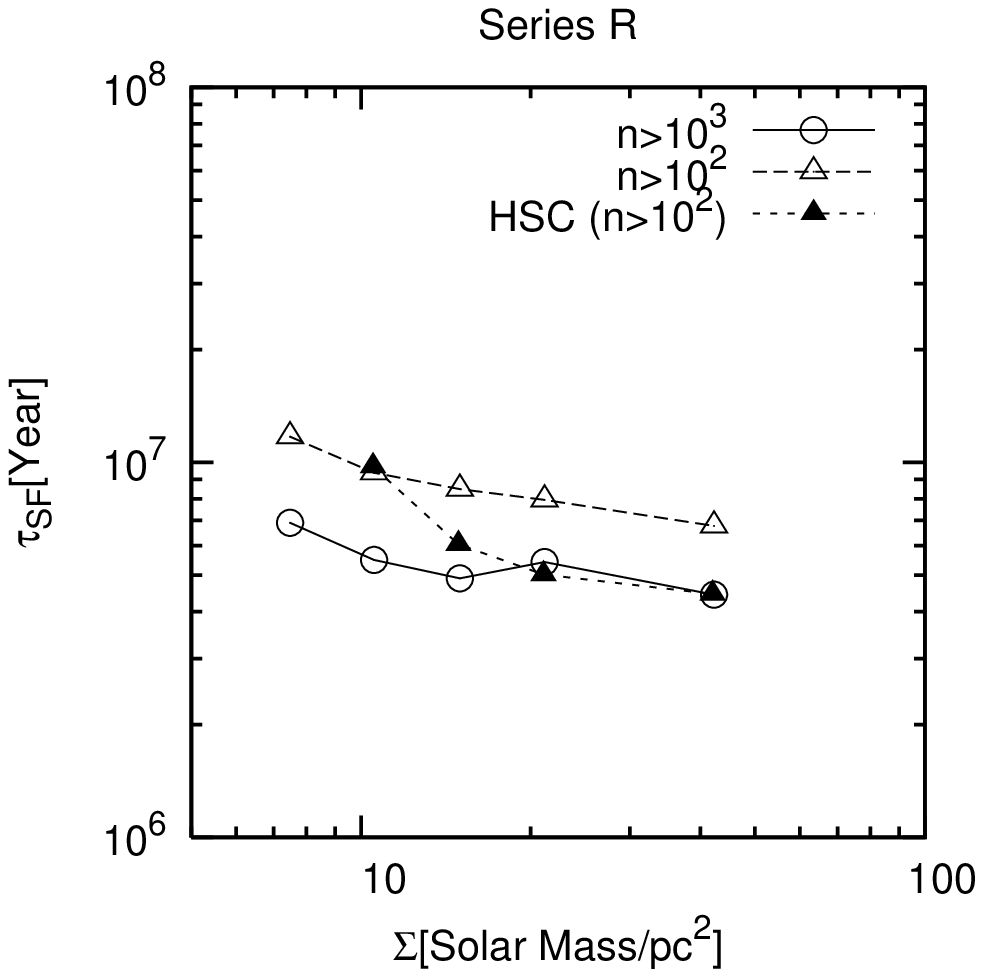}{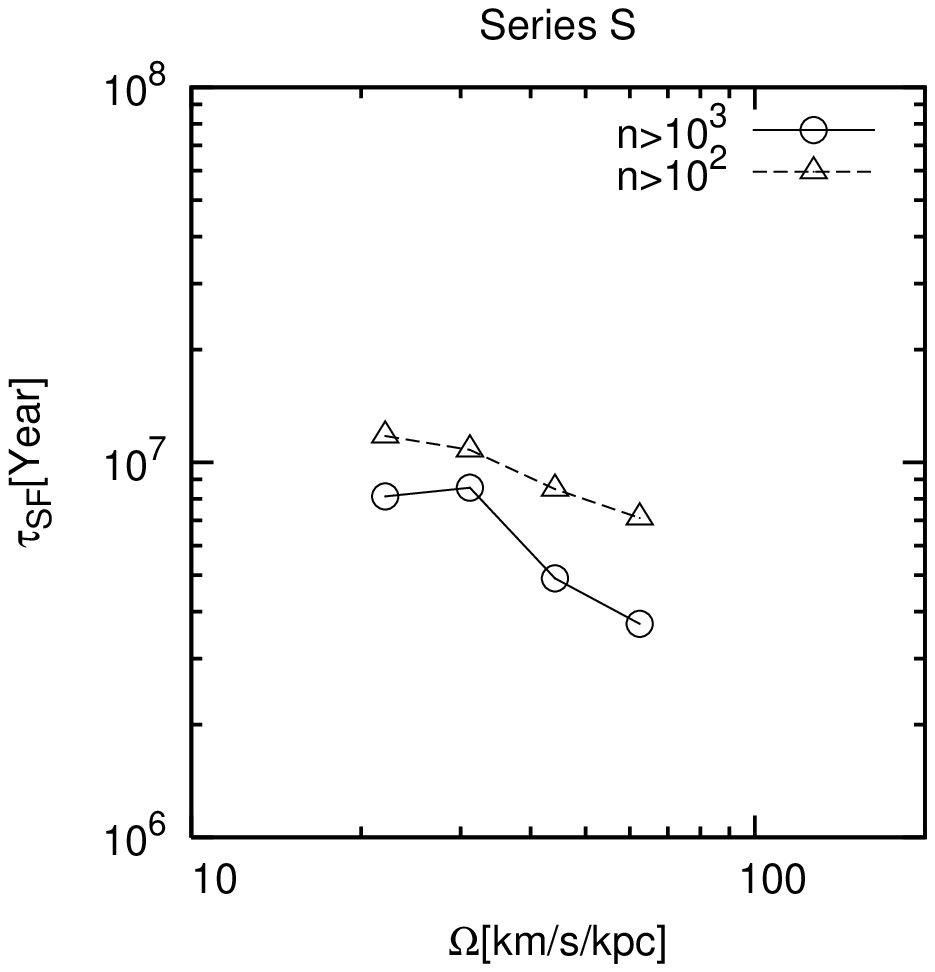}
 \caption{Scaled star formation time, 
$\tau_{\rm SF}= t_{\rm SF}\, \epsilon_{\rm ff} (\rho_{\rm th})
\equiv t_{\rm ff}(\rho_{\rm th})\Sigma/\Sigma(\rho>\rho_{\rm th})
$, 
computed using two different density thresholds 
$n_{\rm th}=\rho_{\rm th}/\mu$.
 Results for dynamical models are shown with open triangles and
 circles; results for hydrostatic models are shown with filled
 triangles.  For Series Q and R (which are most similar to real
 galaxies at varying radius), the two choices of threshold 
would yield consistent values of the star formation time $t_{\rm SF}$
provided that 
the adopted efficiency decreases slightly with density,
$\epsilon_{\rm ff} (n_{\rm th}=100)\sim (1.5 -2)\epsilon_{\rm ff} 
(n_{\rm th}=10^3)$.
}
\label{fig:tSF}
\end{figure}

In numerical simulations, the density threshold for star formation is
an arbitrary parameter; what difference does the choice of this value
make to the resulting SFR?  To address this question, we first compare
values of $\tau_{\rm SF}$ using two different thresholds, 
$n=10^2 ~\cm^{3}$ and $n=10^3 ~\cm^{3}$. Both threshold values are large
enough that the gas at these densities is in gravitationally-bound
structures, based on the results shown in Fig. \ref{fig:virial1}.
Figure \ref{fig:tSF} shows the values of the scaled star formation
time, $\tau_{\rm SF}$, for all models.
For our chosen density thresholds, $\tau_{\rm SF}$ is in the range
$3\times 10^6- 10^7$ yr for all models.
The true star formation time, $t_{\rm SF}$, exceeds $\tau_{\rm SF}$ by a
factor $\epsilon_{\rm ff}^{-1}$; the value of 
$\epsilon_{\rm ff}(\rho_{\rm th})$ must
then be quite small ($<0.01$) for $t_{\rm SF}$ to be $>10^9$ yr.
Also, since $\tau_{\rm SF}$ is larger for the threshold 
choice $n=10^2 ~\cm^{3}$ than $n=10^3 ~\cm^{3}$, the value of 
$\epsilon_{\rm ff}(\rho_{\rm th})$ 
would have to be smaller for the higher density
threshold, in order to yield the same value of $t_{\rm SF}$ at a given
$\Sigma$.  Note,
however, that while the thresholds differ by a factor 10, the values
of $\tau_{\rm SF}$ (and hence required 
$\epsilon_{\rm ff}(\rho_{\rm  th})$) differ by
less than a factor 2.  This reflects the fact that $f$ decreases with
increasing $n$; between $n=10^2 ~\cm^{3}$ and $n=10^3 ~\cm^{3}$, our
results imply a dependence $f(>n)\propto n^{-s}$ with the range of 
$s=0.2-0.5$.  
Alternatively, we can think of our results
requiring a choice for 
$\epsilon_{\rm ff}(\rho_{\rm th})\propto \rho_{\rm th}^{-r}$     
 with the range of $r=0-0.3$
in order for the SFR to be independent of the choice
of threshold at high densities.

Other aspects of the results shown in Figure \ref{fig:tSF} are also
interesting.  First, it is evident that $\tau_{\rm SF}$ depends only
weakly on both surface density (Series Q, K, R) and angular velocity
(Series S).  For Series Q and S, $\tau_{\rm SF}$ decreases with
increasing $\Sigma$.  Interestingly, the hydrostatic models show a
similar range of $\tau_{\rm SF}$ to the dynamic, turbulent models.
The fact that $\tau_{\rm SF}$ is not strongly sensitive to environmental
conditions (total available gas content, local shear rate, level of
turbulence, etc.) may help to explain why empirical SFRs show such a
regular character in observed galaxies, in spite of widely-varying
local conditions.  Conversely, the insensitivity of $\tau_{\rm SF}$ to
conditions within a model has implications for evaluating theoretical
results: successfully reproducing levels of star formation similar to
observations is
not a critical discriminant of how well a simulated galaxy resembles a
real system.  Our hydrostatic models bear minimal resemblance to real
galaxies, yet for a choice of $\epsilon_{\rm ff}(\rho_{\rm th})\sim 0.01$ 
consistent with
observed efficiencies in CO-emitting gas in GMCs (which have
densities in the range $n=10^2-10^3 ~\cm^{3}$), the resulting star
formation times are $\sim 4\times 10^8-10^9$ yr, similar to the
observed range of $t_{\rm SF}$ 
for $\Sigma$ comparable to the range in our models.

\begin{figure}[p]
\epsscale{1.1}
\plottwo{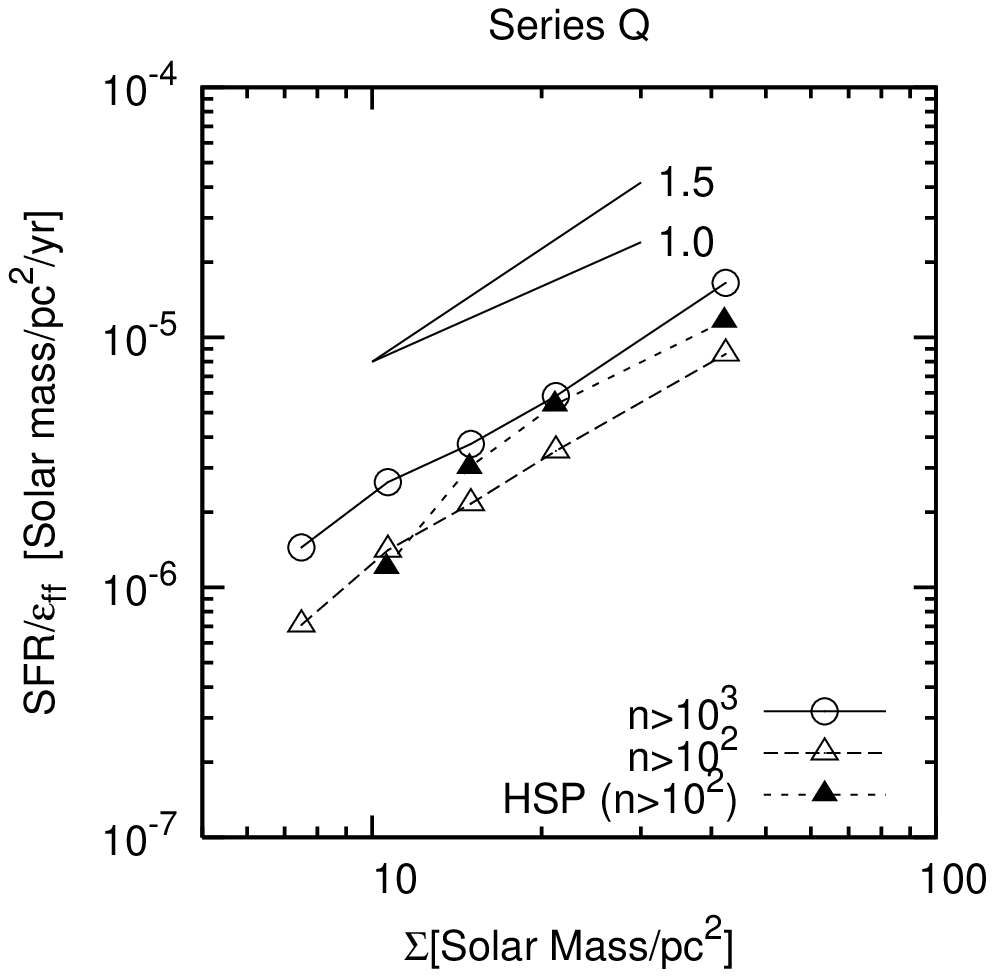}{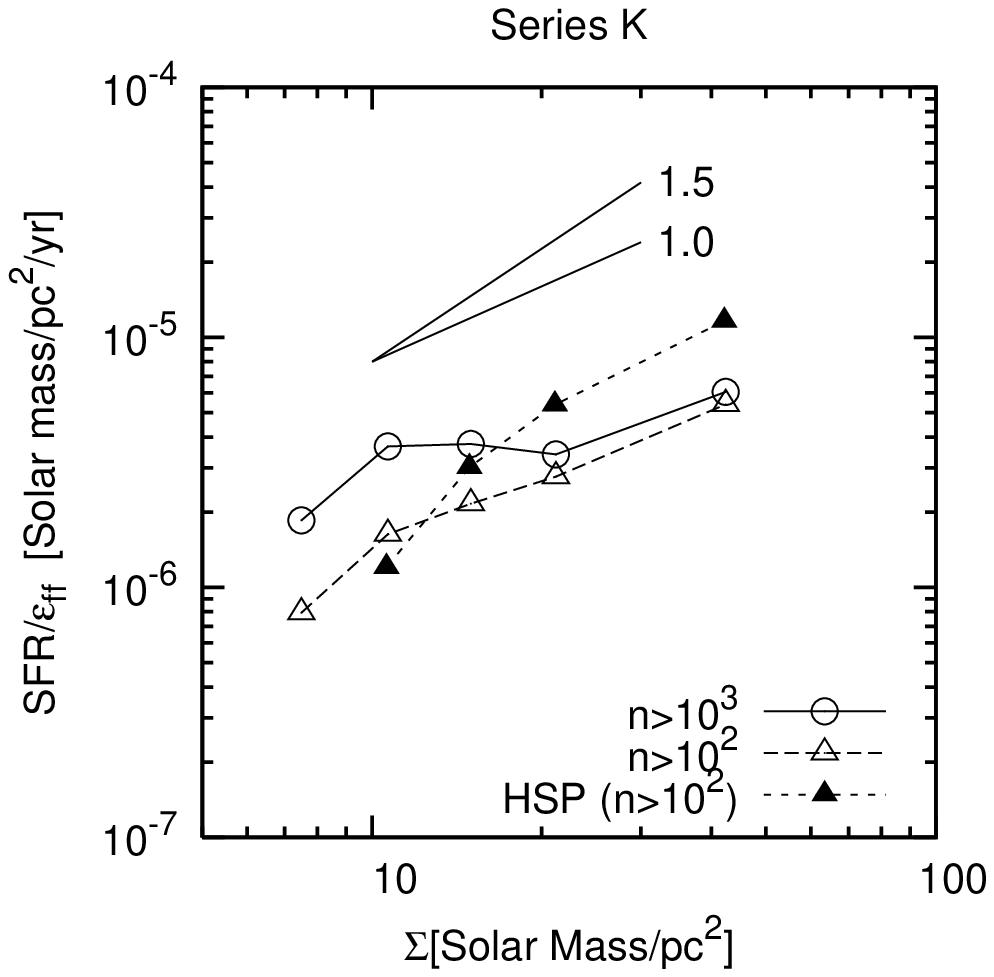}
\plottwo{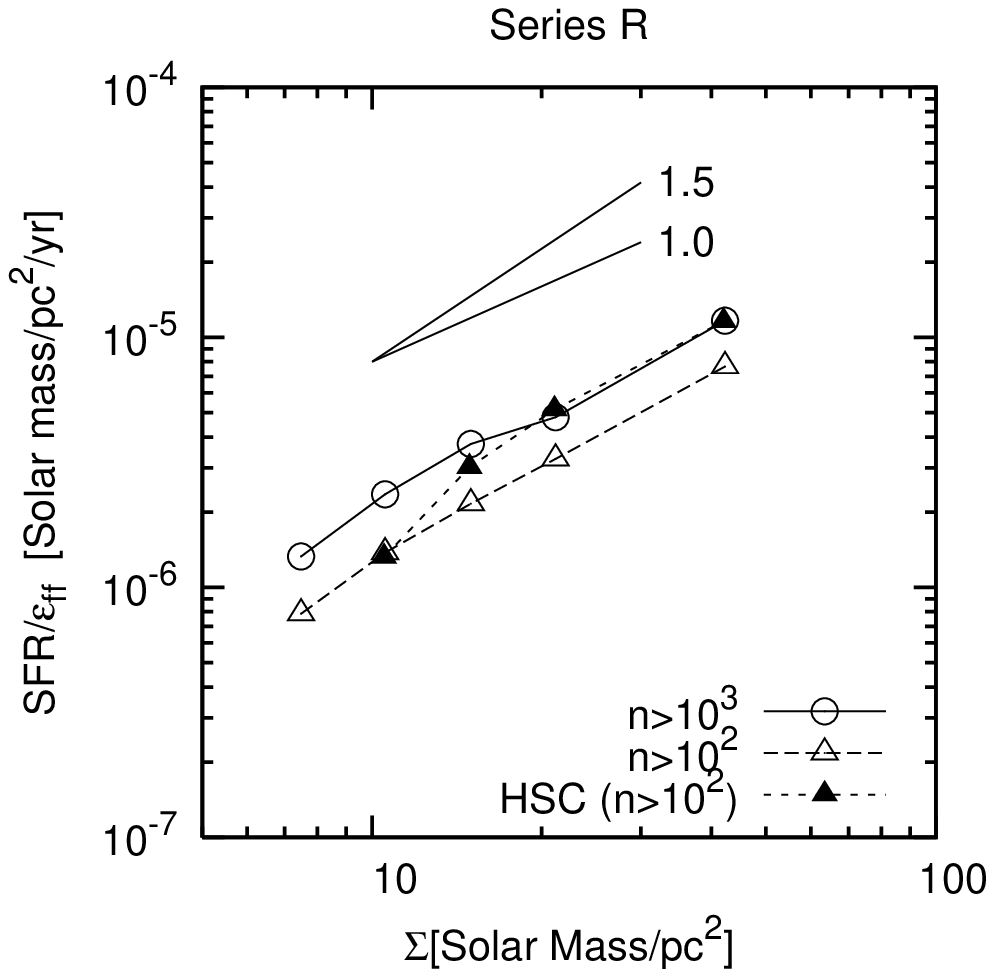}{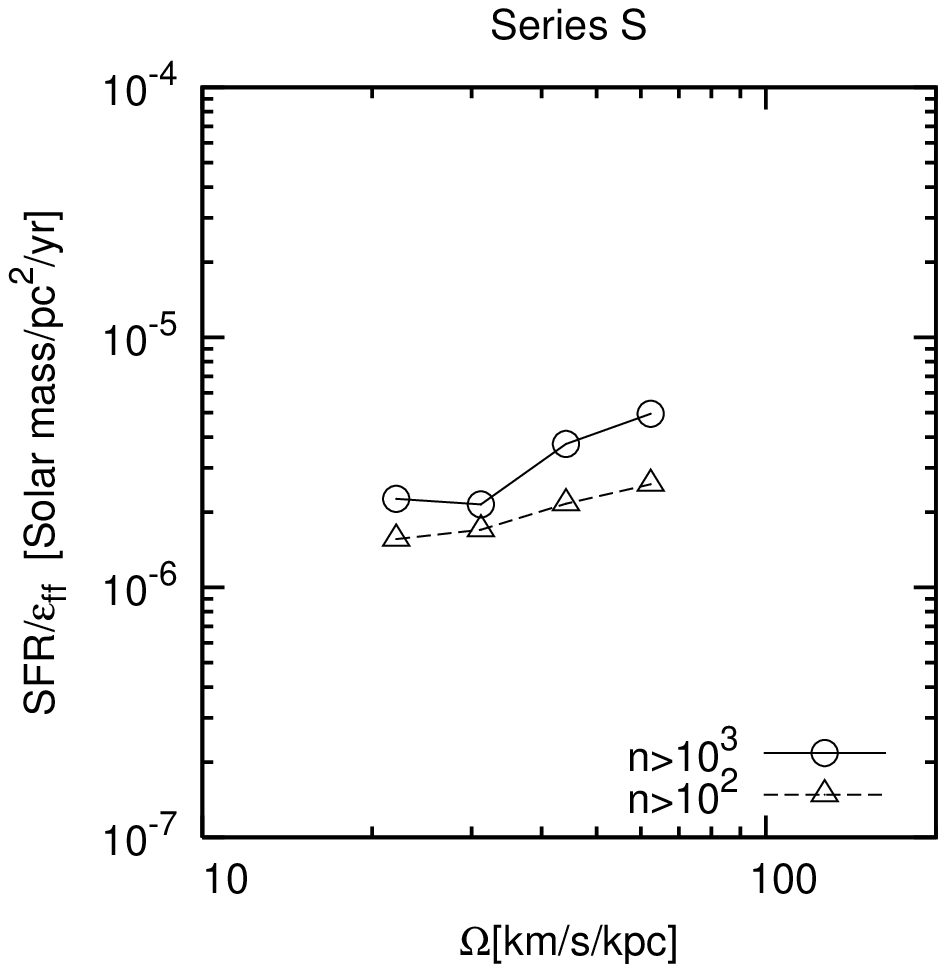}
 \caption{Scaled star formation rates per unit area, 
$\Sigma_{\rm SFR}/\epsilon_{\rm ff}(\rho_{\rm th})\equiv
\Sigma(\rho>\rho_{\rm th})/t_{\rm ff}(\rho_{\rm th})$.
Open triangles and circles are based on threshold
 density of $n_{\rm th}=10^2 ~\cm^{-3}$ and $n_{\rm
   th}=10^3~\cm^{-3}$, 
respectively. 
Filled triangles show the results for hydrostatic models.
The wedge-shape marks indicate reference slopes of 1.0 and 1.5. 
Series Q and R show Kennicutt-Schmidt indices similar to observations.
Efficiency parameters $\epsilon_{\rm ff} \sim 0.001 - 0.01$ for 
 $n_{\rm th}=10^2-10^3 ~\cm^{-3}$ would be required to
match the observed range of $\Sigma_{\rm SFR}$ in this range of gas
surface density $\Sigma$.}
\label{fig:SFR}
\end{figure}

To connect more directly to the way observed SFRs are normally presented,
in Figure \ref{fig:SFR} we show results for scaled surface density of
star formation as a function of
surface density of gas (Series Q, K, R) and angular velocity (Series S).
The scaled SFR per unit area is defined as
$\Sigma_{\rm SFR}/\epsilon_{\rm ff}(\rho_{\rm th})
=\Sigma_{\rm tot}/\tau_{\rm SF}$, where 
the SFR is taken to follow
\begin{eqnarray}
\Sigma_{\rm SFR}\equiv \frac{\Sigma_{\rm tot}}{t_{\rm SF}}
=\frac{\epsilon_{\rm ff}(\rho_{\rm th})
\Sigma_{\rm tot}f(\rho>\rho_{\rm th})}{t_{\rm ff}(\rho_{\rm th})}.
\end{eqnarray}
As before, we compare results based on  two different threshold density
choices, and also show the results from hydrostatic models.
Observations are typically fitted to power laws of the form 
$\Sigma_{\rm SFR} \propto \Sigma^{1+p}$.  For reference, we show
slopes with $1+p=1$ and $1.5$.
For each model series and each value of $\rho_{\rm th}$,
we fit a power-law index.  We find $1+p$ equal to 
 1.32, 1.43 (Series Q for $n=10^2,\, 10^3 ~\cm^{3}$), 
0.94 (Series K for $n=10^2~\cm^{3}$), and 
1.24, 1.19 (Series R for $n=10^2,\, 10^3 ~\cm^{3}$). 
For the hydrostatic cases, the indices are 1.38 (Series HSP) and 1.39
 (Series HSC) at $n=10^2~\cm^{3}$. 
As we shall discuss further in \S \ref{Summary}, 
these results are similar to the
observed ranges of power-law indices that have been reported.  
We note
that Series Q and R show more regular behavior than Series K.  This
reflects the different environmental parameters that are inputs to the
models: in Series K, the epicyclic frequency is held constant, while
in series Q and R we adopt a scaling $\Omega \propto \Sigma$. 
For $\epsilon_{\rm ff}\sim 0.001-0.01$, both the magnitude and scaling
of the $\Sigma_{\rm SFR}$ vs. $\Sigma$ results in Series Q and R 
are similar to observations.

\subsection{Comparison of Timescales}

In \S \ref{sec:SFR}, we investigated the relationship between the mean
large-scale surface density $\Sigma$ and the star formation time based
on the amount of high-density gas (at $n>10^2 ~\cm^{-3}$ within a
zone).  This gas may be considered immediately eligible for star
formation, since it is cold and found in self-gravitating systems.  As
noted in \S \ref{sec:rhoave}, if formation of massive, cold,
gravitationally bound systems is the principal throttle for star
formation, then star formation times would also be expected to vary
with the timescales for GMC formation.

GMC formation is a complex process, and to date no simple formula has
been obtained for the formation rate.  Instead, several different
``large-scale'' dynamical times are commonly invoked to obtain
estimates of the GMC formation time.  These include the free-fall time
at the large-scale mean density, $t_{\rm ff}(\rho_{\rm ave})$, the
Jeans time based on the surface density and the gas velocity
dispersion, $t_{\rm J}$, and the orbital time $t_{\rm orb}$, which is
generally related to the epicyclic and shear times.  It is interesting
to explore how our measurement of $\tau_{\rm SF}$ compares to each of
these times, as a function of the independent parameter in each series.

\begin{figure}
\epsscale{1.1}
\plottwo{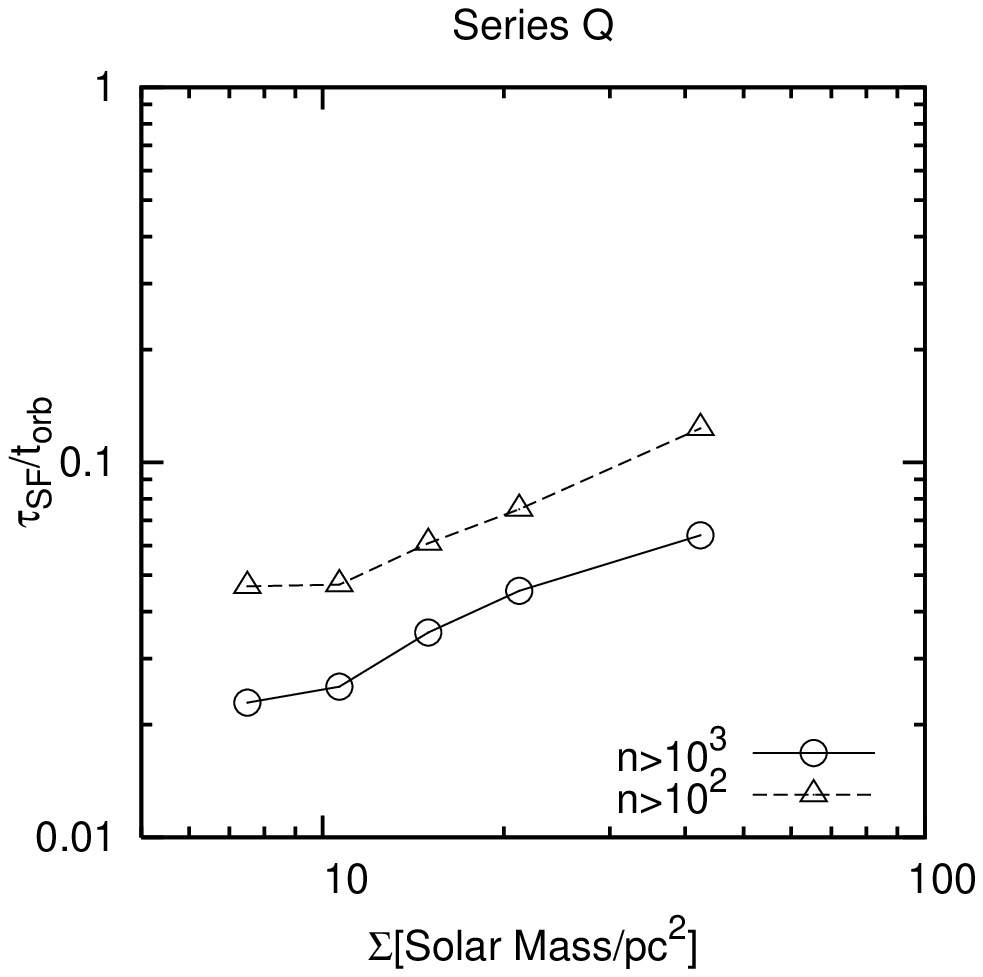}{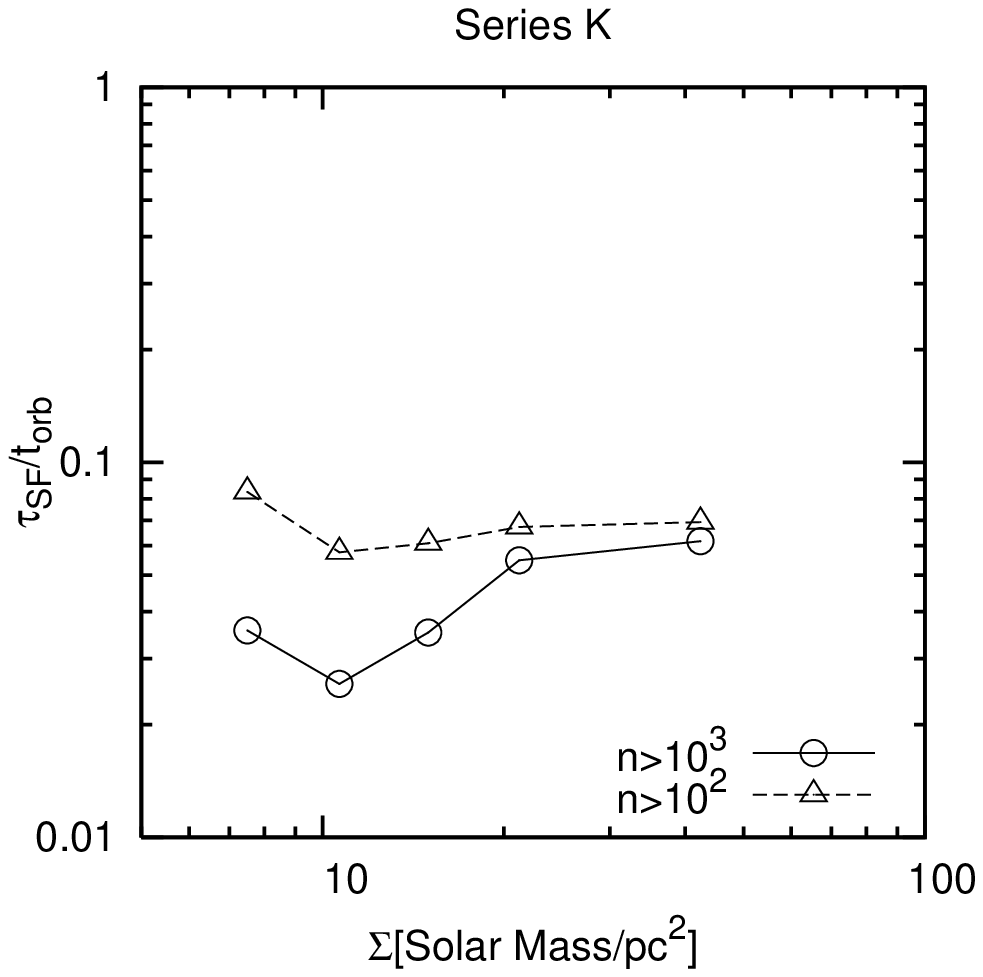}
\plottwo{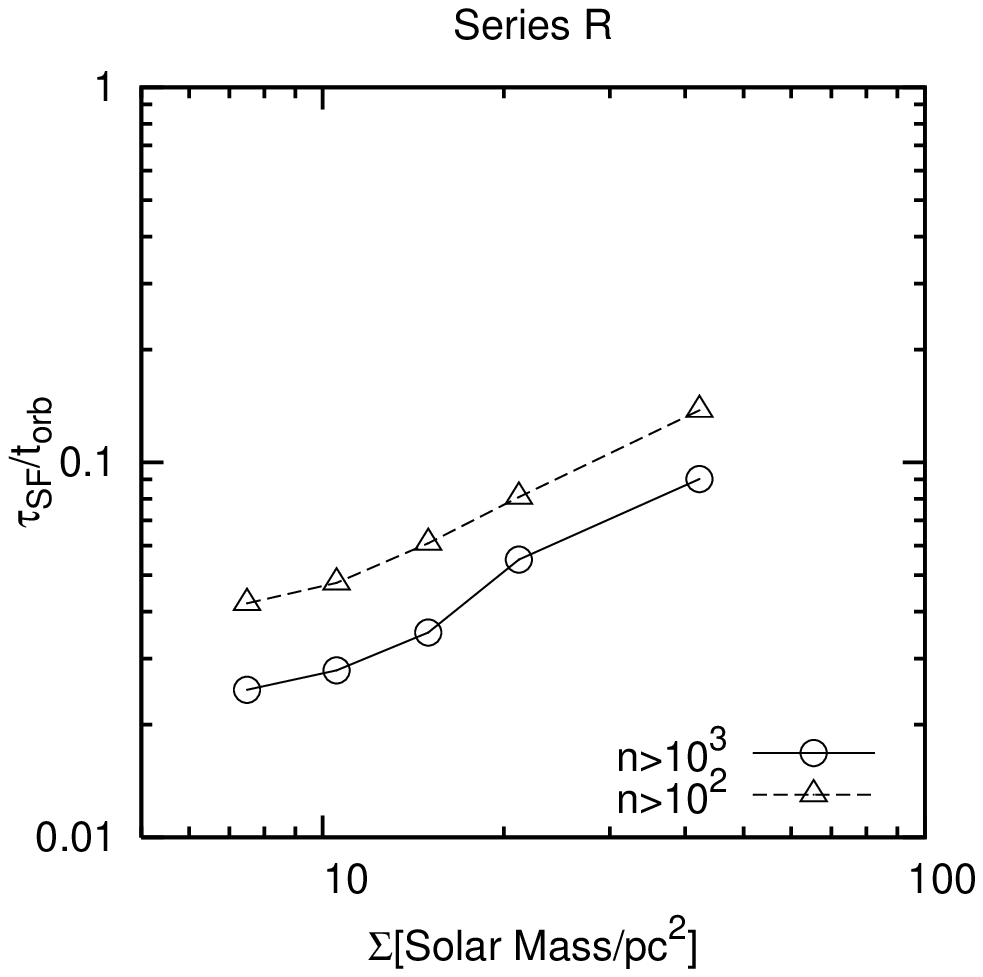}{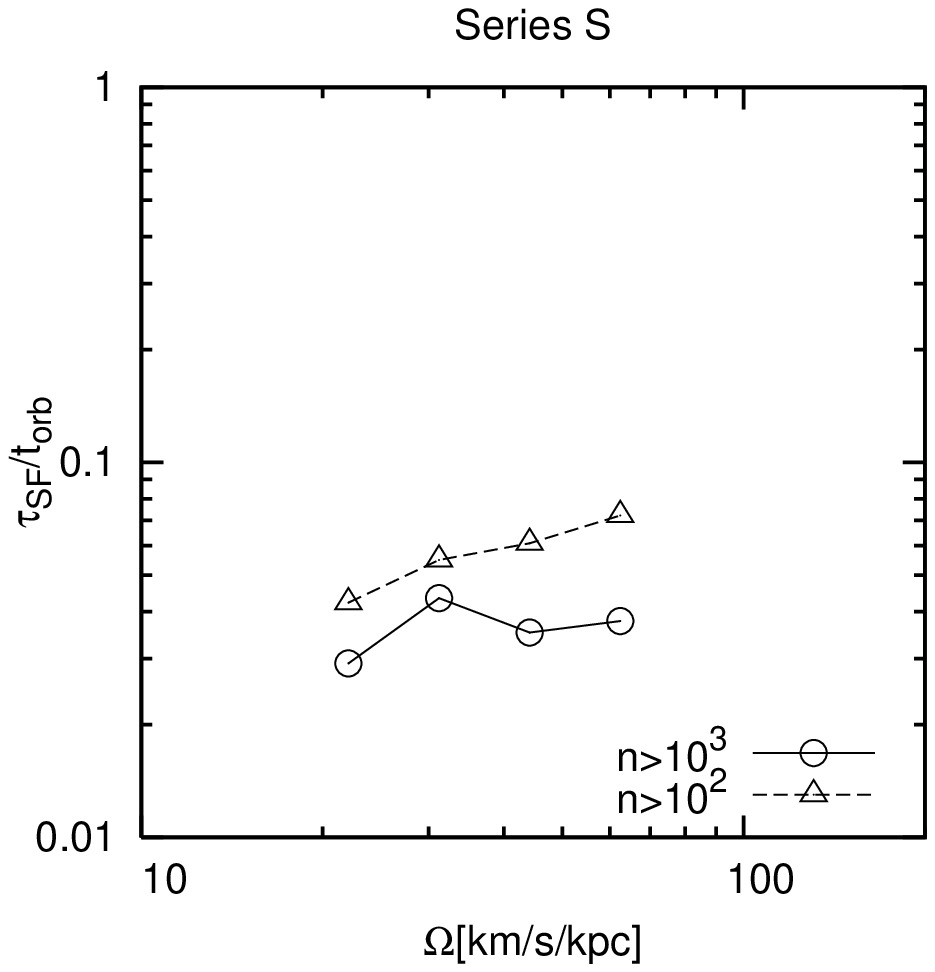}
 \caption{Ratio of scaled star formation time to orbital time,
 indicating that the ratio is not a constant independent of gas
 surface density $\Sigma$.}
\label{fig:tSFtorb}
\end{figure}

We begin with the orbital time, $t_{\rm orb}=2\pi/\Omega$.  Figure
\ref{fig:tSFtorb} shows the ratio between $\tau_{\rm SF}=
\epsilon_{\rm  ff}(\rho_{\rm th})t_{\rm SF}$ (for the two
different density thresholds $\rho_{\rm th}$) 
and the orbital time.  In Series K,
$\Sigma$ is the independent parameter, but $t_{\rm orb}$ is
independent of $\Sigma$; thus the ratio is simply a rescaled version
of $\tau_{\rm SF}$ shown in Fig. \ref{fig:tSF}.  In Series Q and R,
the independent parameter is $\Sigma$, and $t_{\rm orb} \propto
\Sigma^{-1}$, so $\tau_{\rm SF}/t_{\rm orb} \propto \tau_{\rm SF}
\Sigma$.  In Series S, $t_{\rm orb}$ is the independent parameter.
Although the variation with the independent parameter is
moderate in all the series, the ratio is not constant, and for some
series shows secular trends.  Namely, for 
Series Q and R, which showed a trend of decreasing $\tau_{\rm
  SF}$ at increasing $\Sigma$, $\tau_{\rm SF}/t_{\rm orb}$ increases
at larger $\Sigma$. Thus, assuming that the star formation time is
$\propto t_{\rm orb}$ would increasingly overestimate the true star
formation rate (presumed to depend on the amount of dense gas) as
$\Sigma$ increases.

\begin{figure}
\epsscale{1.1}
\plottwo{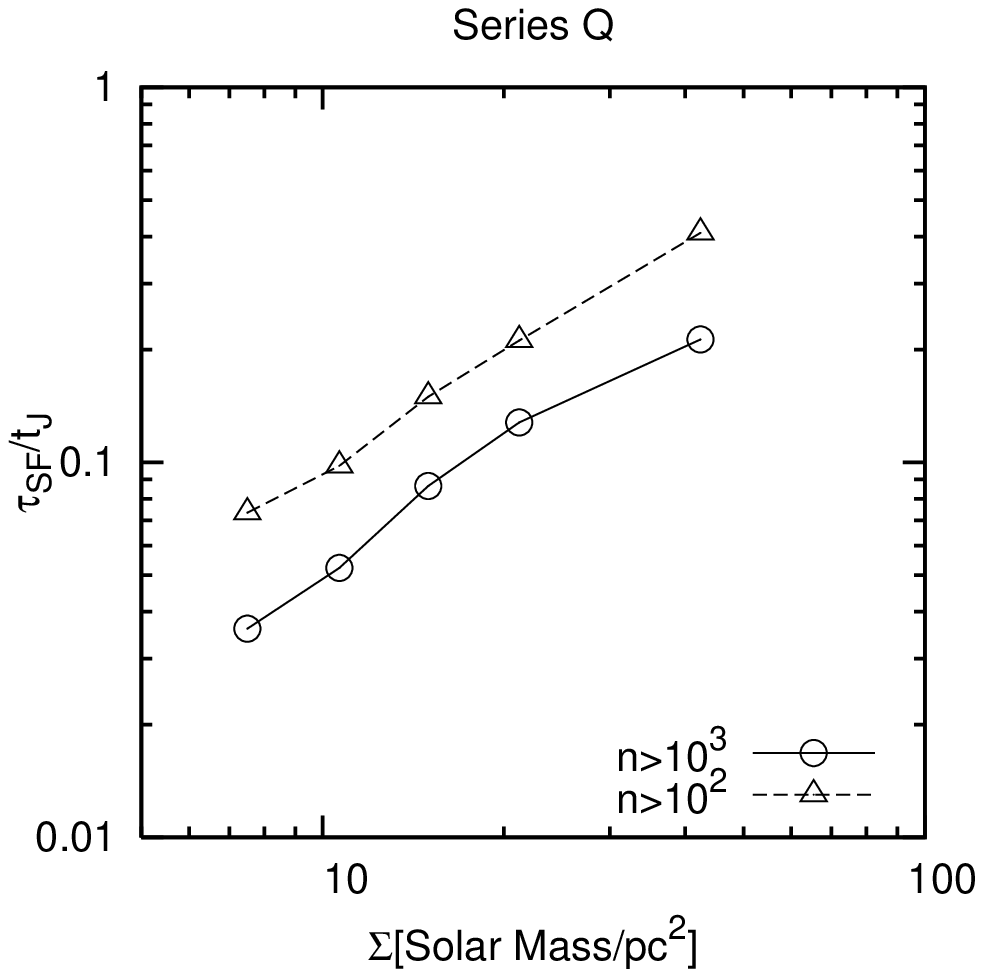}{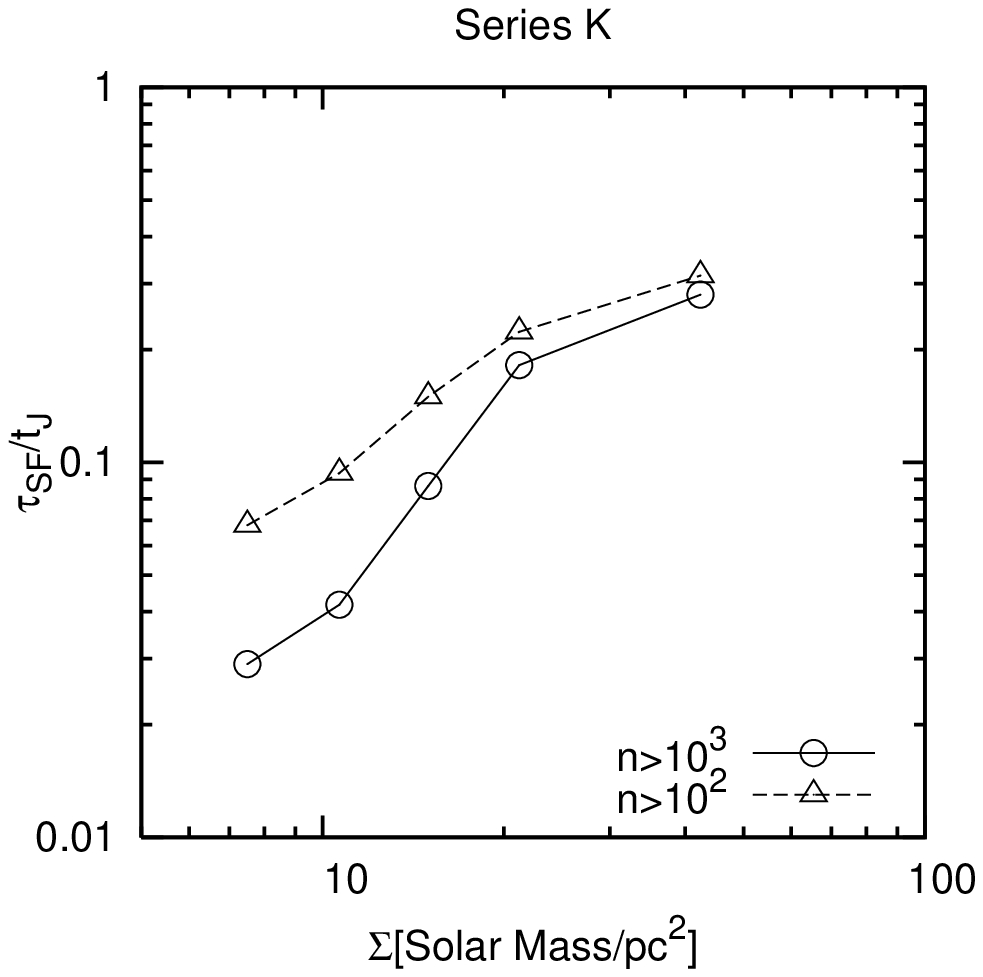}
\plottwo{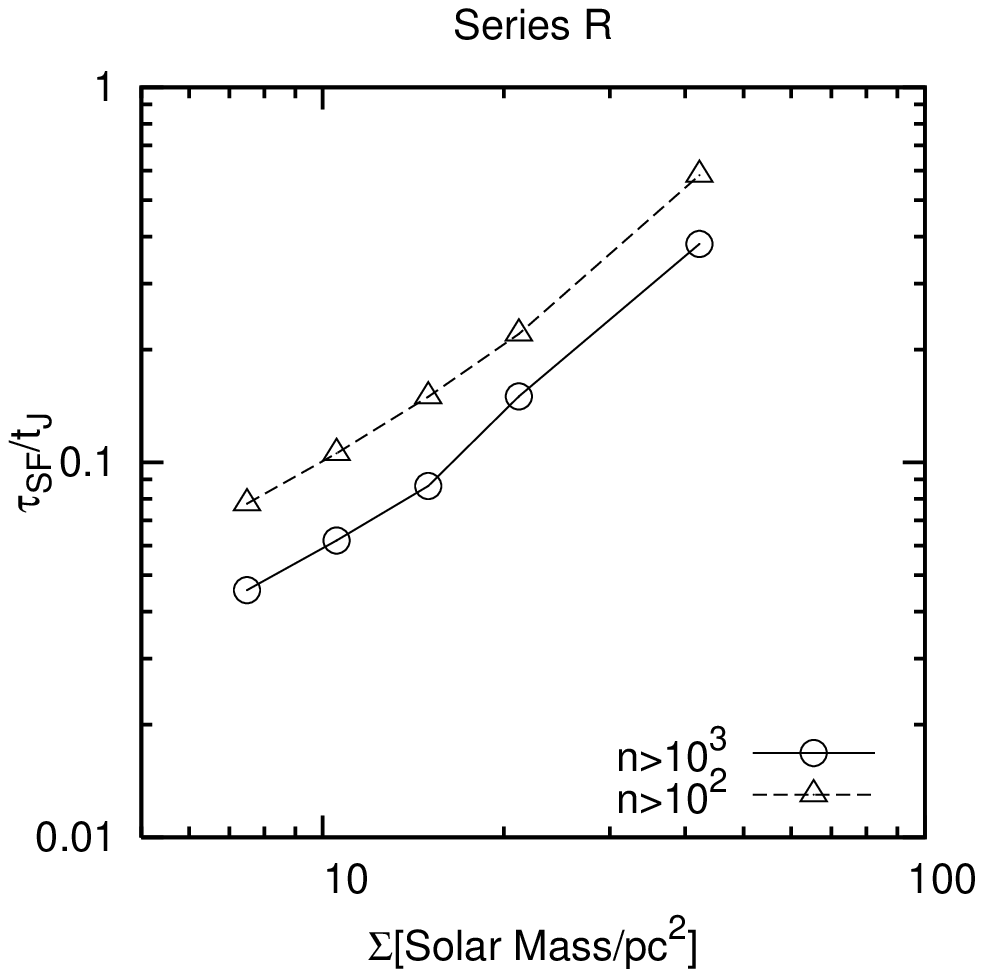}{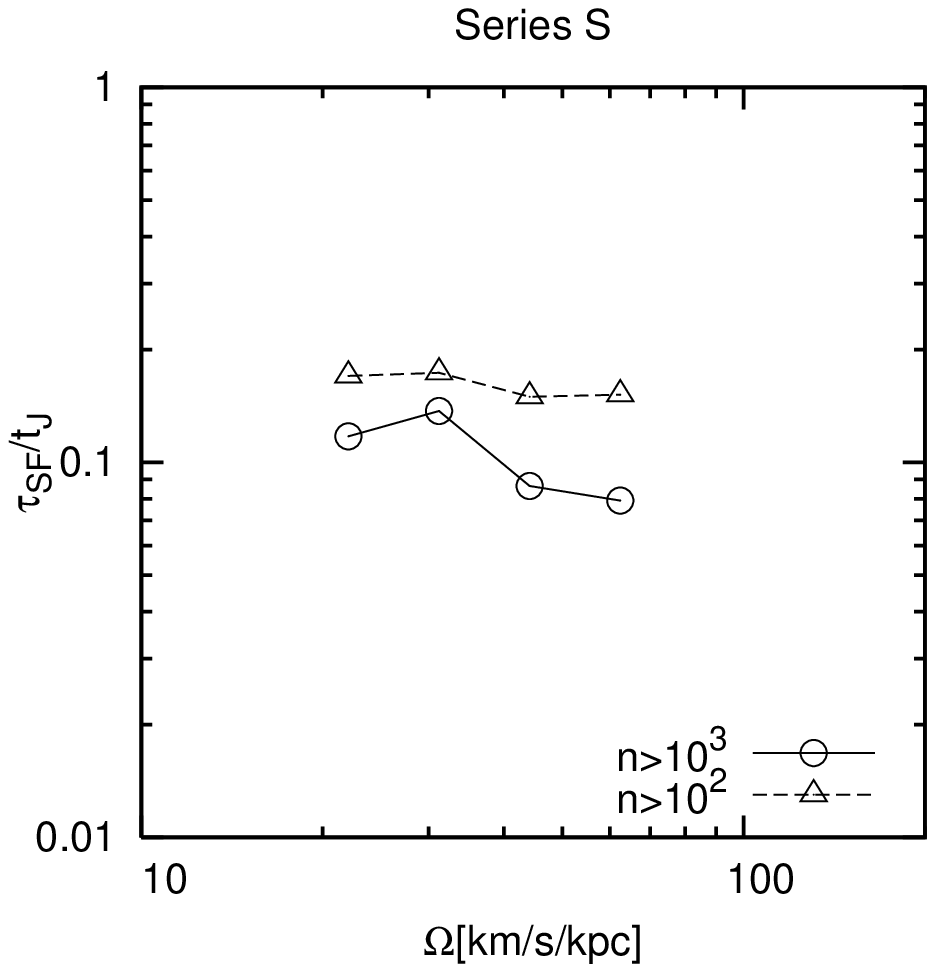}
 \caption{Ratio of scaled star formation time to Jeans time; the Jeans
 time is evidently not a good predictor of the star formation timescale.}
\label{fig:tSFtJ}
\end{figure}

We next consider the Jeans time for a disk, 
$t_{\rm J} \equiv \sigma/(G \Sigma)$,
where $\sigma$ is either the thermal or the total
(radial) turbulent velocity dispersion, $c_s$ or $\sqrt{c_s^2 + v_R^2}$.  
We note that the ratio $t_{\rm J}/t_{\rm orb}$ is given by
$Q/(2\sqrt{2})$, where the Toomre parameter $Q$ is either based on
mean sound speed or the total velocity dispersion (see \S 
\ref{sec:Toomre}).  Figure \ref{fig:tSFtJ} shows the ratio between 
$\tau_{\rm SF}$ and the Jeans time, using 
the total velocity dispersion.
Again, strong secular trends with $\Sigma$ are evident; $t_J$ is not a
good predictor of the star formation time.

\begin{figure}
\epsscale{1.1}
\plottwo{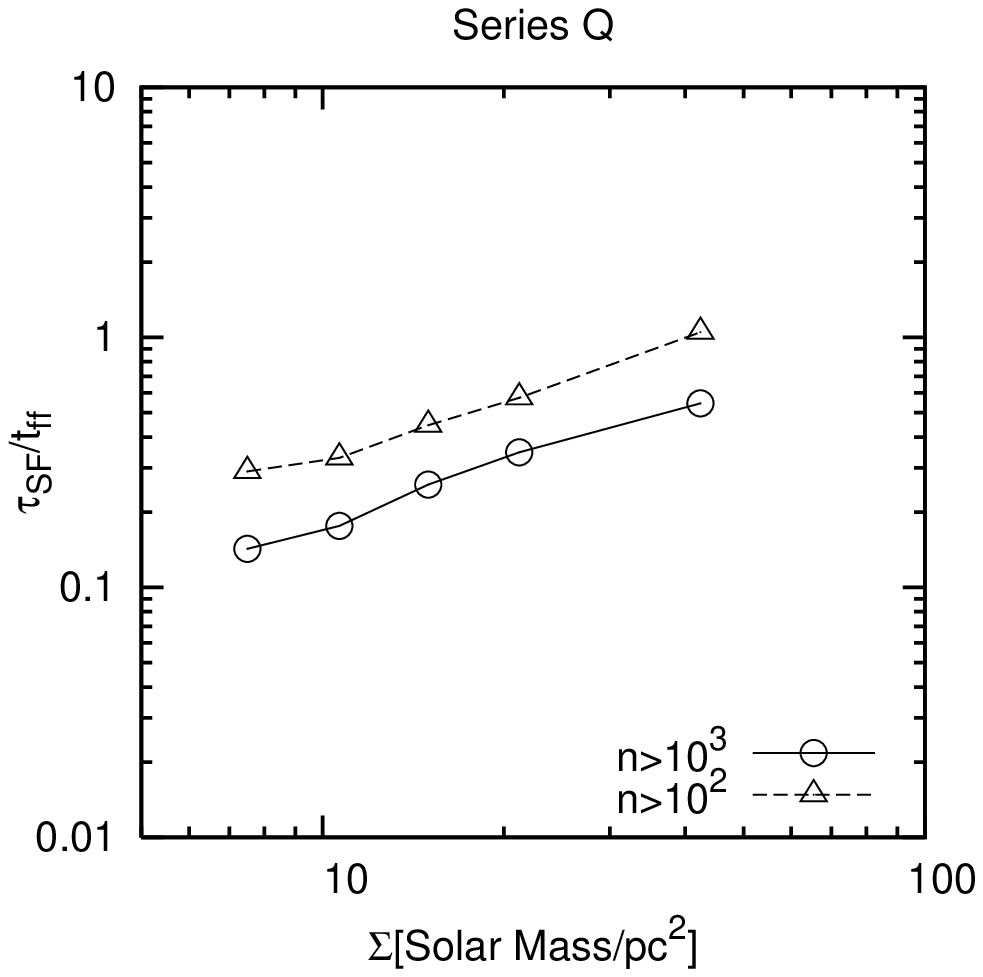}{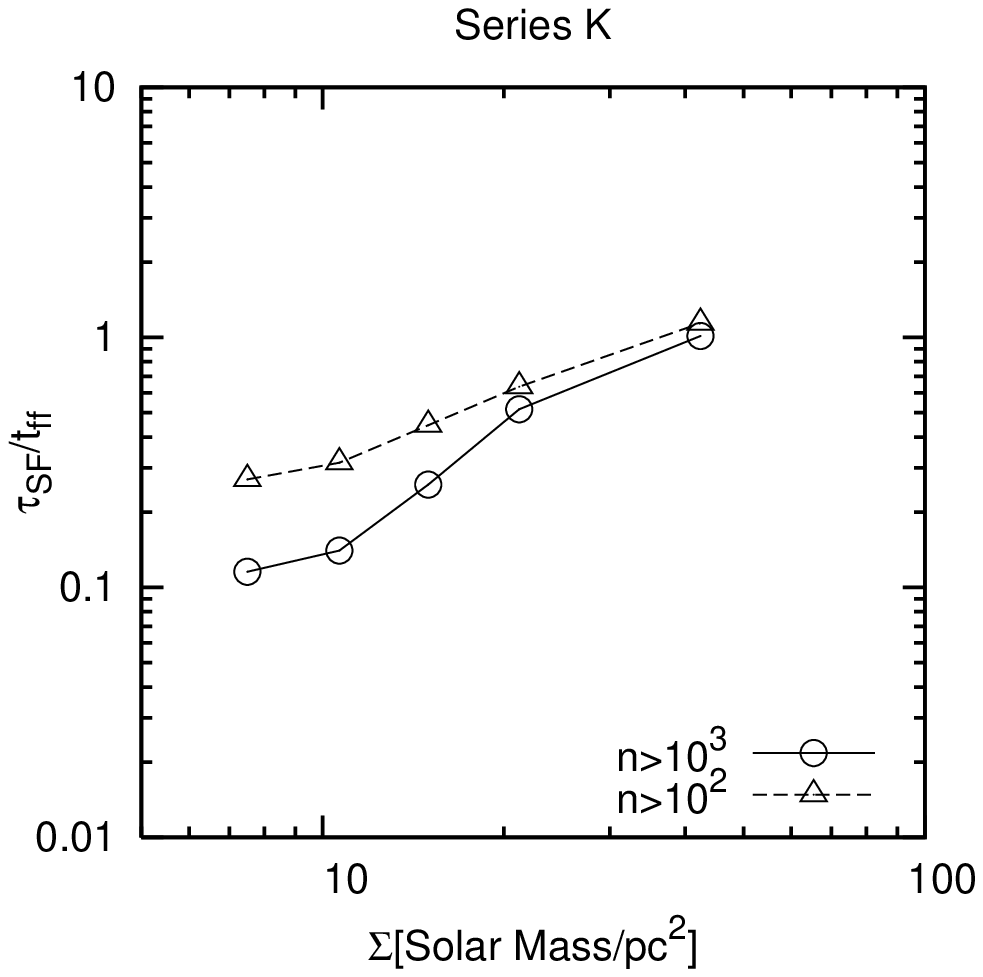}
\plottwo{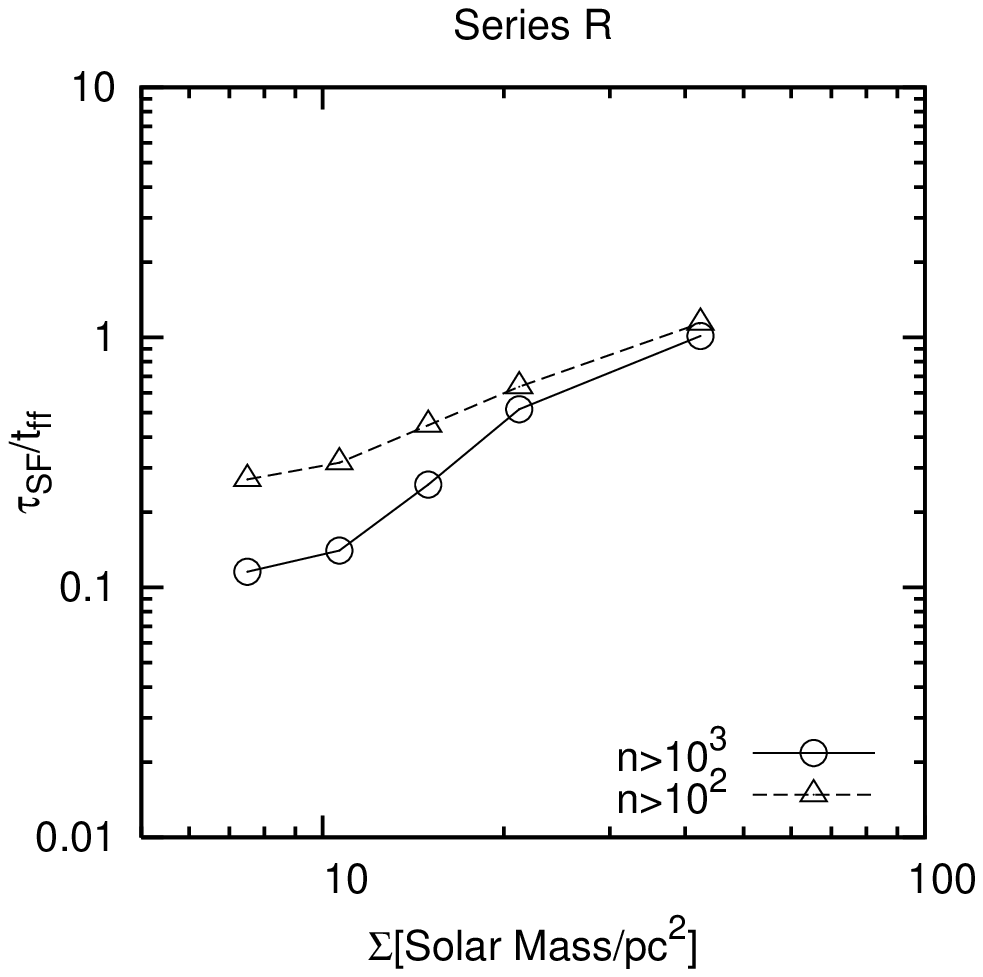}{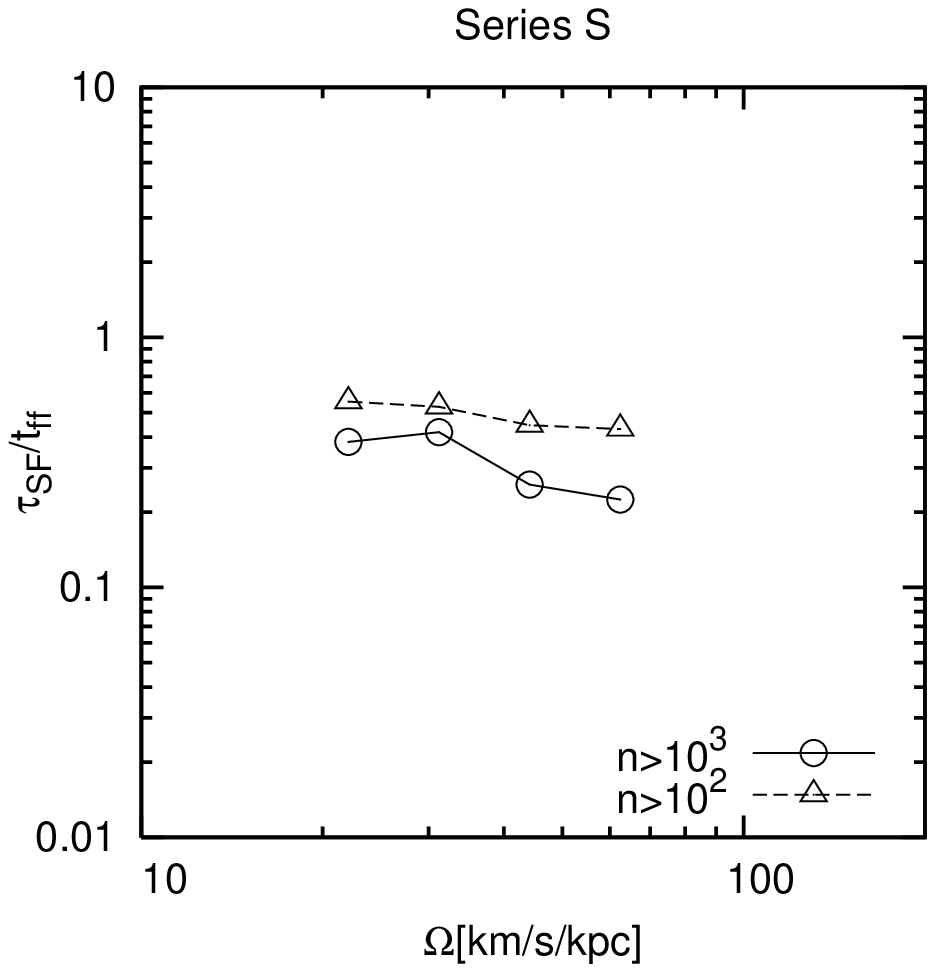}
 \caption{Ratio of scaled star formation time to free fall time at
 large-scale mean density averaged over a disk scale height, 
$t_{\rm ff}(\rho_{\rm ave})$.  An estimated star formation timescale 
proportional to the free-fall time at the  
vertically-averaged density (i.e. the ``vertically-unresolved'' limit)
would increasingly underpredict the true star formation time at 
high $\Sigma$. }
\label{fig:tSFtff}
\end{figure}

Finally, in Figure \ref{fig:tSFtff} we show the ratio between
$\tau_{\rm SF}$ and and the free fall time at the vertically-averaged
large-scale mean density (\S \ref{sec:rhoave}; see Fig. 
\ref{fig:tfftorb}).  Although the values of this ratio are closer to 
unity than $\tau_{\rm SF}/t_{\rm orb}$ and $\tau_{\rm SF}/t_{\rm J}$, we still
see that $\tau_{\rm SF}/t_{\rm ff}(\rho_{\rm ave})$ is not constant as
a function of $\Sigma$.  
When we compare $\Sigma/t_{\rm ff}(\rho_{\rm
  ave})$ to the scaled star formation rates based on high density gas
shown in Fig. (\ref{fig:SFR}), we find 
a steeper rise with $\Sigma$,  close to $\propto \Sigma^2$ in Series Q
and R and slightly shallower in Series K.

If star formation is regulated by the collection of diffuse gas into
self-gravitating regions, then strictly speaking one would expect specific star
formation rates to vary proportional to the fraction of diffuse gas
divided by the GMC formation time (estimated just including the
diffuse gas).  The above comparisons between $\tau_{\rm SF}$ and
timescale estimates based on mean large-scale properties can be corrected
to account for this, yielding the ratios 
$\tau_{\rm SF} (1-f_{\rm dense})/t_{\rm orb}$, 
$\tau_{\rm SF} (1-f_{\rm dense})^{3/2}/t_{\rm ff}(\rho_{\rm ave})$,
and
$\tau_{\rm SF} (1-f_{\rm dense})^2/t_{\rm J}$.  We find, however, that
these ratios also are non-constant in any series, 
although the correction factor does tend to flatten out the secular rise
with increasing $\Sigma$ in the comparison to the free-fall time.

Taking all of our results together, we conclude that several commonly-used 
estimates for galactic star formation timescales based on large-scale
mean galactic properties may have only limited
utility for making detailed predictions of star formation rates.  That
is, the orbital time, the Jeans time, and the free-fall time based on
the vertically-averaged density are not 
proportional to the star formation time based on the amount of dense,
gravitationally bound gas that is present.  Simulations with
insufficient resolution or limited physics may therefore not be able
to provide accurate predictions of star formation rates, if they do
not capture processes at small enough scales to represent dense,
gravitationally-bound structures.

\section{Summary and Discussion
\label{Summary}
} 

We have developed a numerical hydrodynamic code to study the
life-cycle of multiphase, turbulent interstellar gas in disk galaxies;
our model includes gas self-gravity, the vertical gravity associated
with a fixed stellar disk, radiative cooling and heating in the
temperature range of $10 ~\mbox{K} \le T \le 10,000 ~\mbox{K}$,
sheared rotation in the background galactic potential (we adopt a flat
galactic rotation curve), and a prescription for feedback in the form
of HII regions that originate within massive, dense clouds.  Our
simulation domains represent slices in the radial-vertical plane
of a galactic disk.  We focus on scales large enough to include
vertical stratification ($L_z$ up to 410 pc) and significant shear of the
disk angular velocity ($L_x$ up to 1.6 kpc), but small enough
to resolve sub-structure within dense, self-gravitating clouds that
form (typical zone resolution is $\sim 1$pc).  Our models are
2.5-dimensional, in the sense that all three components of the
velocity are time-dependent functions.  For feedback to occur, we
impose thresholds on both the local volume density and on the
gravitational potential, so that HII regions only occur within massive
clouds (consistent with observations).  The expansion of HII regions
drives turbulence in all the components of the gas.  We have performed
a large suite of simulations, covering a factor of six in gas surface
density $\Sigma$. In order to explore the dependence of ISM properties on
galactic environment (in particular, the stellar vertical gravity and
the angular momentum content of the gas), we have considered four
different model series.  In Series Q, we vary $\Sigma$, stellar
volume density $\rho_*$, and the disk rotation rate $\Omega$ in tandem.  
In Series K, we vary $\Sigma$ and $\rho_*$ together while holding
$\Omega$ fixed.  In Series R, we vary $\Sigma$ and $\Omega$ together
while holding $\rho_*$ fixed.  Finally, in Series S the values of
$\Sigma$ and $\rho_*$ are held constant while $\Omega$ is varied.

Our main conclusions, and their relation to other recent work, are as follows:

1. {\it Density, temperature, and pressure distributions}

We find that in spite of time-dependent effects, the density and
temperature distributions of the gas retain bimodal profiles
reminiscent of the classical \citet{1969ApJ...155L.149F} two-phase
model of the ISM.  
Although large-amplitude turbulence 
heats and cools via $P dV$ work and entropy production in shocks, 
most of the gas (by mass) remains near the  
curve in the pressure-density phase plane that is defined
by radiative equilibrium: $n^2 \Lambda - n \Gamma=0$.  This is
possible because the cooling time is generally shorter than the
turbulent dynamical times, for our models.  
If turbulent compressions or expansions
were more extreme in magnitude and also rapid in time compared to radiative
times, then these adiabatic changes would lead to density-temperature
pairs that more strongly departed from thermal equilibrium
(see e.g. \cite{2002ApJ...577..768S,2005A&A...433....1A,2008ApJ...683..786H} 
for a discussion of the dependence on various physical timescales
involved).
\citet{2005A&A...433....1A},
\citet{2005ApJ...629..849P}, and \citet{2007ApJ...663..183P} similarly 
found that even with large-amplitude turbulence -- 
and no direct heating of the gas -- mass-weighted 
density and temperature PDFs have two (broadened) peaks, although 
for very high amplitude turbulence the trough tends to fill in 
(e.g. \citealt{2007A&A...465..431H}).
\citet{2008ApJ...681.1148K} have emphasized the importance of
maintaining sufficient
numerical resolution, as numerical diffusion associated with flow
across the grid broadens warm/cold interfaces, populating the
thermally-unstable regime in the phase diagram.
We note that bimodal character is most easily seen in mass-weighted
rather than volume-weighted PDFs, although many of the results in the
literature show only volume-weighted PDFs.  
Our results on the bimodal thermal distribution of gas 
are consistent with observations of atomic HI
in the midplane of the Milky Way \citep{2003ApJ...586.1067H}, which at
the same time show interesting evidence of out-of-equilibrium gas,
particularly at high latitudes.

2. {\it Turbulence} 

We find that appreciable turbulence can be excited in all components
of the gas.  The values of the velocity dispersion in dense gas 
($n>10^2 \cm^{-3}$) of $\sim 2-4\ \kms$ are similar to, though slightly
lower than, those observed in Milky Way and other local-group 
GMCs (e.g. 
\citealt{1987ApJ...319..730S,
2008ApJ...675..330S,
2008arXiv0807.0009B,
2008arXiv0809.1397H}).  For lower-density gas in
our models, velocity dispersions are slightly higher, but still lower
by a factor two compared to observed velocity dispersions of 
$\sim 7-10~\kms$ of Solar-neighborhood 
warm and cold HI seen in 21 cm emission and absorption 
\citep[e.g.,][]{2003ApJ...586.1067H,2004JApA...25..185M}.

It is not surprising that the turbulence levels in our simulations are
only moderate, given that we have included only one of the many sources of
turbulence that is present in the ISM.  Turbulent driving in the ISM 
has been reviewed by e.g. 
\citet{2004RvMP...76..125M,2004ARA&A..42..211E}. 
Supernova are widely-believed to be the most important
source of turbulence for diffuse gas, and this has been demonstrated
by numerical simulations 
\citep[e.g.,][]{
1999ApJ...514L..99K,  
2005A&A...436..585D}. 
In the outer galaxy
where star formation rates are low, driving by
the magnetorotational instability may, however, dominate 
\citep{1999ApJ...511..660S,2005ApJ...629..849P,2007ApJ...663..183P}.  
Spiral shocks are
also effective in driving turbulence in the warm ISM, 
especially at high latitudes \citep{2008ApJ...681.1148K}.
The interaction between large-scale self-gravity, rotation, and shear 
can drive near-sonic turbulence at large scales
\citep{2002ApJ...577..197W,2007ApJ...660.1232K}, 
although the amplitude of this at
scales less than the disk scale height may be modest.
It is not known how effective these other mechanisms are for driving
turbulence within GMCs, however, which are very dense and therefore present a
small effective crossection to the diffuse ISM.

3. {\it Feedback and the Toomre $Q$ parameter} 

We have measured the Toomre parameter for each of our models, considering 
both the entire medium and just the dense gas, 
and comparing ``thermal-only'' with ``turbulent+thermal'' values. 
For all models, we find that the ``turbulent+thermal'' $Q$-value for
dense gas is in the range $1-2$, and is
much greater than the thermal-only value.  
A further interesting point is that the turbulence level evidently adjusts
with surface density in order to reach a marginally-stable state.
In Series Q and R, which have $\Omega/\Sigma=const.$, 
the velocity dispersions are relatively independent
of $\Sigma$, yielding marginally-stable $Q$ in the cold, dense gas.  
In Series K, which has constant $\Omega$ and
therefore is highly unstable at large $\Sigma$ in the absence of
turbulence, the velocity
dispersions strongly increase with $\Sigma$, as a consequence of much
higher levels of feedback activity.  These higher velocity dispersions
lift $Q$ to near unity. Thus, our simulations give
direct evidence of feedback leading to a self-regulated quasi-steady
state.  We note that $Q$ values vary for different 
components; \citet{2002ApJ...577..197W} similarly found a large range
of $Q$ when measured in local patches within their disk simulations.

Depending on what exactly is included in a model, the
threshold for gravitational instability in previous non-turbulent 
simulations is measured 
to be at $Q\sim 1.5$ (see \citealt{2007ARA&A..45..565M}), which is
similar to the values we find here when turbulence is included in $Q$. 
In the actively-star-forming regions of galaxies, measured values of the
Toomre parameter are not constant, but show a limited range
\citep{2001ApJ...555..301M}.
Evidence for star formation thresholds tied to $Q$ are more mixed
\citep{2001ApJ...555..301M,
2003MNRAS.346.1215B,
2007ApJS..173..524B}, possibly
because star formation in outer disks primarily takes place in spiral
arms 
\citep{1998ApJ...506L..19F,2007ApJS..173..538T,2008ApJ...683L..13B}
which strongly compress the gas above ambient densities.

4. {\it The virial ratio}

We measure the virial ratio $\cal R$ (eq. \ref{eq:vir}) for all our models,
separating into different density regimes.  We find that dense gas 
($n>100\ \cm^{-3}$) generally has $\cal R$ between $1 - 2$, whereas
lower-density gas has large values of $\cal R$.  
$\cal R$ does not vary strongly with $\Sigma$ in any of the series.
In particular, we note that in 
spite of the large range of velocity dispersions in the dense gas in
Series K at different $\Sigma$, $\cal R$ varies only weakly with
$\Sigma$.   This indicates that feedback can
effectively regulate the dynamics within massive, dense clouds,
independent of the larger-scale galactic environment.  
This is consistent with both older studies based on $^{12}$CO
observations \citep{1987ApJ...319..730S}, and recent studies
based on $^{13}$CO observations \citep{2008arXiv0809.1397H}, both of
which find $\cal R$ near $1-2$ for Milky Way GMCs. Although masses based
on CO are less certain in external galaxies, virial ratios also likely
near unity for GMCs observed in the Local Group \citep{2008arXiv0807.0009B}

5. {\it Cloud surface densities}

We estimate the surface density of typical clouds by measuring the
mass-weighted vertically-integrated column of gas; we define this
as $\Sigma_{\rm cloud}$.  For the values of the feedback threshold
that we adopt, we find that $\Sigma_{\rm cloud}$ is in the range
$70-150 ~\msun \pc^{-2}$ for most models.  This is comparable to the
typical GMC surface densities that are observed in the Milky Way and
in Local Group galaxies
\citep{1987ApJ...319..730S,2008ApJ...675..330S,
2008arXiv0807.0009B,
2008arXiv0809.1397H}.  Whether observational selection effects or
physical processes impose a limited range of column densities for GMCs
is an open question.  For example, magnetic fields may impose a minimum surface
density for formation of gravitationally-bound structures in the ISM,
at a value $\Sigma = B/(2 \pi\sqrt{G})= 30 ~ \msun \pc^2 (B/10 \mu {\rm
  G})$ \citep{2007ARA&A..45..565M}.  Here, we find that
altering the volume heated in our HII region
prescription does not appreciably change $\Sigma_{\rm cloud}$, but
changing the gravitational potential threshold for star formation
feedback does:  when potential thresholds are low, $\Sigma_{\rm
  cloud}$ is also low.   Comparison of cloud properties with
observations may turn out to be a much 
more critical test of whether an ISM model is realistic than some
other measures, such as the star formation rate.

6. {\it Dependence of $\Sigma_{\rm SFR}$ on $\Sigma$} 

We obtain estimates of the dependence of surface star formation rate
$\Sigma_{\rm SFR}$ on gas surface density $\Sigma$
in our models by assuming that the timescale for star
formation is proportional to the free-fall time in gas above some
density threshold $n_{\rm th}=\rho_{\rm th}/\mu$.  We
compare results for two different threshold densities, $n_{\rm th} = 100\
\cm^{-3}$ and $n_{\rm th} = 10^3\ \cm^{-3}$.  For Series Q and R
(which most resemble real galaxies), we find (for
either choice of threshold) 
relations that are well-described by power laws:
$\Sigma_{\rm SFR}\propto \Sigma^{1+p}$ with $1+p=1.2-1.4$.  In Series
K, a power law is a less-good fit; the slope is also shallower (closer
to unity).

These results are consistent with empirical Kennicutt-Schmidt laws
\citep{1959ApJ...129..243S,1963ApJ...137..758S,1998ARA&A..36..189K},
which show similar values of $1+p$ when all the gas is included in $\Sigma$  
(e.g. 
\citealt{1989ApJ...344..685K,1998ApJ...498..541K,2002ApJ...569..157W,
2007A&A...461..143S,2007ApJ...671..333K}).  
Recent work has suggested that $1+p$
is close to unity if just CO-emitting molecular gas is included
\citep{Bigiel2008}
; this implies all CO-emitting gas (most of which
is at $n=10^2-10^3 \cm^{-3}$)
has the same star formation rate independent of galactic environment.
Our prescription that the star formation rate
per unit mass of dense gas is constant is equivalent to empirically
finding $1+p=1$ if only molecular gas is included.

It is encouraging that the results we find for Kennicutt-Schmidt
relations in our disk models with feedback are compatible with observations.
We also find, however, that star formation rates predicted from
hydrostatic models are in fact similar to those predicted from the 
hydrodynamic models,
with similar slopes at large $\Sigma$.  This is true even though the
hydrostatic models are not at all like real galaxies. Thus, one must
be cautious in considering a numerical model successful if it yields
reasonable star formation rates, since this can simply be a
consequence of choosing reasonable initial conditions
in a simulation.  Indeed, a number of recent
numerical studies have found results similar to observed
Kennicutt-Schmidt laws, regardless of the detailed physics that they
included in the models (e.g. 
\citealt{2006ApJ...639..879L,2006ApJ...641..878T,
2008ApJ...673..810T,2008ApJ...680.1083R}).
\citet{2008MNRAS.383.1210S} have also recently emphasized that
reproducing empirical star formation scaling laws is not by itself a 
critical test of an ISM model.

7. {\it Density-dependence of star formation efficiency}

The star formation efficiency per free-fall time can be defined
locally as a 
function of threshold density by $\epsilon_{\rm ff}(\rho_{\rm th}) =
t_{\rm ff}(\rho_{\rm th}) \Sigma_{\rm SFR}/\Sigma(\rho> \rho_{\rm
th})$; corresponding global measures can also be obtained.  For a
given true star formation rate, 
$\epsilon_{\rm ff}(\rho_{\rm th}) \propto \tau_{\rm SF}$
where the scaled star formation time 
$\tau_{\rm SF}$ is shown for two different threshold densities
in Fig. \ref{fig:tSF}. We find that $\tau_{\rm SF}$ 
(or $\epsilon_{\rm ff}(\rho_{\rm th})$)
decreases with increasing $\rho_{\rm th}$.   This is not a strong
effect, however: it is less than a factor 2 for an order of magnitude
difference in $\rho_{\rm th}$.  In Series Q and R, the 
ratio of efficiencies at different 
threshold densities are also independent of $\Sigma$ (although this is
not true for Series K).  

\citet{2007ApJ...654..304K} recently compiled a range of observations
of $\epsilon_{\rm ff}$ (which they refer to as SFR$_{\rm ff}$).  They
point out that for threshold densities above $\sim 100 \cm^{-3}$, the 
value of $\epsilon_{\rm ff}$ does not vary strongly with density.  They also
  find a smaller value for gas traced by HCN than for gas traced by
  CO, which since HCN has a higher critical density than CO is
  consistent with our finding that $\epsilon_{\rm ff}$ decreases with
  increasing $\rho_{\rm th}$.

Even though scaled star formation times 
at high density vary together independent of $\Sigma$, the free-fall time at
the vertically-averaged density is {\it not} proportional to
$\tau_{\rm SF}$.  Instead, we find that $\tau_{\rm SF}/t_{\rm
  ff}(\rho_{\rm ave})$ increases at increasing $\Sigma$.  This implies
that a prescription for star formation based on the mean density
within one ``average'' 
scale height in a disk (or from a simulation that does not 
resolve high-density gas)
would increasingly overestimate the star formation rate at high $\Sigma$.
The same is true for the orbital time and the Jeans time:  an
assumption that the star formation rate varies $\propto \Sigma/t_{\rm
  orb}$ or $\propto \Sigma/t_{\rm J}$ would increasingly overestimate 
$\Sigma_{\rm SFR}$ at high $\Sigma$ compared to the value obtained
from measuring the mass of gas in dense, gravitationally-bound
regions.  Thus, ISM models must resolve self-gravitating structures at
scales less than the disk thickness in
order to make accurate predictions of the star formation rate.

\begin{acknowledgements}
\vspace{12pt}

We are grateful to the referee for a thorough reading and thoughtful
set of comments that have helped us to clarify our presentation.  
This research was supported by grant NNG-05GG43G from NASA.  Numerical
computations were carried out on the OIT High Performance Computing
Cluster (HPCC) and CTC cluster in the Department of Astronomy, at
the University of Maryland.

\end{acknowledgements}

\bibliographystyle{apj}


\clearpage
\appendix
\section{Cartesian Disk Potential via Fourier Transforms}

In this section, we provide details for our method of solving
Poisson's equation in Cartesian disk geometry.
We generalize to the three-dimensional case.
Thus, the solution is applicable to problems in which periodic
boundary conditions in the horizontal (x-y) directions are assumed,
and vacuum boundary conditions are required in the vertical (z)
direction.
Following Binney \& Tremaine (1987) and \citet{1987PThPh..78.1273M},
the gravitational potential at the location $(x_a,x_b,x_c)$ on a 
regular Cartesian grid with dimensions $(N_x,N_y,N_z)$ for integer
indices $(a,b,c)$ can be written as  
\begin{eqnarray}
\Phi(x_a,y_b,z_c)=-\frac{4\pi G}{N_xN_y} 
\sum_{m=0}^{N_x-1} \sum_{n=0}^{N_y-1} \sum_{j=0}^{N_z-1} 
e^{\displaystyle -2\pi i\left(\frac{am}{N_x}+\frac{bn}{N_y}\right)}
\hat{\rho}_{m,n}(z_j)G_{m,n}(z_c,z_j).
\end{eqnarray}
Here $G_{m,n}$ is the Green function of the Poisson equation for a
horizontal sheet sinusoidal source charge and is written as 
\begin{eqnarray}
G_{m,n}(z,z^{\prime})&=&\frac{L_z}{N_z}
\left\{
\begin{array}{ll}
\displaystyle \frac{1}{2k_{m,n}}\exp(-k_{m,n}|z-z^{\prime}|)
\quad & (k_{m,n}\ne 0),\\
\displaystyle -\frac{1}{2}|z-z^{\prime}| &(k_{m,n}=0),
\label{eq:green}
\end{array}
\right.
\end{eqnarray}
where
\begin{eqnarray}
k_{m,n}&=&\left[
\left(\frac{2\pi m}{L_x}\right)^2
+\left(\frac{2\pi n}{L_y}\right)^2\right]^{1/2}.
\end{eqnarray}
Note that up to a constant, the $k_{m,n}=0$ solution is equal to the
limit of the general solution for $k_{m,n}\to 0$. 
The coefficients $\hat{\rho}_{m,n}(z_j)$ are given by the discrete
Fourier transforms of the density in the horizontal plane $z=z_j$,
i.e., 
\begin{eqnarray}
\hat{\rho}_{m,n}(z_j)=\sum_{a=0}^{N_x-1}\sum_{b=0}^{N_y-1}
e^{\displaystyle 2\pi i\left(\frac{am}{N_x}+\frac{bn}{N_y}\right)}
\rho(x_a,y_b,z_j).
\end{eqnarray}
Since $z-z^{\prime}$ in the Green function takes on values between
$-L_z$ and $L_z$, we can apply Fourier transforms on the domain
$(-L_z,L_z)$ to obtain
\begin{eqnarray}
G_{m,n}(z_c,z_j)=\frac{1}{2N_z}\sum_{\ell=0}^{2N_z-1}
e^{\displaystyle -2\pi i \frac{(c-j)\ell}{2N_z}}\hat{G}_{m,n,\ell}
\end{eqnarray}
where
\begin{eqnarray}
\hat{G}_{m,n,\ell}\equiv \sum_{p=0}^{2N_z-1}
e^{\displaystyle 2\pi i \frac{(p-N_z)\ell}{2N_z}}
G_{m,n}\left(z-z^{\prime}=\frac{(p-N_z)L_z}{N_z}\right)
\end{eqnarray}
Substitution of equation (\ref{eq:green}) and evaluation of the sum shows that 
\begin{eqnarray}
\hat{G}_{m,n,\ell}=
\frac{1-\exp\left(\pm i \ell \pi - k_{m,n}L_z\right)}
{k_{m,n}^2+\left(\frac{\pi\ell}{L_z}\right)^2}.
\end{eqnarray}
If we now substitute into the original expression for $\Phi$, 
we obtain
\begin{eqnarray}
\Phi(x_a,x_b,x_c)=-\frac{4\pi G}{N_xN_y(2N_z)}
\sum_{m=0}^{N_x-1}
\sum_{n=0}^{N_y-1}
\sum_{\ell=0}^{2N_z-1}
\frac{e^{\displaystyle -2\pi i 
\left(\frac{am}{N_x}+\frac{bn}{N_y}+\frac{c\ell}{2N_z}\right)}
}{k_{m,n,\ell}^2}
\hat{\rho}_{m,n,\ell}^{\rm corr}
\end{eqnarray}
where
\begin{eqnarray}
k_{m,n,\ell}^2=
\left(\frac{2\pi m}{L_x}\right)^2
+\left(\frac{2\pi n}{L_y}\right)^2
+\left(\frac{2\pi \ell}{2L_z}\right)^2
\end{eqnarray}
and
\begin{eqnarray}
\hat{\rho}_{m,n,\ell}^{\rm corr}\equiv \sum_{j=0}^{2N_z-1}
e^{\displaystyle 2\pi i \frac{j\ell}{2N_z}}
\hat{\rho}_{m,n}^{\rm corr}(z_j).
\end{eqnarray}
Here,
\begin{eqnarray}
\hat{\rho}_{m,n}^{\rm corr}\equiv
\left\{\begin{array}{ll}
\hat{\rho}_{m,n}(z_j) & \mbox{for ~ $j=0,N_z-1$} \\
-\hat{\rho}_{m,n}(z_{j-N_z})e^{-k_{m,n}L_z} & \mbox{for ~ $j=N_z,2N_z-1$}
\end{array}
\right.
\end{eqnarray}
That is, in addition to using the horizontal Fourier components of the
original domain, we must define an image density in the augmented
domain, based on the $\hat{\rho}_{m,n}$ values a distance $L_z$ away. 
After the corrected density is defined, a Fourier transform in $z$ is
taken in the usual way.

The complete procedure for obtaining the gravitational potential is
therefore as follows:
First compute two-dimensional ($x-y$) Fourier components for each
horizontal layer in $z$.
Next define $\hat{\rho}_{m,n}^{\rm corr}(z_j)$ and take the Fourier
transform in $z$ to obtain $\hat{\rho}_{m,n,\ell}^{\rm corr}$.
Finally, multiply by the Poisson kernel $-4\pi G/k_{m,n,\ell}^2$,
and take the three-dimensional inverse transform.
We note that even and odd terms in the sum on $\ell$
may also be combined, such that the forward and inverse transforms
are both three dimensional; the Poisson kernel then is adapted to
incorporate terms from the image charge.

Since we use FFTs (via the Package FFTW) for all transforms the
computational expense scales as $O(N\log N)$, which is superior to
the Green function method using direct summation.


\begin{table*}
\begin{center}
\caption{Dynamical Model Parameters}
\label{tbl:models}
\begin{tabular}{llrrrrrrrrrrc}
\tableline\tableline
model &&
$\Sigma$&
$\Omega$&
$t_{\rm orb}$&
$t_{\rm final}$ &
$\rho_{\ast}$&
$L_z$&
$\bar{n}$&
$K/K_0$&
$\Delta z$& 
$\epsilon$ & 
resolution \\
\tableline
 &   & [a] & [b] & [c] & [d] & [e] & [f] & [g] & [h] & [i] & [j] &
 $N_x\times N_z$ \\ 
\tableline
QQQ8 &$\cdots$ &  7.50 &  22.1 & 2.79 & 5.57  & 0.035 & 410 & 0.525
 & 0.1   & 1.07 & 0.2 & $1536\times 384$ \\
Q8 &$\cdots$ &  $\vdots$ & $\vdots$ & $\vdots$ & $\vdots$ & $\vdots$ & $\vdots$ & $\vdots$
 & 0.4   & 2.13 & 0.2 & $768\times 192$ \\
Q8e0.02
&$\cdots$&$\vdots$&$\vdots$&$\vdots$&$\vdots$&$\vdots$&$\vdots$&$\vdots$
 &$\vdots$&$\vdots$&0.02 & $\vdots$ \\
QQQ11
&$\cdots$    & 10.60 &  31.2 & 1.97 & 3.94  & 0.07  & 290 & 1.05
 & 0.056    & 0.75 & 0.2 & $1536\times 384$ \\
Q11
&$\cdots$ & $\vdots$ & $\vdots$ & $\vdots$ & $\vdots$ & $\vdots$ & $\vdots$ & $\vdots$
 & 0.2    & 1.51 & 0.2 & $768\times 192$ \\
Q11e0.02
&$\cdots$&$\vdots$&$\vdots$&$\vdots$&$\vdots$&$\vdots$&$\vdots$&$\vdots$
 &$\vdots$&$\vdots$&0.02 & $\vdots$ \\
QQQ15&$\cdots$ & 15.00 &  44.12 & 1.39 & 2.79  & 0.14   & 205 & 2.1
 & 0.025    & 0.53 & 0.2 & $1536\times 384$ \\
Q15&$\cdots$ & $\vdots$ &  $\vdots$ & $\vdots$ & $\vdots$ & $\vdots$ & $\vdots$ & $\vdots$
 & 0.1    & 1.07 &0.2 & $768\times 192$ \\
Q15e0.02
&$\cdots$&$\vdots$&$\vdots$&$\vdots$&$\vdots$&$\vdots$&$\vdots$&$\vdots$
 &$\vdots$ & $\vdots$ & 0.02 & $\vdots$ \\
Q21&$\cdots$ & 21.20 &  62.40 & 0.99 & 1.97  & 0.28   & 145 & 4.2
 & 0.05   & 0.75 & 0.2 & $\vdots$ \\
Q21e0.02
&$\cdots$&$\vdots$&$\vdots$&$\vdots$&$\vdots$&$\vdots$&$\vdots$&$\vdots$
 &$\vdots$&$\vdots$&0.02 & $\vdots$ \\
Q42&$\cdots$ & 42.40 & 124.80 & 0.49 & 0.99  & 1.12   & 72.5 & 16.8
 & 0.0125 & 0.38 & 0.2 & $\vdots$ \\
Q42e0.02
&$\cdots$&$\vdots$&$\vdots$&$\vdots$&$\vdots$&$\vdots$&$\vdots$&$\vdots$
 &$\vdots$&$\vdots$&0.02 & $\vdots$ \\
\tableline
KKKK8 &$\cdots$ &  7.50 &  44.12 & 1.39 & 2.79  & 0.035 & 410 & 0.525
 & 0.056   & 0.80 & 0.2 & $2048\times 512$ \\
KKK11&$\cdots$ & 10.60 &$\vdots$&$\vdots$& $\vdots$& 0.07  & 290 & 1.05
 & 0.05   & 0.75 & $\vdots$ & $1536\times 384$ \\
K21&$\cdots$ & 21.20 &$\vdots$&$\vdots$& $\vdots$& 0.28   & 145 & 4.2
 & 0.05   & 0.75 &$\vdots$& $768\times 192$ \\ 
K42&$\cdots$ & 42.40 &$\vdots$&$\vdots$& $\vdots$& 1.12   & 72.5 & 16.8
 & 0.0125 & 0.38 &$\vdots$ & $\vdots$ \\
\tableline
RR8 &$\cdots$ &  7.50 &  22.1 & 2.79 & 5.57 & 0.14 & 205 & 1.05 &
 0.056 & 0.79 & 0.2 & $1024\times 256$ \\
RR11&$\cdots$ & 10.60 &  31.2 & 1.97 & 3.94 &$\vdots$&$\vdots$&
 1.49 & $\vdots$ & $\vdots$ & $\vdots$& $\vdots$ \\ 
R21&$\cdots$ & 21.20 &  62.40 & 0.99 & 1.97 &$\vdots$&$\vdots$&
 2.98 & 0.1 & 1.07 &$\vdots$ & $768\times 192$ \\ 
R42&$\cdots$ & 42.40 & 124.80 & 0.49 & 0.99 &$\vdots$&$\vdots$&
 5.97 &$\vdots$&$\vdots$&$\vdots$& $\vdots$\\ 
\tableline
SS22 &$\cdots$ & 15.00  &  22.1 & 2.79 & 5.57 & 0.14 & 205 & 2.1 &
 0.056 & 0.79 & 0.2 & $1024\times 256$ \\
SS31 &$\cdots$ &$\vdots$&  31.2 &
 2.30 &$\vdots$&$\vdots$&$\vdots$&$\vdots$&$\vdots$ & $\vdots$&$\vdots$ & $\vdots$ \\ 
S44 &$\cdots$ &$\vdots$&  44.12 &
 1.18 &$\vdots$&$\vdots$&$\vdots$&$\vdots$& 0.1 & 1.07 &$\vdots$ &
 $768\times 192$ \\ 
\tableline
\end{tabular}
\tablenotetext{}{[a] Gas surface density, in $\msun \pc^{-2}$}
\tablenotetext{}{[b] Orbital angular velocity, in $\rm km~s^{-1}~kpc^{-1}$}
\tablenotetext{}{[c] Orbital time, $t_{\rm orb}=2\pi/\Omega$, in units
  $10^8$ yr}
\tablenotetext{}{[d] Duration of simulation, in units $10^8$ yr}
\tablenotetext{}{[e] Stellar density at midplane, in $\msun \pc^{-3}$}
\tablenotetext{}{[f] Vertical size of domain, in pc. Horizontal size
  is $L_x=4L_z$.}
\tablenotetext{}{[g] Average number density, $\bar n= \Sigma/(\mu L_z)$,
  in cm$^{-3}$  }
\tablenotetext{}{[h] Thermal conduction in units of 
$K_0=6.91\times 10^{7}$ erg cm$^{-1}$ s$^{-1}$ K$^{-1}$}
\tablenotetext{}{[i] Grid scale, in pc }
\tablenotetext{}{[j] Control parameter for photoheating region
(see eq.\ref{eq:HII_eps})}
\end{center}
\end{table*}

\begin{center}
\begin{table*}
\caption{Hydrostatic Model Parameters}
\label{tbl:HS}
\begin{tabular}{llrrrrrr}
\tableline\tableline
model &&
$\Sigma$&
$\rho_{\ast}$&
$L_z$&
$\bar{n}$&
$K/K_0$&
$\Delta z$\\
\tableline
HSP8 &$\cdots$ &  7.50 &  0.035 & 410  & 0.525 & 0.4   & 0.40 \\
HSP11&$\cdots$ & 10.60 &  0.07  & 290  & 1.05 & 0.2    & 0.28 \\
HSP15&$\cdots$ & 15.00 &  0.14  & 205  & 2.1  & 0.1    & 0.20 \\
HSP21&$\cdots$ & 21.20 &  0.28  & 145  & 4.2  & 0.05   & 0.14 \\
HSP42&$\cdots$ & 42.40 &  1.12  & 72.5 & 16.8 & 0.0125 & 0.070 \\
\tableline
HSC8 &$\cdots$ &  7.50 &   0.14 &   353  & 1.05 &   0.1  & 0.20 \\
HSC11&$\cdots$ & 10.60 &$\vdots$&$\vdots$& 1.48 &$\vdots$&$\vdots$ \\
HSC21&$\cdots$ & 21.20 &$\vdots$&$\vdots$& 2.97 &$\vdots$&$\vdots$ \\
HSC42&$\cdots$ & 42.40 &$\vdots$&$\vdots$& 5.93 &$\vdots$&$\vdots$ \\
\tableline
\end{tabular}
\tablecomments{All parameters and units are as in Table \ref{tbl:models}.}
\end{table*}
\end{center}

\clearpage

\clearpage

\clearpage

\clearpage

\clearpage

\clearpage

\clearpage

\clearpage

\clearpage

\clearpage

\clearpage

\clearpage

\clearpage

\clearpage

\clearpage

\clearpage

\clearpage

\clearpage

\clearpage

\clearpage

\end{document}